\documentclass[twocolumn]{aa}  

    \usepackage{graphicx}
    \usepackage{txfonts}
    \usepackage[colorlinks = true, allcolors=blue]{hyperref}
    \usepackage[normalem]{ulem}
    \usepackage{booktabs}
    \newcommand\arcdeg{\mbox{$^\circ$}}%
    \newcommand\farcdeg{\mbox{$.\!\!^\circ$}}%
    \newcommand{\kms}{km s$^{-1}$}%
    \newcommand{\Msun}{\hbox{M$_\odot$}}
\newcommand{\orcidlink}[1]{\protect\href{https://orcid.org/#1}{\protect\includegraphics[width=8pt]{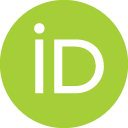}}}    
    
\begin{document}

       \title{The spin, expansion and contraction of open star clusters}
       % \subtitle{}
       \titlerunning{Open Cluster Spin and Expansion}
       % \authorrunning{Jadhav}
       
       \author{Vikrant V. Jadhav\orcidlink{0000-0002-8672-3300}\inst{1},
              Pavel Kroupa\orcidlink{0000-0002-7301-3377}\inst{1,2},
              Wenjie Wu\inst{1}, Jan Pflamm-Altenburg\inst{1},
              Ingo Thies\inst{1}}
    
       \institute{
            Helmholtz-Institut für Strahlen- und Kernphysik, Universität Bonn, Nussallee 14-16, D-53115 Bonn, Germany\\
            \email{vjadhav@astro.uni-bonn.de}
            \and
            Astronomical Institute, Faculty of Mathematics and Physics, Charles University, V Holešovičkách 2, CZ-180 00 Praha 8, Czech Republic
            }
    
       \date{Received January 13, 2023; accepted June 27, 2023}
    
    % \abstract{}{}{}{}{} 
    % 5 {} token are mandatory
     
      \abstract
      % context heading (optional)
       {Empirical constraints on the internal dynamics of open clusters are important for understanding their evolution and evaporation. High precision astrometry from \textit{Gaia} DR3 are thus useful to observe aspects of the cluster dynamics.
       }
      % aims heading (mandatory)
       {This work aims to identify dynamically peculiar clusters such as spinning and expanding clusters. We also quantify the spin frequency and expansion rate and compare them with $N$-body models to identify the origins of the peculiarities. 
       }
      % methods heading (mandatory)
       {We used the latest \textit{Gaia} DR3 and archival spectroscopic surveys to analyse the radial velocities and proper motions of the cluster members in 1379 open clusters. A systematic analysis of synthetic clusters is performed to demonstrate the observability of the cluster spin along with effects of observational uncertainties. $N$-body simulations were used to understand the evolution of cluster spin and expansion for initially non-rotating clusters.
       }
      % results heading (mandatory)
       {We identified spin signatures in 10 clusters (and 16 candidates). Additionally, we detected expansion in 18 clusters and contraction in 3 clusters. The expansion rate is compatible with previous theoretical estimates based on expulsion of residual gas. The orientation of the spin axis is independent of the orbital angular momentum.
       }
      % conclusions heading (optional)
       {
       The spin frequencies are much larger than what is expected from simulated initially non-rotating clusters. This indicates that $>1$\% of the clusters are born rotating and/or they have undergone strong interactions. 
       Higher precision observations are required to increase the sample of such dynamically peculiar clusters and to characterise them.
       }
    
    
       \maketitle
    %
    %-------------------------------------------------------------------

\section{Introduction} \label{sec:introduction}
% Abbreviations: OC, RV, PM
% regex: \b(?:[A-Z]){2,}

Almost all stars begin life in an embedded cluster within a giant molecular cloud \citep{Kroupa1995MNRAS.277.1491K, Kroupa1995MNRAS.277.1507K, Lada2003ARA&A..41...57L, Dinnbier2022A&A...660A..61D}. The dynamics of the contracting embedded cloud core should be imprinted on the resultant stellar population, particularly on a surviving star cluster. One of the understudied dynamical properties in star clusters is their spin.
Rotation signatures have been observed in molecular clouds and embedded clusters \citep{Kutner1977ApJ...215..521K, Rosolowsky2003ApJ...599..258R, Henault2012A&A...545L...1H, Chen2019ApJ...886..119C}. Similarly, spin has been detected in globular clusters \citep{Anderson2003AJ....126..772A, van2006A&A...445..513V, Bellini2017ApJ...844..167B, Bianchini2018MNRAS.481.2125B, Kamann2018MNRAS.473.5591K, Sollima2019MNRAS.485.1460S, Vasiliev2021MNRAS.505.5978V, Szigeti2021MNRAS.504.1144S}. These studies used proper motions (PMs), radial velocities (RVs) and their combinations to measure the spin orientation.

In this work, we investigate the presence of spin and its possible origin in open clusters (OCs). In contrast to globular clusters, there have been only a few attempts to identify spin in OCs (\citealt{Kuhn2019ApJ...870...32K}; \citealt{Loktin2020AN....341..638L}\footnote{The method used to detect rotation in Praesepe is erroneous and is affected by the projection effects due to the solar motion}; \citealt{Guilherme2023A&A...673A.128G}).
This has been challenging because the smaller population in OCs leads to poor statistics and the unavailability of accurate PMs and RVs for a significant number of cluster members. 
The PMs of the cluster also inhabit signatures of expansion \citep{Guilherme2023A&A...673A.128G}, which is discussed in previous theoretical and observational work \citep{Tutukov1978A&A....70...57T, Kroupa2001MNRAS.321..699K, Pfalzner2013A&A...559A..38P, Dinnbier2020A&A...640A..84D, Dinnbier2020A&A...640A..85D}.
The \textit{Gaia} mission \citep{Gaia2016A&A...595A...1G} has significantly increased the availability and precision of PMs. In addition, \textit{Gaia} DR3 \citep{Gaia2022arXiv220800211G} has provided the latest and largest catalogue of RVs. The six-dimensional coordinates are required to characterise the internal dynamics of clusters.

The star clusters could be spinning due to various reasons: 
i) The parent molecular cloud may have primordial spin \citep{Braine2020A&A...633A..17B}. 
ii) An anisotropy during the cloud collapse/fragmentation. 
iii) Any asymmetry in the number and direction of supernovae kicks from the early type stars \citep{Hobbs2005MNRAS.360..974H}. 
iv) The axisymmetric nature of the Galactic potential through which all OCs orbit. 
v) The asymmetric stellar evaporation evident through asymmetric tidal tails of the clusters \citep{Jerabkova2021A&A...647A.137J, Kroupa2022MNRAS.517.3613K}. 
vi) A collisional/flyby interaction with a massive object \citep{Piatti2022MNRAS.511L...1P}.
vii) Disc shocking \citep{Moreno2014ApJ...793..110M} and dynamical friction \citep{Moreno2022MNRAS.510.5945M}.
viii) Dynamical effects of the bar and spiral arms, and interaction with bar and/or spiral arms resonances \citep{Moreno2021MNRAS.506.4687M}.
Any one or more of the above phenomena could lead to a spinning cluster. 

This work aims to detect and quantify the rotation signatures in OCs. For a given spin axis with the spin frequency of $\nu$, we can use the RVs to calculate the spin component in the sky plane ($\nu$ cos$i$) and the PMs to measure the spin component along the line of sight ($\nu$ sin$i$). We also use $N$-body simulations to test the origin and impacts of various factors on clusters' spin.
\S~\ref{sec:data} introduces the \textit{Gaia} data and the methods used to simulate and detect cluster spin. \S~\ref{sec:results} and \S~\ref{sec:discussion} describe and discuss the results, respectively. And \S~\ref{sec:conclusions} gives the conclusions.

\section{Data and analysis} \label{sec:data}

\subsection{\textit{Gaia} data}

\citet{Hunt2023A&A...673A.114H} produced an OC catalogue based on a blind all-sky search of the \textit{Gaia} DR3 catalogue \citep{Gaia2022arXiv220800211G}. 
The high-quality clusters are selected with the following criteria: cluster detection significance ($CST \geq 5$), results of CMD-based classifier ($CMDClass\_50 \geq 0.5$), reliable RV measurement ($n\_RV \geq 10$), and the type of the stellar group ($Type = o$, where $o$ is an OC). Among the 4105 reliable clusters in the catalogue, we selected OCs with at least 10 stars with RV measurements, resulting in 1379 OCs.
The \citet{Hunt2023A&A...673A.114H} catalogue also provides a list of cluster members, including candidates outside the tidal radius and in the tidal tails. Only the stars within the tidal radius and with the probability of $\geq$0.5 are considered for the numerical analysis.

The \textit{Gaia} DR3 parallaxes are representative of the distance, but just inverting the parallax to get distance is non-trivial in case of large uncertainties and negative parallax. Hence, we used the distances calculated with a probabilistic approach by \citet{Bailer2021AJ....161..147B} ($r\_med\_geo$) instead of \textit{Gaia} DR3 parallaxes. We only considered the stars with $3\sigma$ confidence in their distance for further analysis.
The PM and RV for cluster members were corrected for projection effects due to the solar motion (\S~\ref{sec:projection_correction}). Table~\ref{tab:param_definitions} gives the definitions and formulae for various observed and derived parameters.

\subsection{Radial velocity data}
The \textit{Gaia} DR3 provides RVs for most stars brighter than 14.5 Gmag. However, the \textit{Gaia} DR3 RVs are limited by identification of spectral lines within the narrow neighbourhood of the Calcium triplet (845--872 nm). To compliment the \textit{Gaia} RVs, we included archival RV data from Survey of Surveys (SoS; \citealt{Tsantaki2022A&A...659A..95T}). SoS is a homogeneously merged catalogue of RVs combining \textit{Gaia} DR2 \citep{Gaia2018A&A...616A...1G}, APOGEE DR16 \citep{Ahumada2020ApJS..249....3A}, GALAH DR2 \citep{Zwitter2018MNRAS.481..645Z}, \textit{Gaia}-ESO DR3 \citep{Gilmore2012Msngr.147...25G}, RAVE DR6 \citep{Steinmetz2020AJ....160...82S} and LAMOST DR5 \citep{Deng2012RAA....12..735D}. \citet{Tsantaki2022A&A...659A..95T} used \textit{Gaia} DR2 as the reference for homogenising the RVs and errors. We have ignored all \textit{Gaia} DR2 RVs given in the SoS catalogue as they are superseded by \textit{Gaia} DR3.

For the RV based analysis (\S~\ref{sec:spin_using_rv}), we created a subset of OCs with at least 100 stars having \textit{Gaia} DR3 RVs. There are 19753 stars with \textit{Gaia} DR3 RVs in 106 clusters. We crossmatched all these members with SoS and found additional 6929 stars with RVs obtained using non-\textit{Gaia} surveys. For the RV analysis, we removed any RV outliers using 3$\sigma$ clipping.

\begin{figure*}
    \centering
    \includegraphics[width=0.98\textwidth]{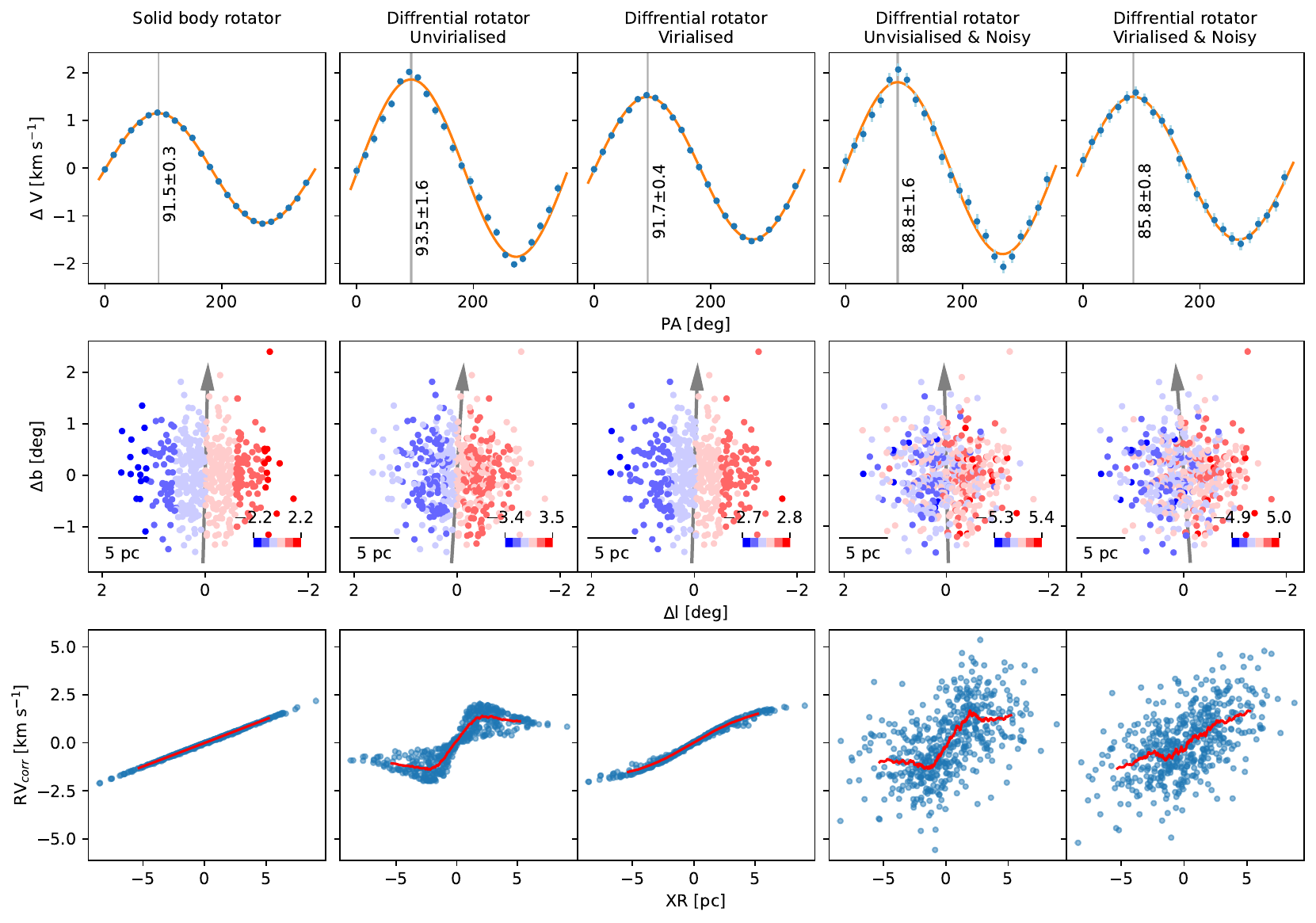}
    \caption{Spin detection using RV. All these clusters have 90\arcdeg\ inclination.
    \textit{The first row}: Variation of the $\Delta V$ across various $PA$. The fitted sin curve is shown in orange, while the grey bar shows the $PA_{peak}$. 
    \textit{The second row}: The spatial positions ($\Delta l$,$\Delta b$) of the stars coloured according to their RV. The identified spin axis (corresponding to $PA_{peak}$) is shown in grey. 
    \textit{The third row}: The variation of $RV_{corr}$ with $XR$. The red curve shows the rolling average of the $RV_{corr}$ values.}
    \label{fig:demo_rv}
\end{figure*}

\subsection{Projection effect correction} \label{sec:projection_correction}
The PM and RVs of individual stars are affected by the projection effect due to relative motion and position between the cluster and the Sun. Not correcting the projection artefacts could lead to false rotation and contraction signatures in nearby star clusters.
The PM ($\mu_{\alpha*,i},\mu_{\delta*,i}$) and RV ($RV_{i}$) resulting from the projection effect of the cluster motion can be calculated as follows \citep{van2009A&A...497..209V}:

\begin{equation}
\begin{split}
RV_{i} &\approx 
RV_{0}+\dfrac{\kappa}{\varpi_0}
\left(\cos\delta_0\,\Delta\alpha_i\,\mu_{\alpha*,0}+\Delta\delta_i\,\mu_{\delta,0}\right)\\
\mu_{\alpha*,i} &\approx 
\frac{\varpi_i}{\varpi_0}\left(\mu_{\alpha*,0}-\dfrac{\Delta\alpha_i\,\cos\delta_0\,RV_{0}\,\varpi_0}{\kappa}+
\Delta\alpha_i\,\sin\delta_0\,\mu_{\delta,0}
\right)\\
\mu_{\delta,i} &\approx 
\frac{\varpi_i}{\varpi_0}\left(\mu_{\delta,0}-\dfrac{\Delta\delta_i\,RV_{0}\,\varpi_0}{\kappa} - \Delta\alpha_i\,\sin\delta_0\,\mu_{\alpha*,0}\right)\\
\end{split}
\end{equation}
Here, 
$RV_{0}$ is the cluster RV, 
$\varpi_0$ is its parallax,
$\varpi_i$ the stellar parallax,
$\delta_0$ the cluster declination,
($\mu_{\alpha*,0}$, $\mu_{\delta,0}$) the cluster's PM,
($\Delta\alpha_i$, $\Delta\delta_i$) the stellar angular separation from the cluster centre in radians and
$\kappa$ is 4.74047 (the conversion factor for 1 mas yr$^{-1}$ at 1 kpc to 1 \kms).

The PM ($\mu_{\alpha*, corr}$, $\mu_{\delta, corr}$) and RV ($RV_{corr}$) corrected for the bulk cluster motion and projection effect are as follows:
\begin{equation}
    \begin{split}
        RV_{corr} & \approx RV - RV_i\\
        \mu_{\alpha*, corr} & \approx \mu_{\alpha*} - \mu_{\alpha*,i}\\
        \mu_{\delta, corr} & \approx \mu_{\delta} - \mu_{\delta,i}\\
    \end{split}
\end{equation}
where,
$RV$ is the stellar RV and
($\mu_{\alpha*}$, $\mu_{\delta}$) is the stellar PM.

\begin{figure*}
    \centering
    \includegraphics[width=0.98\textwidth]{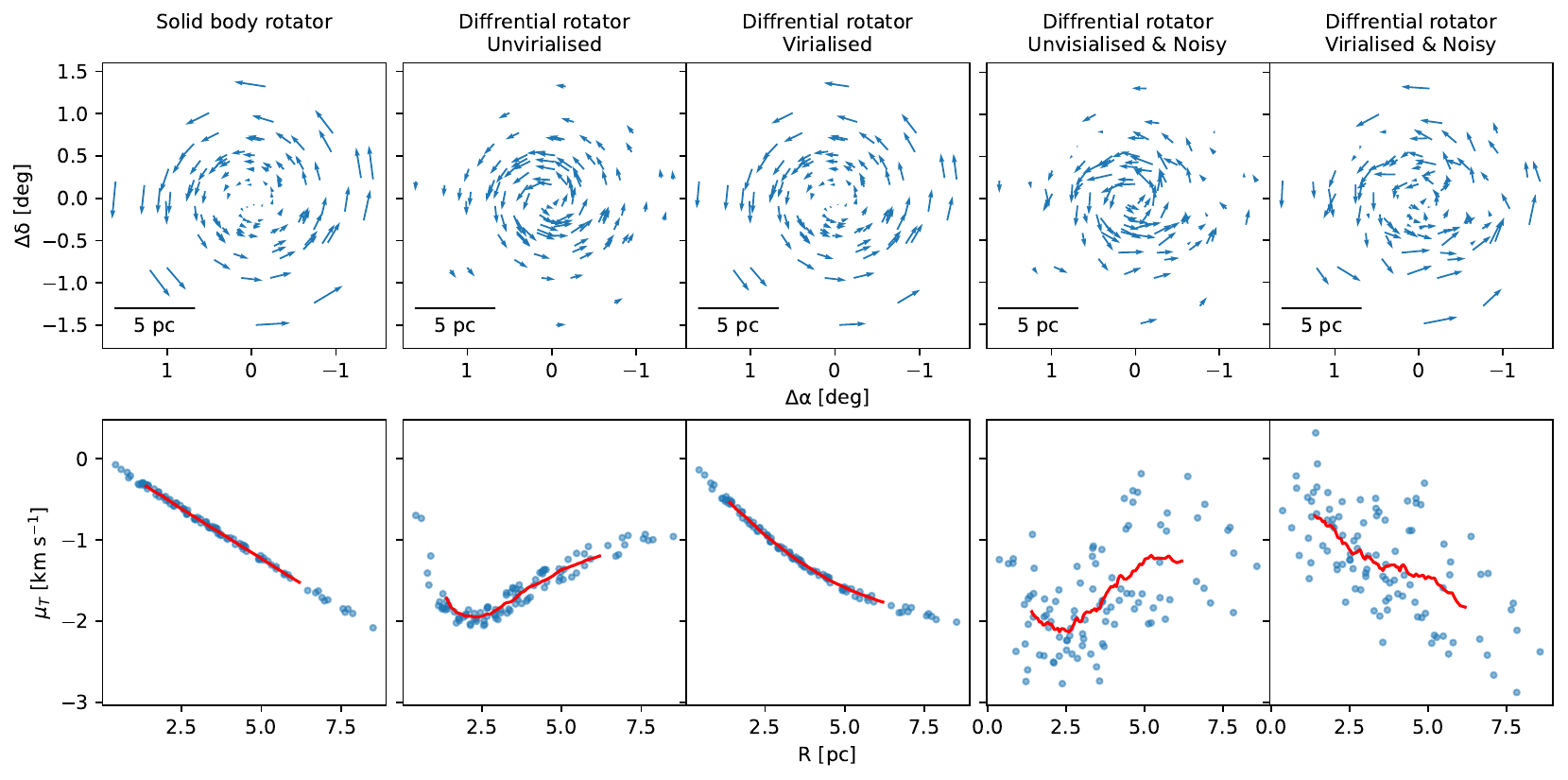}
    \caption{Spin detection using PM. All these clusters have 0\arcdeg\ inclination. \textit{The first row}: The spatial positions ($\Delta \alpha$,$\Delta \beta$) of the stars, along with the arrows indicating the PM of each star. \textit{The second row}: Variation of the $\mu_{T}$ with the radius. The red curve shows the rolling average of the $\mu_{T}$ values.}
    \label{fig:demo_pm}
\end{figure*}

\subsection{Synthetic clusters}

\subsubsection{Demo clusters}
We created demo clusters to visualise the various observational aspects of cluster spin and the effects of observational errors on these observables. 

The simplest synthetic cluster is a solid body rotator with a token rotational frequency of 40 cycles per gigayear (cy Gyr$^{-1}$). These synthetic clusters have stars distributed in a sphere where the positions follow random normal distributions with $\sigma=3$ pc. The initial rotation axis is assumed to be along the Z direction, which could be changed later depending on the required inclination.

However, real stellar systems have differential rotation. Hence, we created realistic cylindrically-rotating clusters with the following velocity law \citep{Lynden1967MNRAS.136..101L, Lanzoni2018ApJ...861...16L}:
\begin{equation}
V_{rot} = \frac{2 A_{peak}}{XR_{peak}} \frac{XR}{1+XR^2/XR_{peak}^2}  
\end{equation}
where $XR$ is the projected distance measured from the spin axis, 
$XR_{peak}$ corresponds to the rough radius up to which the cluster is virialised,
$A_{peak}$ is the maximum amplitude of the RV.
Typical $A_{peak}$ values observed in globular clusters are 1--3 \kms, while $XR_{peak}$ values are 2--5 pc \citep{Mackey2013ApJ...762...65M, Kacharov2014A&A...567A..69K, Lanzoni2018ApJ...861...16L}. Unfortunately, no such parameters are available for OCs; hence, we have assumed that the $A_{peak}$ of the demo clusters are similar to globular clusters. As some OCs may be virialised systems, we have used two values of $R_{peak}$: 2 pc for an unvirialised cluster and 10 pc for a virialised cluster. $R_{peak}$ of 2 pc is based on globular clusters and is useful to demonstrate the rotation curve of a partially virialised cluster. $R_{peak}$ of 10 pc means most of the OC is virialised and the rotation curve can be approximated to a linear trend (see bottom row of Fig.~\ref{fig:demo_rv}).

Optionally, we also add noise in the data to simulate the observables as seen through the \textit{Gaia} telescope. The details of the used synthetic clusters and the added noise are given in Table~\ref{tab:demo_clusters}.

\subsubsection{$N$-body simulated clusters}
The demo clusters are non-physical due to the artificial assignment of velocities and positions. Hence, we also selected a variety of $N$-body simulations to understand the presence of spin in realistic clusters.

\begin{itemize}
    \item \textit{Kroupa+2022:} We used archival $N$-body simulations of star clusters from \citet{Kroupa2022MNRAS.517.3613K}. They calculated the dynamical evolution of a grid of star clusters with total initial masses ranging from 2500\,$M_\odot$ to 20000\,$M_\odot$ with 0.1\Msun\ stellar particles on a circular orbit with a Galactocentric distance of 8.3\,kpc using the \textsc{Phantom of Ramses} code \citep{lueghausen2015a}. In this code subroutines are added to the \textsc{Ramses} code \citep{Teyssier2002A&A...385..337T} in order to solve the MoNDian gravitational field equation in the QUMOND formulation \citep{milgrom2010a}. As a collision-less field method requires a sufficient particle density in order to calculate a reliable potential, the 2500\,$M_\odot$ mass model already shows artificial effects of limited resolution.
    
    \item \textit{PeTar:} We simulated clusters with 5000 particles in Newtonian dynamics and using the direct $N$-body code \textsc{PeTar} \citep{Wang2020MNRAS.497..536W}. We created two simulations, first with no primordial binaries and second with 100\% primordial binaries.
    For both models, the initial mass spectra are sampled from the \citet{Kroupa2001MNRAS.322..231K} canonical initial mass function and using the Plummer model with a half-mass radius $r_{\rm h}=3$ pc. We employ the Milky-Way potential as an external field \citep{Bovy2015ApJS..216...29B}. Both demo clusters are located at 8 kpc to the Galactic centre and with a circular orbit along the disk. We also enable stellar evolution by using the updated \textsc{sse/bse} code \citep{Hurley2000MNRAS.315..543H, Hurley2002MNRAS.329..897H, Banerjee2020A&A...639A..41B}, which is embedded in \textsc{PeTar}. The remnant mass is based on the rapid supernovae scenario from \citet{Fryer2012ApJ...749...91F}. The calculation also includes the pair-instability \citep{Belczynski2016A&A...594A..97B} and electron capture supernova \citep{Belczynski2008ApJS..174..223B}. We assume a solar metallicity $Z=0.02$ for both models \citep{von2016ApJ...816...13V}. Table \ref{tab:petarbseparameter} lists the relevant stellar evolution parameters and the used values for both calculations.
    
    \item \textit{Milgromian Law Dynamics (MLD):} As QUMOND \citep{milgrom2010a} and AQUAL \citep{bekenstein1984a} are non-linear field theoretical formulations of MoND, stellar dynamical codes for the evolution of discrete systems are very difficult to construct and currently not available. As a first step the standard Hermite scheme is extended to solve Milgrom's law \citep{Milgrom1983ApJ...270..365M} for a discrete $N$-body system (Pflamm-Altenburg, submitted to A\&A). The \textsc{MLD} code includes smoothing of the inter-particle forces in the Newtonian regime in order to suppress Newtonisation of centre of mass motions of close compact sub-systems. We simulated a Hyades-like cluster with 2000 particles with 0.5 \Msun\ each. This stellar system of 2000 particles requires a direct $N$-body solver because simulations performed with a particle-mesh code (for example PoR) would be strongly affected by resolution limitations.
\end{itemize}

\begin{table}[htbp]
    \centering
    \begin{tabular}{cccc}
    \toprule
Parameter	&	Value	&	Parameter	&	Value	\\ \midrule
\texttt{bse-alpha}	&	3	&	\texttt{bse-sigma}	&	265	\\
\texttt{bse-beta}	&	0.125	&	\texttt{bse-xi}	&	1	\\
\texttt{bse-bhwacc}	&	1.5	&	\texttt{bse-bhflag}	&	2	\\
\texttt{bse-bwind}	&	0	&	\texttt{bse-ceflag}	&	0	\\
\texttt{bse-eddfac}	&	1	&	\texttt{bse-ecflag}	&	1	\\
\texttt{bse-epsnov}	&	0.001	&	\texttt{bse-kmech}	&	1	\\
\texttt{bse-gamma}	&	-1	&	\texttt{bse-nsflag}	&	3	\\
\texttt{bse-hewind}	&	1	&	\texttt{bse-psflag}	&	1	\\
\texttt{bse-lambda}	&	0.5	&	\texttt{bse-tflag}	&	1	\\
\texttt{bse-metallicity}	&	0.02	&	\texttt{bse-wdflag}	&	1	\\
\texttt{bse-neta}	&	0.5	&		&		\\ \bottomrule

    \end{tabular}
    \caption{Relevant \textsc{sse/bse} parameters for simulations with \textsc{PeTar}.}
    \label{tab:petarbseparameter}
\end{table}

The six-dimensional astrometry ($\alpha, \delta, \varpi, \mu_{\alpha},\mu_{\delta},RV$) of the synthetic clusters is then used to perform synthetic observations and create diagnostic plots similar to real OCs as mentioned in \S~\ref{sec:results}. To avoid the contamination from tidal tails along the line of sight, we kept the position of the observer (Sun) along the line connecting the cluster centre to the Galactic centre. The overall results do not change due to using the corotating reference frame instead of the inertial frame.

\subsection{Spin detection using radial velocity} \label{sec:spin_using_rv}

As the Galactic potential in the solar neighbourhood is approximately axisymmetric around the Galactic Z axis, we use the Galactic coordinates ($l$, $b$) as the base coordinates for this particular step\footnote{The 0\farcdeg1 difference between the Galactic midplane and the $b=0$ plane does not affect the results of this study}. First, we projected the cluster on a tangential plane and measured relative Galactic coordinates ($\Delta l$, $\Delta b$). Then, we divided the cluster into 2 sectors for an arbitrary position angle ($PA$; as measured clockwise from the Galactic East direction, in the sky plane). The difference between the mean RV of the two halves ($\Delta V$) will increase if the assumed $PA$ is along the spin axis. We shifted the $PA$ from 0\arcdeg\ to 360\arcdeg\ in steps of 15\arcdeg. The $PA$--$\Delta V$ distribution is then fitted with a sin wave, which gives the $PA_{peak}$ with maximum $\Delta V$ and the $\Delta V$ amplitude ($\approx$ twice the velocity dispersion).

We used the same process to measure the best fit $PA$ for different annuli. A rotating cluster should have similar $PA$ for the whole cluster and the different annuli. The central region contains the most noisy data due to relatively higher velocity dispersion compared to the differential RV. Hence, we removed the central 25th percentile data for our final classification to bring out the rotation signature. 
To further analyse the cluster, we created a new coordinate system ($XR$,$YR$) where $YR$ is along the $PA$ vector while $XR$ is perpendicular to it.

The first row in Figure~\ref{fig:demo_rv} shows examples of how the $PA$--$XR$ plots for synthetic clusters. All the demo clusters have the rotation axis along the Galactocetric Z axis (parallel to b). The maximum of the sin curve gives $PA_{peak}$, shown as an arrow in the second row, along with the RV distribution along the sky.
The third row shows the resulting $XR$--$RV_{corr}$ variation. This relation is linear for a solid body rotator and roughly linear for a virialised cluster. 
The last two columns of Figure~\ref{fig:demo_rv} show the effect of adding noise to the data due to internal velocity dispersion and \textit{Gaia}-like observational errors.
The observed $XR$--$RV_{corr}$ curves in these noisy demo clusters only retain a rough increasing trend, where the differences due to possible virialisation are hidden due to the noise.

\subsection{Spin detection using proper motion} \label{sec:spin_using_pm}
The first Figure~\ref{fig:demo_pm} shows the views of rotating clusters as seen along the spin axis. 
The tangential components of PM ($\mu_T$) contain the rotation signature of the cluster. The $\mu_T$ will increase with projected radius ($R$) if the cluster rotates (assuming the spin axis is along the light of sight).

In an ideal case with a solid body rotation, one expects a linear distribution in the $\mu_T$ and $R$ as seen in the first column of Figure~\ref{fig:demo_pm}. However, in the case of a differential rotator, the rotation speed varies across the radius. Thus, one would expect an increase in $\mu_T$ till $R \approx R_{peak}$ and then a slight dip. However, as shown by the last column of Figure~\ref{fig:demo_pm}, a virialised OC should have a linear trend in the $R$--$\mu_T$ plot. 

\subsection{Expansion/contraction using proper motion}

The radial component of PM ($\mu_R$) shows the radial motion of the stars. The $\mu_R$ should not change with radius in a stable cluster. However, in an expanding cluster, $\mu_R$ will increase with the radius and in a contracting cluster, $\mu_R$ will decrease with the radius. To identify the possible expansion/contraction, we created $R$--$\mu_R$ plot for each cluster. 
The slope of the $R$--$\mu_R$ distribution gives the expansion rate of the cluster.

\subsection{Cluster orbits}
We used the cluster position and motion to calculate the cluster's orbit around the Milky Way using \textsc{astropy} \citep{astropy:2013,astropy:2018,astropy:2022} and \textsc{galpy} \citep{Bovy2015ApJS..216...29B}. We have assumed the \texttt{MWPotential2014} Milky Way potential, solar position ($-$8000, 0, 15 pc) and velocity (10, 235, 7 \kms) \citep{Bovy2015ApJS..216...29B}. The cluster orbits plotted in diagnostic plots are of arbitrary time used for analytical assistance.

\subsection{Statistical tests and effect of observational errors} \label{sec:statistical_tests}

\begin{figure}
    \centering
    \includegraphics[width=0.48\textwidth]{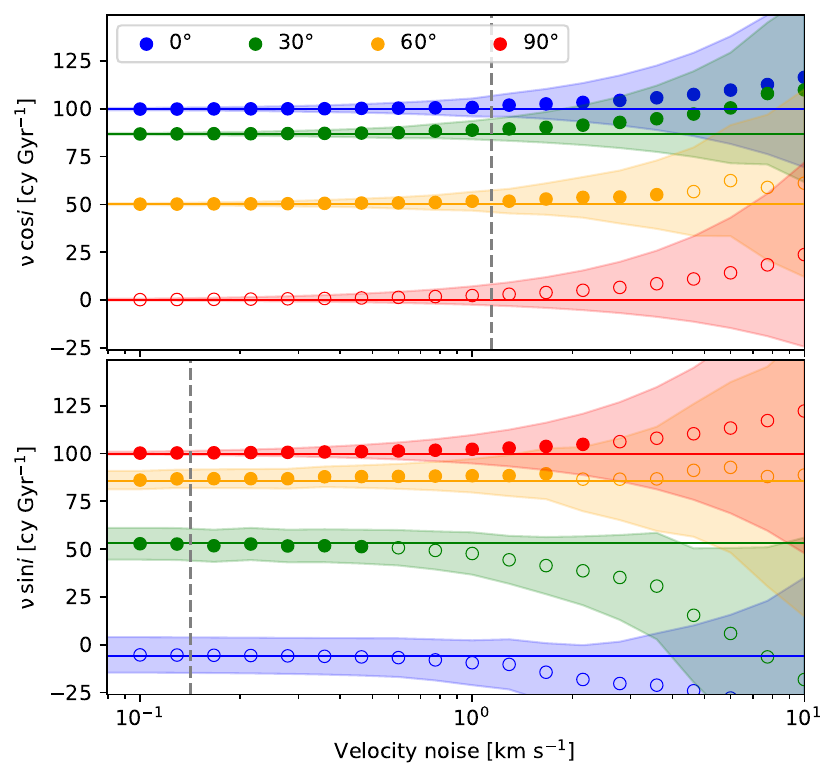}
    \caption{Effect of observed velocity error on the measurement of $\nu\,$cos$i$ and $\nu\,$sin$i$ on a solid body rotator with $\nu$ = 100 cy Gyr$^{-1}$. The statistically significant results (see \S\ref{sec:statistical_tests}) are filled circles, and the corresponding 1$\sigma$ errors are the shaded areas. The horizontal lines show the measurements with no error. The typical errors in \textit{Gaia} RV (1.3 \kms) and PM (0.02 \kms) are shown as dashed grey lines.}
    \label{fig:error_effect}
\end{figure}

\begin{figure*}
    \centering
    \includegraphics[width=0.98\textwidth]{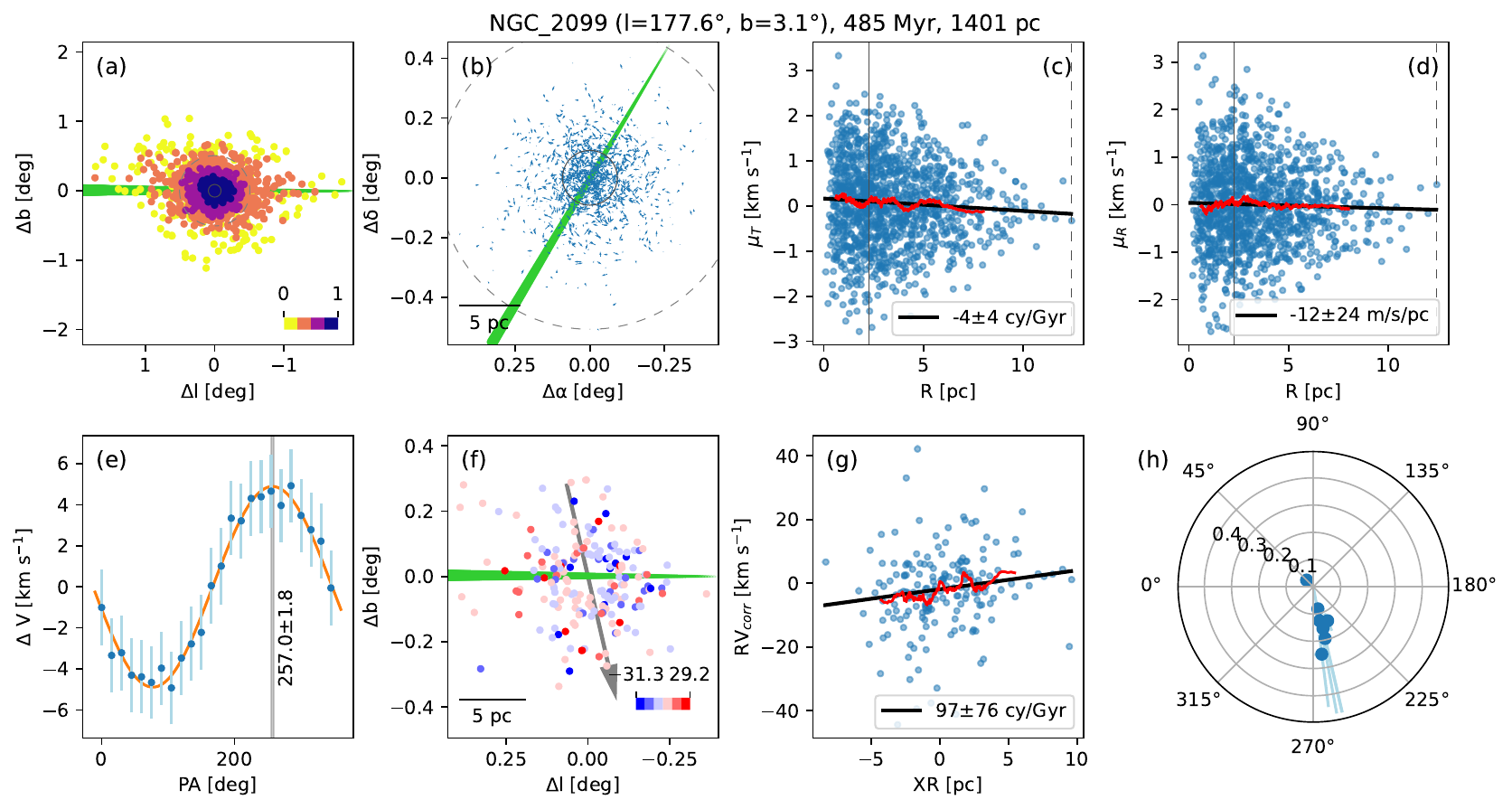}
    \caption{Diagnostic plots for NGC 2099. 
    (a) Spatial distribution of cluster members and candidates. The stars are coloured according to the membership probability.
    (b) The spatial distribution of stars along with arrows indicating their PM. 
    (c) Variation of $\mu_T$ with radius. 
    (d) Variation of $\mu_R$ with radius. 
    (e) Variation of $\Delta V$ with $PA$. The $PA_{peak}$ is shown as a grey band, and the fitted sin curve is orange.
    (f) Spatial distribution of stars coloured according to $RV_{corr}$. The orbital axis corresponding to $PA_{peak}$ is shown as the grey arrow.
    (g) Distribution of $RV_{corr}$ with $XR$. 
    (h) Variation of $PA_{peak}$ for different radial slices of the cluster. The radial distance used here is the average radius of the stars within the selection, and the radial error bars show the minimum--maximum radius within the radial slice. 
    The green wedge points towards the cluster's orbital motion in (a), (b) and (f). The red curves show the rolling average in (c), (d) and (g). The black lines show the linear fits in (c), (d) and (g).
    The core radii (grey solid circle/line) and tidal radii (grey dashed circle/line) are shown in (a), (b) and (c).}
    \label{fig:comb_2099}
\end{figure*} %comb_2099

\begin{sidewaysfigure*}
    \centering
    \includegraphics[width=0.95\textwidth]{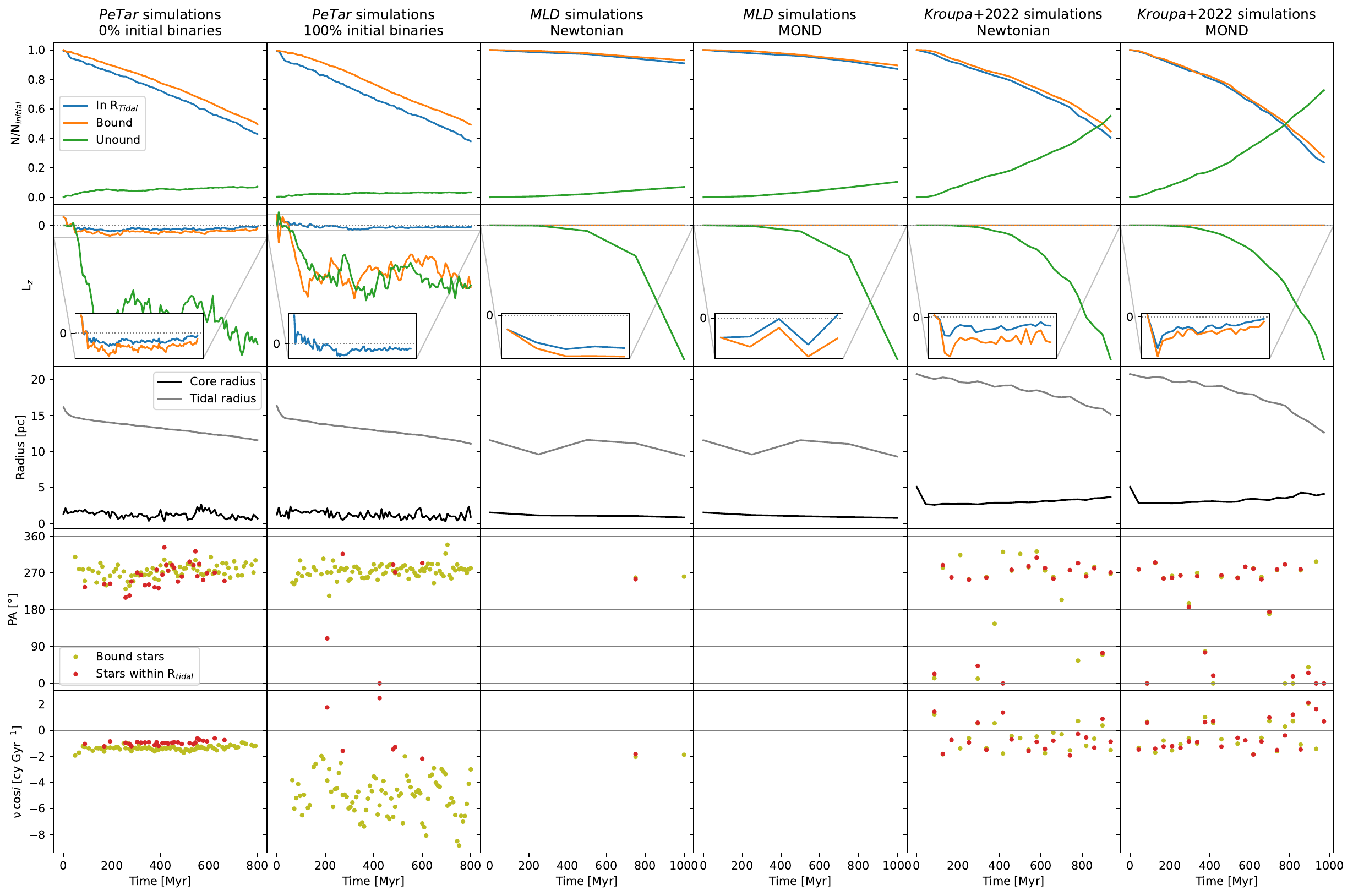}
    \caption{Evolution of the properties of simulated clusters. \textit{First row:} The evolution of the number of stars which are bound (orange), unbound (green) and are within the tidal radius (blue). Here, the bound/unbound means the sum of the kinetic and potential energy of the particle is negative/positive. The numbers are normalised with respect to the total population: 5000, 5000, 2000, 2000, 50000, and 50000 for the columns from left to right, respectively.
    \textit{Second row:} Spin angular momentum in the centre of density frame ($\mathcal{L}_z$). The $\mathcal{L}_z$ values are arbitrarily scaled for visual clarity. A negative value means the cluster is spinning in the same direction as it's orbit i.e. orbital and spin vectors are aligned.
    The colours are the same as the first row. \textit{Third row:} Core (black) and tidal (grey) radius. \textit{Fourth row:} PA (90$\equiv$counter-rotation, 270$\equiv$aligned rotation). The PA measurement for bound stars is shown in olive, while the stars within the tidal radius are shown in red. \textit{Fifth row:} Spin frequency, $\nu\,$cos$i$, for the corresponding PA.}
    \label{fig:properties_simulation}
\end{sidewaysfigure*} %properties_simulation

The $XR$--$RV_{corr}$, $R$--$\mu_T$ and $R$--$\mu_R$ distributions are fitted with a linear function to calculate the RV based rotation, PM based rotation and expansion, respectively. Before fitting the line, we removed the 3$\sigma$ outliers. The observational uncertainties in astrometric parameters were propagated while measuring the slope and it's uncertainty. The possible covariances between the astrometric parameters were ignored.
The measurements of $PA$ requires averaged velocities of two halves of the cluster. The uncertainties used while finding the $PA_{peak}$ are Poissonian in nature based on the number of stars in each half (roughly $>$50 stars for the half-cluster and $>$12 for each radial half-slice mentioned in \S~\ref{sec:spin_using_rv} and Fig.~\ref{fig:roi}). These Poisson errors are used while fitting the sin curve and the uncertainty in the $PA_{peak}$.

The RV analysis gives the $XR$--$RV_{corr}$ distribution. We performed the KS test to check whether the two populations in the two halves of the cluster are significantly different ($p<0.05$). We also fitted a line to this distribution, whose slope gives the $\nu\,$cos$i$ component of the spin axis in ideal solid body rotation. The line fitting included outlier rejection and accounted for parameter uncertainties.
We consider the $\nu\,$cos$i$ measurement to be statistically significant if it passes the KS test and the fitted measurement of $\nu\,$cos$i$ is better than 1$\sigma$.

For the PM analysis, a Spearman rank-order correlation test is applied to the $R$--$\mu_T$ distribution to identify any monotonicity. Due to the noisy data, we used a Monte Carlo approach to create multiple realisations with values shifted with a Gaussian noise proportional to the errors. The median \textit{statistic} ($\geq$0.4) and \textit{p} ($\leq$0.05) of the resulting Spearman tests are used to identify monotonic trends in the distributions. The slope of the $R$--$\mu_T$ plot gives the value of $\nu\,$sin$i$ for a solid body rotator. We consider the $\nu\,$sin$i$ measurement to be statistically significant if it passes the Spearman test and the fitted measurement of $\nu\,$cos$i$ is better than 1$\sigma$.

The noisy Spearman rank-order correlation test is also applied to the $R$--$\mu_R$ plots to identify monotonicity. We consider the expansion rate measurement to be statistically significant if it passes the Spearman test and the fitted slope measurement is better than 1$\sigma$.

Figure~\ref{fig:error_effect} shows the effect of observational errors on the measurements of $\nu\,$cos$i$ and $\nu\,$sin$i$ in a solid body rotator. The markers are also filled according to their statistical significance. The figure shows that most of the $\nu\,$cos$i$ or $\nu\,$sin$i$ measurements lie within 1$\sigma$ of the original value. The low inclination systems show measurable and significant $\nu\,$cos$i$ values, which could be overestimated in the presence of noise. The high inclination systems show measurable and significant PM $\nu\,$sin$i$ values which could be overestimated in the presence of noise. 
Overall, the RV method (top panel in Fig.~\ref{fig:error_effect}) is more noise-resistant than the PM method (bottom panel in Fig.~\ref{fig:error_effect}).

\section{Results} \label{sec:results}
\subsection{Open clusters using \textit{Gaia} data}

\begin{table*}
    \centering
    \begin{tabular}{l rrr rrr rr}
    \toprule
Name	&	l	&	b	&	log($age$)	&	Distance	&	$\nu\,$cos$i$	&	$\sigma_{\nu\,\text{cos}i}$	&	PA	&	$\sigma_{PA}$	\\
	&	[deg]	&	[deg]	&	[yr]	&	[pc]	&	[cy Gyr$^{-1}$]	&	[cy Gyr$^{-1}$]	&	[$^{\circ}$]	&	[$^{\circ}$]	\\ \hline
\multicolumn{9}{c}{\textit{RV based gold sample}}																	\\
Collinder\_69	&	195.108	&	-11.993	&	6.7	&	394	&	-13	&	6	&	306	&	8	\\
NGC\_1912	&	172.272	&	0.680	&	8.4	&	1085	&	-66	&	65	&	234	&	5	\\
NGC\_2099	&	177.648	&	3.089	&	8.7	&	1402	&	-97	&	76	&	257	&	2	\\
NGC\_2437	&	231.887	&	4.068	&	8.6	&	1555	&	109	&	96	&	30	&	2	\\
NGC\_6633	&	36.054	&	8.357	&	8.7	&	390	&	61	&	49	&	86	&	9	\\
Trumpler\_5	&	202.821	&	1.019	&	9.3	&	2905	&	17	&	17	&	159	&	5	\\ \hline
\multicolumn{9}{c}{\textit{RV based silver sample}}																	\\
ASCC\_113	&	82.825	&	-6.669	&	8.3	&	557	&	-33	&	29	&	300	&	4	\\
Blanco\_1	&	14.758	&	-79.127	&	8.2	&	234	&	-63	&	58	&	279	&	4	\\
COIN-Gaia\_11	&	162.518	&	-5.774	&	8.5	&	643	&	53	&	32	&	56	&	28	\\
HSC\_2636	&	311.904	&	17.550	&	7.0	&	134	&	-91	&	58	&	225	&	4	\\
IC\_2602$^{\alpha}$	&	289.687	&	-4.853	&	7.4	&	151	&	-34	&	29	&	295	&	7	\\
IC\_4725	&	13.688	&	-4.437	&	8.0	&	645	&	115	&	101	&	91	&	12	\\
Melotte\_20$^{\alpha}$	&	147.120	&	-6.389	&	7.7	&	174	&	16	&	13	&	84	&	21	\\
NGC\_2477$^{\alpha}$	&	253.569	&	-5.838	&	8.8	&	1384	&	-15	&	26	&	191	&	7	\\
NGC\_2547	&	264.446	&	-8.589	&	7.3	&	382	&	-42	&	33	&	265	&	14	\\
NGC\_2632$^{\alpha}$	&	205.916	&	32.474	&	8.5	&	183	&	9	&	18	&	87	&	4	\\
NGC\_3532	&	289.538	&	1.398	&	8.4	&	472	&	-33	&	21	&	193	&	4	\\
NGC\_6087	&	327.731	&	-5.401	&	8.0	&	938	&	167	&	106	&	18	&	6	\\
NGC\_6124	&	340.707	&	6.023	&	7.9	&	612	&	-78	&	73	&	298	&	2	\\
Ruprecht\_147$^{\alpha}$	&	21.012	&	-12.780	&	8.9	&	303	&	18	&	11	&	54	&	7	\\
Ruprecht\_171	&	16.454	&	-3.090	&	9.2	&	1482	&	-16	&	53	&	288	&	9	\\
Stock\_2$^{\alpha}$	&	133.387	&	-1.580	&	8.5	&	370	&	17	&	25	&	129	&	8	\\ \bottomrule

    \end{tabular}
    \begin{tabular}{l rrr rrr}
Name	&	l	&	b	&	log($age$)	&	Distance	&	$\nu\,$sin$i$	&	$\sigma_{\nu\,\text{sin}i}$	\\
	&	[deg]	&	[deg]	&	[yr]	&	[pc]	&	[cy Gyr$^{-1}$]	&	[cy Gyr$^{-1}$]	\\ \hline
\multicolumn{7}{c}{\textit{PM based gold sample}}													\\
CWNU\_1092	&	199.756	&	-6.635	&	7.3	&	411	&	13	&	10	\\
HSC\_1840	&	232.763	&	-11.853	&	8.2	&	273	&	10	&	8	\\
HSC\_2133	&	264.730	&	31.008	&	8.3	&	379	&	-13	&	8	\\
Ruprecht\_53	&	245.842	&	3.591	&	8.5	&	993	&	-16	&	9	\\ \bottomrule

    \end{tabular}
    \caption{Properties of spinning clusters. The sign of $\nu\,$cos$i$ and $\nu\,$sin$i$ denotes whether the spin axis is coming out (positive) or going into (negative) the Galactic plane. $^{\alpha}$ These results are based on SoS-only RV data.}
    \label{tab:spin_properties}
\end{table*} %spin properties

\begin{table*}
    \centering
    \begin{tabular}{l rrr rrr}
\toprule
Name	&	l	&	b	&	log($age$)	&	distance	&	expansion rate	&	$\sigma_{expansion\ rate}$	\\
	&	[deg]	&	[deg]	&	[yr]	&	[pc]	&	[m s$^{-1}$ pc$^{-1}$] 	&	[m s$^{-1}$ pc$^{-1}$] 	\\ \hline
Alessi\_13	&	237.427	&	-55.757	&	7.4	&	104	&	31	&	4	\\
CWNU\_1111	&	192.429	&	-8.954	&	7.3	&	267	&	35	&	11	\\
Collinder\_69	&	195.108	&	-11.993	&	6.7	&	394	&	160	&	6	\\
ESO\_332-13	&	344.661	&	1.656	&	6.6	&	1593	&	162	&	28	\\
HSC\_151	&	11.615	&	-16.013	&	7.9	&	470	&	-112	&	18	\\
HSC\_598	&	75.954	&	-35.583	&	8.2	&	184	&	-30	&	8	\\
HSC\_1262	&	159.531	&	-14.770	&	6.7	&	386	&	105	&	13	\\
HSC\_1318	&	168.824	&	-15.665	&	6.7	&	129	&	340	&	39	\\
HSC\_1633	&	206.754	&	-21.732	&	6.8	&	374	&	90	&	16	\\
HSC\_1766	&	224.388	&	-23.579	&	7.5	&	112	&	84	&	33	\\
HSC\_1807	&	228.873	&	-2.162	&	8.3	&	690	&	-39	&	21	\\
HSC\_1897	&	237.592	&	0.497	&	8.0	&	637	&	46	&	21	\\
HSC\_1900	&	237.706	&	-31.790	&	7.4	&	58	&	111	&	30	\\
HSC\_2247	&	276.783	&	-24.060	&	7.1	&	413	&	43	&	9	\\
HSC\_2931	&	354.271	&	19.740	&	6.8	&	133	&	95	&	29	\\
NGC\_7429	&	108.969	&	0.197	&	7.9	&	417	&	48	&	14	\\
OCSN\_96	&	350.274	&	22.202	&	6.7	&	141	&	105	&	12	\\
OC\_0470	&	260.197	&	-10.144	&	6.8	&	389	&	109	&	5	\\
OC\_0479	&	263.432	&	-10.261	&	6.9	&	399	&	105	&	19	\\
Sigma\_Orionis	&	206.837	&	-17.320	&	6.6	&	396	&	611	&	69	\\
Theia\_1918	&	266.058	&	-4.353	&	7.4	&	405	&	38	&	6	\\ \bottomrule
    \end{tabular}
    \caption{Properties of expanding or contracting (negative expansion rate) clusters.}
    \label{tab:expansion_properties}
\end{table*} %expansion_properties

\begin{figure*}
    \centering
    \includegraphics[width=0.98\textwidth]{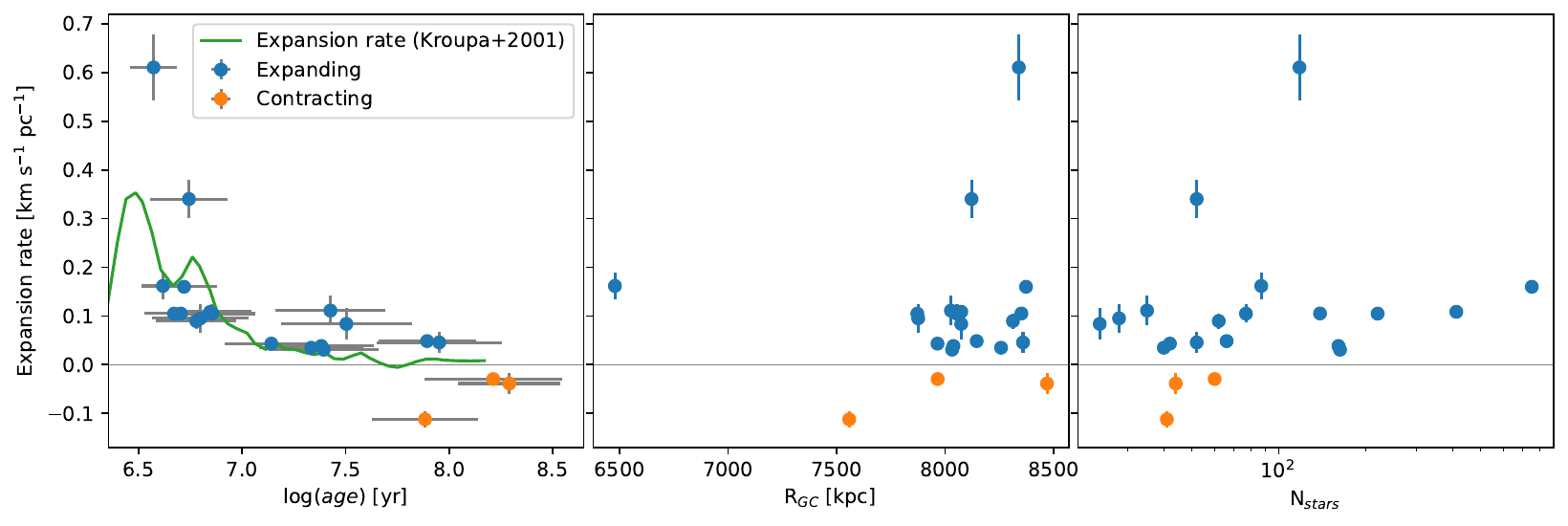}
    \caption{Relations between cluster expansion rate and cluster properties such as the age, Galactocentric distance (R$_{GC}$) and number of stars in the cluster (N$_{stars}$). The first panel shows the expansion rate (in green) of the 40\% Lagrangian radius in an 8340 \Msun\ cluster $N$-body model which undergoes gas expulsion and mass loss through evolving stars and reproduces the Orion Nebula Cluster at the time of 1 Myr and the Pleiades cluster at time 100 Myr.}
    \label{fig:properties_expanding}
\end{figure*} %properties_expanding

\begin{figure*}
    \centering
    \includegraphics[width=0.98\textwidth]{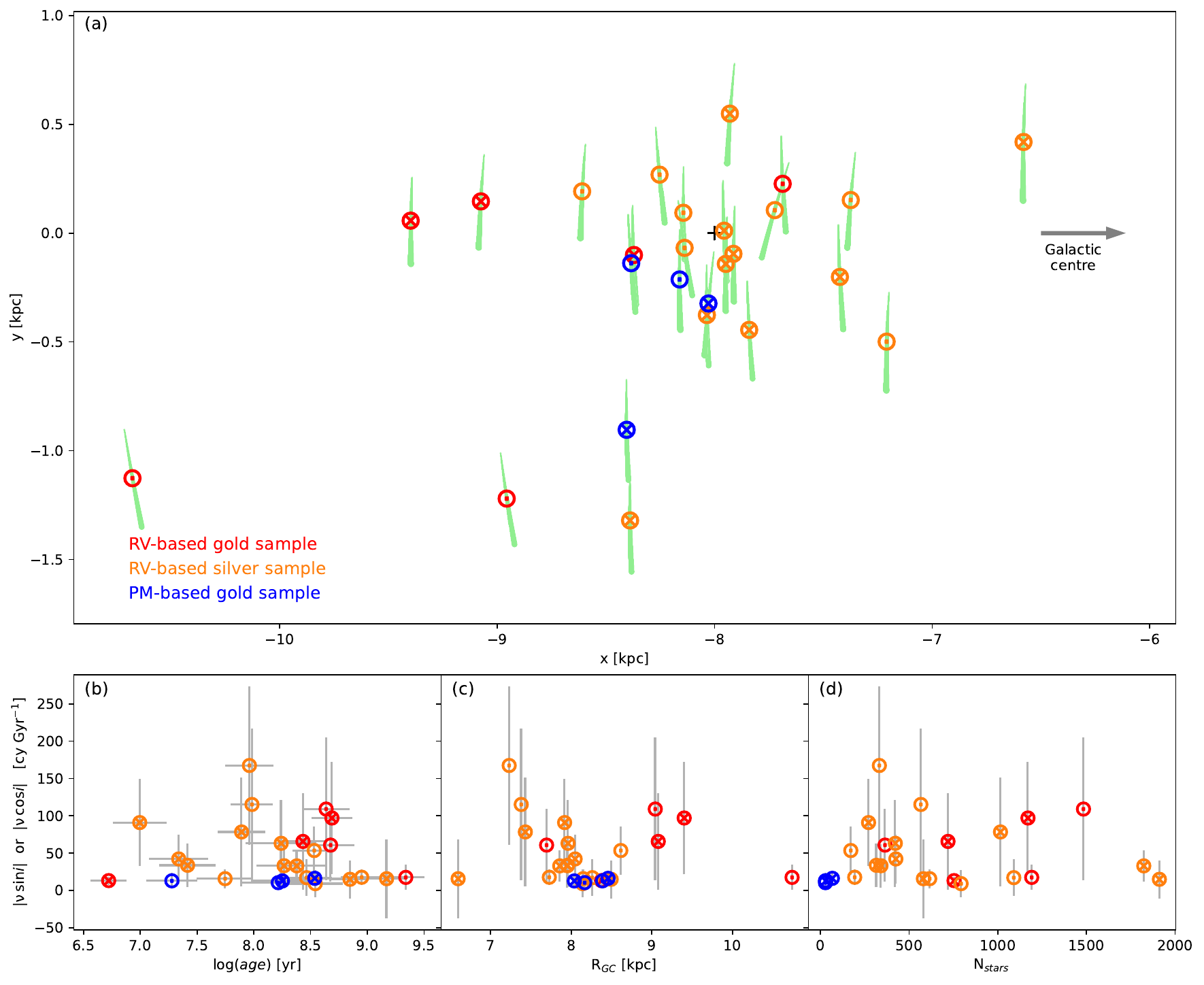}
    \caption{Diagnostic plots for analysing the relation between the spin orientation and cluster properties.
    The RV-based gold sample, RV-based silver sample and PM-based sample are shown red, orange and blue markers, respectively.
    A spin axis going into and coming out of the plane is shown as $\otimes$ and $\odot$, respectively.
    (a) Galactocentric positions of clusters as seen from the Galactic north pole. The cluster orbits are shown in light green, and the Sun's position is the black cross.
    (b) Distribution of spin frequency with the cluster age. 
    The RV-based method gives the $\nu\,$cos$i$ component of the spin frequency ($\nu$), while the PM-based method gives the $\nu\,$sin$i$ component.
    (c) Distribution of spin frequency with the Galactocentric radius.
    (d) Distribution of spin frequency with the number of cluster members within the tidal radius with robust distance measurements.}
    \label{fig:properties_spinning}
\end{figure*} %properties_spinning

Observationally, the inclination of the cluster's orbit is not known. Hence, we created diagnostic plots to identify spin signatures using both RV and PM. Figure~\ref{fig:comb_2099} shows an example of such a diagnostic plot. Here, spatial plots show the variation of RV and PM across the cluster members. The $R$--$\mu_T$ and $XR$--$RV_{corr}$ plots are the primary indicators of the spin. 
The polar plot shows the values of identified $PA$ using different radial slices of the cluster (as shown in Figure~\ref{fig:roi}). Ideally, all the $PA$ values should be similar. 
We have removed the central 25 percentile stars while analysing the RVs to avoid the noisy velocities near the cluster centre. Including the central stars increases the errors of the parameters; however, this does not change the overall results for the gold sample defined later.

As seen in Figure~\ref{fig:demo_rv} and \ref{fig:demo_pm}, a virialised cluster produces linear trends in the $XR$--$RV_{corr}$ and $R$--$\mu_T$ plots. 
The two-body relaxation time for OCs ranges from a few Myr to a Gyr \citep{Jadhav2021MNRAS.507.1699J}. The distribution in the $R$--$\mu_T$ and $XR$--$RV_{corr}$ plots vary with different stages of relaxation and tend towards linearity for virialised clusters. We have fitted these distributions with lines to gauge the extent of the spin. The slope of the $XR$--$RV_{corr}$ data gives $\nu$ cos$i$ in ideal conditions, while the slope of the $R$--$\mu_T$ data gives the $\nu$ sin$i$. However, the noise and state of relaxation change the slope values. We report the values of the slopes in Table~\ref{tab:spin_properties}.
However, it has to be noted that the values are rough estimates of the real spin frequency, and the quoted errors are formal errors based on line fitting (without accounting for virialisation or internal velocity dispersion).
The individual diagnostic plots for the spinning clusters using the RV and PM methods are given in Figure~\ref{fig:comb_rv} and ~\ref{fig:comb_pm}, respectively. The diagnostic plots for expanding and contracting clusters are given in Figure~\ref{fig:comb_expanding} and \ref{fig:comb_contracting}, respectively.

Overall, we identified 22 clusters with some indication of rotation using RV based analysis. 6 of the clusters passed KS-test and $>1\sigma$ slope significance test (see \S~\ref{sec:statistical_tests}) and are referred as the \textit{RV-based gold sample}. 10 of these clusters passed at least one of the test and are classified as \textit{RV-based silver sample}. The remaining 6 clusters did not show any significant trends in the \textit{Gaia} DR3 + SoS data, however they did show significant trends in SoS-only analysis. These clusters are also classified as \textit{RV-based silver sample} and the use of SoS-only data is mentioned in the corresponding tables and figures.
The PM analysis showed spin signatures in 4 OCs (referred as \textit{PM-based gold sample}). The $\mu_R$ analysis identified 18 expanding clusters and 3 contacting clusters.

% Overall, the RV analysis showed 5 clusters (referred as RV-based gold sample in Table~\ref{tab:spin_properties}) with spin signature and 10 more candidates (referred as RV-based silver sample), which passed either the KS test or the slope test. 

\subsection{$N$-body simulations}
We analysed the evolution of cluster properties (angular momentum, $\nu\,$cos$i$, radius) for the various $N$-body simulations. 
Figure~\ref{fig:properties_simulation} shows their evolution in the simulated clusters. All the clusters revolve clockwise about the Galaxy, as seen from the Galactic north pole. Thus, their orbital angular momentum in the Z direction, as measured from the Galactic centre, is negative.
We calculated the spin angular momentum ($\mathcal{L}_z$) in the corotating reference frame, assuming the centre of density as the origin. The evolution of $\mathcal{L}_z$ is shown in the second row of Figure~\ref{fig:properties_simulation}. 
In order to identify the cluster members, we calculated the kinetic and potential energies of the simulated particles with \textsc{clustertools} \citep{Webb2023JOSS....8.4483W} using the positions and velocities of individual particles. Similarly, we measured the tidal radius based on the Galactic potential \citep{Bovy2015ApJS..216...29B} and \citet{Bertin2008ApJ...689.1005B} formalism.
The $\mathcal{L}_z$ is plotted for three populations: (i) stars within the \textit{tidal radius}; (ii) \textit{bound stars}, which have negative total energy (kinetic+potential); (iii) \textit{unbound stars}, which have positive total energy. The stars within the tidal radius best represent the observed cluster members for real clusters. 

The angular momentum of the whole system changes with time due to escaping particles, stellar evolution and external forces. However, all simulations show significantly negative $\mathcal{L}_z$ for unbound particles (i.e. the tidal tails), indicating that the spin axis of the tidal tail particles is aligned with the revolution axis.
The stars within the tidal radius collectively do not have much angular momentum. However, the sign of this minimal $\mathcal{L}_z$ is almost always negative. The only exceptions are the very early stages of the simulations. 

The synthetic observations of these simulations provide the PA and $\nu\,$cos$i$ values for each snapshot. The \textit{PeTar} simulations show that the cluster members predominantly rotate with their spin axis aligned with the revolution axis. 7 of 8 snapshots in \textit{MLD} simulations do not show any rotation. The \textit{Kroupa+2022} clusters seem to have some rotation signatures; however, their spin axis alignment keeps fluctuating. 

In addition, the spin is much stronger for \textit{bound stars}. 
The \textit{bound stars} span a larger radius, and the outer bound stars are not considered cluster members from an observational perspective. However, the $\mathcal{L}_z$ and $\nu\,$cos$i$ measurements are much more significant for this collection. The bound stars are mostly forming spinning population after $\approx25$ Myr in almost all simulations. Moreover, the majority of the time, the spin axis is aligned with the orbital angular momentum.

Overall, some simulated clusters have a detectable spin with $\nu\,$cos$i$ ranging from $-2$ to $2$ cy Gyr$^{-1}$. However, the spin orientation is not always aligned with the orbital angular momentum. Additionally, none of the simulated clusters showed statistically significant expansion signatures within the tidal radius of individual snapshots.

\section{Discussion} \label{sec:discussion}

\subsection{Cluster expansion/contraction}
Figure~\ref{fig:properties_expanding} shows the variation of expansion rate with the cluster properties. 18 clusters younger than 100 Myr show expansion signatures (3.6\% of the 492 OCs with similar ages).
Panel (a) shows that the expansion rate decreases with age becoming negative near 100--300 Myr. None of the older clusters showed detectable expansion/contraction. The theoretical expansion rate of the 40\% Lagrangian radius \citep{Kroupa2001MNRAS.321..699K} closely matches the observed expansion rate. No relation is found with the Galactocentric radius or the number of stars in the clusters. However, it is important to note that not all young clusters showed expansion.

\subsection{Cluster spin}

Figure~\ref{fig:properties_spinning} depicts the relation between the cluster properties and their spin. Panel (a) shows the cluster orbits and spin axis orientation with respect to the Galactic plane. All clusters are orbiting the Galactic centre clockwise (as seen from the North pole with the Sun at $-8$ kpc). However, there is no correlation between the spin axis of the clusters and the Z direction.
Panels (b)--(d) show the distribution of spin frequency with the cluster's age, Galactocentric distance and the number of members within the tidal radius. There is no significant trend seen in all three distributions. The overall value of $\nu\,$cos$i$ ranges from 2--170 cy Gyr$^{-1}$ and $\nu\,$sin$i$ are all around 15 cy Gyr$^{-1}$. 
The analysis of our demo clusters showed that measurement of both $\nu\,$sin$i$ and $\nu\,$sin$i$ is possible in non-noisy clusters and can be reliably used to calculate the inclination and the true spin frequency. However, as real clusters are not perfect rotators and none of the observed clusters showed rotation signatures in both RV and PM methods, we cannot calculate the true frequency of the observed clusters. 

The simulated clusters showed minimal spin frequencies ($-2$ to 2 cy Gyr$^{-1}$), mostly aligned with the the orbital angular momentum. This slight spin could be due to the asymmetric evaporation of stars as seen by \citet{Kroupa2022MNRAS.517.3613K} who noticed that the leading tail was more dense than the trailing tail. However, the observed clusters have much higher spin frequencies ($\nu\,$cos$i_{\text{gold sample}} \in [-84,30]$, $\nu\,$cos$i_{\text{silver sample}} \in [-77,167]$ and $\nu\,$sin$i \in [-16,13]$ cy Gyr$^{-1}$).

The simulated clusters were set up without any spin. Hence, the result shows that the cluster spin due to internal dynamics and interaction with a smooth Galactic potential is minimal. Thus, primordial spin within the parent molecular cloud or a significantly strong dynamical interaction (such as collisions, flybys, and/or mergers of clusters) is required to explain the detected spin frequencies (up to 170 cy Gyr$^{-1}$).

According to \citet{Braine2020A&A...633A..17B}, the RV gradient for molecular clouds in the galaxy M51 goes up to 0.2 \kms\ pc$^{-1}$ (equivalent to $\nu\,$cos$i \leq$ 32 cy Gyr$^{-1}$). This is consistent with the spin frequencies observed in the OCs.
The initial spin would also explain the age independence of spin frequency and the randomness of the $PA$.

\citet{Toomre1972ApJ...178..623T, Read2006MNRAS.366..429R, Kroupa2022MNRAS.517.3613K} have shown that stars in prograde orbits within a stellar system are preferentially lost compared to stars in retrograde orbits. This could have been interpreted as losing prograde angular momentum resulting in a star system spinning in the opposite direction to its orbit. However, the $N$-body simulations in this work show that the cluster's spin is majorly aligned with the orbital angular momentum. Additionally, the observations show that the spin does not correlate with the orbital angular momentum, which is likely a result of external effects such as interactions or initial spin.

All the $N$-body simulations used in this study assume that the Galactic potential is axisymmetric. However, the presence of bar and spiral arms changes the Galactic potential and the var/spiral arm rotation makes the potential evolve with time.
\citet{Rossi2015MNRAS.454.1453R} demonstrated that the Galactic bar has negligible effect on the cluster evolution for Galactocentric distance of more than 4 kpc. As all of the peculiar clusters in this study lie well beyond 6 kpc, the effect of Galactic bar should be minimal. However, more detailed simulations including a bar, spiral arms, primordial spin and collisions with star clusters/molecular clouds would be helpful to further study the present day spins seen in the OCs.

\subsection{Comparison with literature on open clusters}

\citet{Kamann2019MNRAS.483.2197K} detected rotation in NGC 6791 and absence of rotation in NGC 6819 using a RV based analysis. We could not detect rotation in both the clusters.
\citet{Healy2020ApJ...903...99H} detected signs of contraction in NGC 2516, however we could not confirm any radial motion in the cluster.
\citet{Loktin2020AN....341..638L} used \textit{Gaia} DR2 data and detected rotation in Praesepe. However, they did not correct for the projection effect which leads to erroneous results. 
\citet{Healy2021ApJ...923...23H} detected rotation in Praesepe and absence of rotation in Pleiades and M35 (NGC 2168).
Similarly, \citet{Hao2022ApJ...938..100H} and \citet{Hao2024arXiv240212160H} detected rotation in Praesepe, Pleiades, Alpha Persei and Hyades. 
Our analysis, after correcting for the projection effects, shows a slight rotation signature in Praesepe (NGC 2632) and Alpha Persei (Melotte 20). However, we could not detect any rotation in the Pleiades (Melotte 22), Hyades (Melotte 25) and NGC 2168.
\citet{Guilherme2023A&A...673A.128G} used PM based analysis to identify 8 spinning (and 9 candidates), 14 expanding (and 15 candidates) and 2 contracting (and 1 candidate) clusters. They used vector field reconstruction to identify kinematic structures in $R$--$\mu_T$ and $R$--$\mu_R$ plots. However, they did not quantify the spin/expansion rates. 
We have 2 of the spinning clusters in our sample (Ruprecht 147 and Stock 2). In addition, one of their spinning cluster, Alessi 13, had a significant slope in the $R$--$\mu_T$ plane, however it did not pass the Spearman test for monotonicity. 
Among their 14 expanding clusters, 2 are common with our list (Alessi 13 and Collinder 69). The differences in the classification method and in the parent sample lead to the differences in the list of clusters.

\section{Conclusions} \label{sec:conclusions}
We analysed the internal dynamics of OCs using \textit{Gaia} DR3 data, specifically the spin and expansion of the clusters. We also used synthetic observations of $N$-body simulations to validate the results. The major conclusions of the work are as follows:
\begin{itemize}
    \item Among the 1379 OCs identified with \textit{Gaia} DR3, we could detect spin signatures in 10 (0.7\%) clusters with 16 more candidate OCs. The spin frequencies are roughly 9 to 167 cy Gyr$^{-1}$. Of these 26, 12 spin in the same direction as they orbit the Galaxy.
    \item The $N$-body simulations of similar clusters have a spin frequency of $\approx$2--5 cy Gyr$^{-1}$.
    \item The observed spin frequencies are much larger than expected for initially non-rotating clusters. Thus, we conclude that their parent molecular clouds must have initial spin, and/or the clusters have undergone strong tidal interactions with other massive objects in the Milky Way. 
    \item The spin orientation in $N$-body simulations majorly aligns with the cluster's orbit, contrary to previous conjectures based on preferential tidal stripping of stars with prograde orbits. 
    Observationally, the spin orientation is not correlated with the orbital angular momentum. Any spin imparted due to the tidal stripping is dominated by random sources of spin (such as initial spin and interactions).
    \item From the same sample, we could identify 18 (3.6\% of the OCs younger than 100 Myr) expanding and 3 contracting clusters. The expansion rate of these young clusters is compatible with theoretical models of cluster expansion due to gas expulsion, stellar mass loss, and dynamical heating.
\end{itemize}

The RV and parallax precision are the major hurdles in identifying more spinning/expanding OCs. Recent attempts, such as the SoS, are improving the RV homogeneity and precision using multiple spectroscopic surveys. Upcoming spectroscopic surveys with better precision and containing fainter stars would be necessary to understanding the 3D dynamics of the OCs. In addition, simulations with primordial spin and external interactions would be required to understand the spinning nature and origin of OCs thoroughly.

\begin{acknowledgements}
    We thank the anonymous referee for constructive comments. 
    VJ thanks the Alexander von Humboldt Foundation for their support.
    PK acknowledges support through the DAAD Eastern European Exchange Programme.
    This work has made use of data from the European Space Agency (ESA) mission {\it Gaia} (\url{https://www.cosmos.esa.int/gaia}), processed by the {\it Gaia} Data Processing and Analysis Consortium (DPAC, \url{https://www.cosmos.esa.int/web/gaia/dpac/consortium}). Funding for the DPAC has been provided by national institutions, in particular the institutions participating in the {\it Gaia} Multilateral Agreement.
\end{acknowledgements}
\bibliographystyle{aa} % style aa.bst
\bibliography{references}

\begin{thebibliography}{80}
\expandafter\ifx\csname natexlab\endcsname\relax\def\natexlab#1{#1}\fi

\bibitem[{{Ahumada} {et~al.}(2020){Ahumada}, {Allende Prieto}, {Almeida},
  {Anders}, {Anderson}, {Andrews}, {Anguiano}, {Arcodia}, {Armengaud},
  {Aubert}, {Avila}, {Avila-Reese}, {Badenes}, {Balland}, {Barger},
  {Barrera-Ballesteros}, {Basu}, {Bautista}, {Beaton}, {Beers}, {Benavides},
  {Bender}, {Bernardi}, {Bershady}, {Beutler}, {Bidin}, {Bird}, {Bizyaev},
  {Blanc}, {Blanton}, {Boquien}, {Borissova}, {Bovy}, {Brandt}, {Brinkmann},
  {Brownstein}, {Bundy}, {Bureau}, {Burgasser}, {Burtin}, {Cano-D{\'\i}az},
  {Capasso}, {Cappellari}, {Carrera}, {Chabanier}, {Chaplin}, {Chapman},
  {Cherinka}, {Chiappini}, {Doohyun Choi}, {Chojnowski}, {Chung}, {Clerc},
  {Coffey}, {Comerford}, {Comparat}, {da Costa}, {Cousinou}, {Covey}, {Crane},
  {Cunha}, {Ilha}, {Dai}, {Damsted}, {Darling}, {Davidson}, {Davies}, {Dawson},
  {De}, {de la Macorra}, {De Lee}, {Queiroz}, {Deconto Machado}, {de la Torre},
  {Dell'Agli}, {du Mas des Bourboux}, {Diamond-Stanic}, {Dillon}, {Donor},
  {Drory}, {Duckworth}, {Dwelly}, {Ebelke}, {Eftekharzadeh}, {Davis Eigenbrot},
  {Elsworth}, {Eracleous}, {Erfanianfar}, {Escoffier}, {Fan}, {Farr},
  {Fern{\'a}ndez-Trincado}, {Feuillet}, {Finoguenov}, {Fofie},
  {Fraser-McKelvie}, {Frinchaboy}, {Fromenteau}, {Fu}, {Galbany}, {Garcia},
  {Garc{\'\i}a-Hern{\'a}ndez}, {Garma Oehmichen}, {Ge}, {Geimba Maia},
  {Geisler}, {Gelfand}, {Goddy}, {Gonzalez-Perez}, {Grabowski}, {Green},
  {Grier}, {Guo}, {Guy}, {Harding}, {Hasselquist}, {Hawken}, {Hayes}, {Hearty},
  {Hekker}, {Hogg}, {Holtzman}, {Horta}, {Hou}, {Hsieh}, {Huber}, {Hunt}, {Ider
  Chitham}, {Imig}, {Jaber}, {Jimenez Angel}, {Johnson}, {Jones},
  {J{\"o}nsson}, {Jullo}, {Kim}, {Kinemuchi}, {Kirkpatrick}, {Kite}, {Klaene},
  {Kneib}, {Kollmeier}, {Kong}, {Kounkel}, {Krishnarao}, {Lacerna}, {Lan},
  {Lane}, {Law}, {Le Goff}, {Leung}, {Lewis}, {Li}, {Lian}, {Lin}, {Long},
  {Longa-Pe{\~n}a}, {Lundgren}, {Lyke}, {Mackereth}, {MacLeod}, {Majewski},
  {Manchado}, {Maraston}, {Martini}, {Masseron}, {Masters}, {Mathur},
  {McDermid}, {Merloni}, {Merrifield}, {M{\'e}sz{\'a}ros}, {Miglio}, {Minniti},
  {Minsley}, {Miyaji}, {Mohammad}, {Mosser}, {Mueller}, {Muna},
  {Mu{\~n}oz-Guti{\'e}rrez}, {Myers}, {Nadathur}, {Nair}, {Nandra}, {Correa do
  Nascimento}, {Nevin}, {Newman}, {Nidever}, {Nitschelm}, {Noterdaeme},
  {O'Connell}, {Olmstead}, {Oravetz}, {Oravetz}, {Osorio}, {Pace}, {Padilla},
  {Palanque-Delabrouille}, {Palicio}, {Pan}, {Pan}, {Parker}, {Paviot},
  {Peirani}, {Ram{\'r}ez}, {Penny}, {Percival}, {Perez-Fournon},
  {P{\'e}rez-R{\`a}fols}, {Petitjean}, {Pieri}, {Pinsonneault}, {Poovelil},
  {Povick}, {Prakash}, {Price-Whelan}, {Raddick}, {Raichoor}, {Ray}, {Rembold},
  {Rezaie}, {Riffel}, {Riffel}, {Rix}, {Robin}, {Roman-Lopes},
  {Rom{\'a}n-Z{\'u}{\~n}iga}, {Rose}, {Ross}, {Rossi}, {Rowlands}, {Rubin},
  {Salvato}, {S{\'a}nchez}, {S{\'a}nchez-Menguiano}, {S{\'a}nchez-Gallego},
  {Sayres}, {Schaefer}, {Schiavon}, {Schimoia}, {Schlafly}, {Schlegel},
  {Schneider}, {Schultheis}, {Schwope}, {Seo}, {Serenelli}, {Shafieloo},
  {Shamsi}, {Shao}, {Shen}, {Shetrone}, {Shirley}, {Silva Aguirre}, {Simon},
  {Skrutskie}, {Slosar}, {Smethurst}, {Sobeck}, {Sodi}, {Souto}, {Stark},
  {Stassun}, {Steinmetz}, {Stello}, {Stermer}, {Storchi-Bergmann},
  {Streblyanska}, {Stringfellow}, {Stutz}, {Su{\'a}rez}, {Sun},
  {Taghizadeh-Popp}, {Talbot}, {Tayar}, {Thakar}, {Theriault}, {Thomas},
  {Thomas}, {Tinker}, {Tojeiro}, {Toledo}, {Tremonti}, {Troup}, {Tuttle},
  {Unda-Sanzana}, {Valentini}, {Vargas-Gonz{\'a}lez}, {Vargas-Maga{\~n}a},
  {V{\'a}zquez-Mata}, {Vivek}, {Wake}, {Wang}, {Weaver}, {Weijmans}, {Wild},
  {Wilson}, {Wilson}, {Wolthuis}, {Wood-Vasey}, {Yan}, {Yang}, {Y{\`e}che},
  {Zamora}, {Zarrouk}, {Zasowski}, {Zhang}, {Zhao}, {Zhao}, {Zheng}, {Zheng},
  {Zhu}, \& {Zou}}]{Ahumada2020ApJS..249....3A}
{Ahumada}, R., {Allende Prieto}, C., {Almeida}, A., {et~al.} 2020, \apjs, 249,
  3

\bibitem[{{Anderson} \& {King}(2003)}]{Anderson2003AJ....126..772A}
{Anderson}, J. \& {King}, I.~R. 2003, \aj, 126, 772

\bibitem[{{Astropy Collaboration} {et~al.}(2022){Astropy Collaboration},
  {Price-Whelan}, {Lim}, {Earl}, {Starkman}, {Bradley}, {Shupe}, {Patil},
  {Corrales}, {Brasseur}, {N{"o}the}, {Donath}, {Tollerud}, {Morris},
  {Ginsburg}, {Vaher}, {Weaver}, {Tocknell}, {Jamieson}, {van Kerkwijk},
  {Robitaille}, {Merry}, {Bachetti}, {G{"u}nther}, {Aldcroft},
  {Alvarado-Montes}, {Archibald}, {B{'o}di}, {Bapat}, {Barentsen}, {Baz{'a}n},
  {Biswas}, {Boquien}, {Burke}, {Cara}, {Cara}, {Conroy}, {Conseil}, {Craig},
  {Cross}, {Cruz}, {D'Eugenio}, {Dencheva}, {Devillepoix}, {Dietrich},
  {Eigenbrot}, {Erben}, {Ferreira}, {Foreman-Mackey}, {Fox}, {Freij}, {Garg},
  {Geda}, {Glattly}, {Gondhalekar}, {Gordon}, {Grant}, {Greenfield}, {Groener},
  {Guest}, {Gurovich}, {Handberg}, {Hart}, {Hatfield-Dodds}, {Homeier},
  {Hosseinzadeh}, {Jenness}, {Jones}, {Joseph}, {Kalmbach}, {Karamehmetoglu},
  {Ka{l}uszy{'n}ski}, {Kelley}, {Kern}, {Kerzendorf}, {Koch}, {Kulumani},
  {Lee}, {Ly}, {Ma}, {MacBride}, {Maljaars}, {Muna}, {Murphy}, {Norman},
  {O'Steen}, {Oman}, {Pacifici}, {Pascual}, {Pascual-Granado}, {Patil},
  {Perren}, {Pickering}, {Rastogi}, {Roulston}, {Ryan}, {Rykoff}, {Sabater},
  {Sakurikar}, {Salgado}, {Sanghi}, {Saunders}, {Savchenko}, {Schwardt},
  {Seifert-Eckert}, {Shih}, {Jain}, {Shukla}, {Sick}, {Simpson},
  {Singanamalla}, {Singer}, {Singhal}, {Sinha}, {Sip{H{o}}cz}, {Spitler},
  {Stansby}, {Streicher}, {{{S}}umak}, {Swinbank}, {Taranu}, {Tewary},
  {Tremblay}, {Val-Borro}, {Van Kooten}, {Vasovi{'c}}, {Verma}, {de Miranda
  Cardoso}, {Williams}, {Wilson}, {Winkel}, {Wood-Vasey}, {Xue}, {Yoachim},
  {Zhang}, {Zonca}, \& {Astropy Project Contributors}}]{astropy:2022}
{Astropy Collaboration}, {Price-Whelan}, A.~M., {Lim}, P.~L., {et~al.} 2022,
  apj, 935, 167

\bibitem[{{Astropy Collaboration} {et~al.}(2018){Astropy Collaboration},
  {Price-Whelan}, {Sip{\H{o}}cz}, {G{\"u}nther}, {Lim}, {Crawford}, {Conseil},
  {Shupe}, {Craig}, {Dencheva}, {Ginsburg}, {Vand erPlas}, {Bradley},
  {P{\'e}rez-Su{\'a}rez}, {de Val-Borro}, {Aldcroft}, {Cruz}, {Robitaille},
  {Tollerud}, {Ardelean}, {Babej}, {Bach}, {Bachetti}, {Bakanov}, {Bamford},
  {Barentsen}, {Barmby}, {Baumbach}, {Berry}, {Biscani}, {Boquien}, {Bostroem},
  {Bouma}, {Brammer}, {Bray}, {Breytenbach}, {Buddelmeijer}, {Burke},
  {Calderone}, {Cano Rodr{\'\i}guez}, {Cara}, {Cardoso}, {Cheedella}, {Copin},
  {Corrales}, {Crichton}, {D'Avella}, {Deil}, {Depagne}, {Dietrich}, {Donath},
  {Droettboom}, {Earl}, {Erben}, {Fabbro}, {Ferreira}, {Finethy}, {Fox},
  {Garrison}, {Gibbons}, {Goldstein}, {Gommers}, {Greco}, {Greenfield},
  {Groener}, {Grollier}, {Hagen}, {Hirst}, {Homeier}, {Horton}, {Hosseinzadeh},
  {Hu}, {Hunkeler}, {Ivezi{\'c}}, {Jain}, {Jenness}, {Kanarek}, {Kendrew},
  {Kern}, {Kerzendorf}, {Khvalko}, {King}, {Kirkby}, {Kulkarni}, {Kumar},
  {Lee}, {Lenz}, {Littlefair}, {Ma}, {Macleod}, {Mastropietro}, {McCully},
  {Montagnac}, {Morris}, {Mueller}, {Mumford}, {Muna}, {Murphy}, {Nelson},
  {Nguyen}, {Ninan}, {N{\"o}the}, {Ogaz}, {Oh}, {Parejko}, {Parley}, {Pascual},
  {Patil}, {Patil}, {Plunkett}, {Prochaska}, {Rastogi}, {Reddy Janga},
  {Sabater}, {Sakurikar}, {Seifert}, {Sherbert}, {Sherwood-Taylor}, {Shih},
  {Sick}, {Silbiger}, {Singanamalla}, {Singer}, {Sladen}, {Sooley},
  {Sornarajah}, {Streicher}, {Teuben}, {Thomas}, {Tremblay}, {Turner},
  {Terr{\'o}n}, {van Kerkwijk}, {de la Vega}, {Watkins}, {Weaver}, {Whitmore},
  {Woillez}, {Zabalza}, \& {Astropy Contributors}}]{astropy:2018}
{Astropy Collaboration}, {Price-Whelan}, A.~M., {Sip{\H{o}}cz}, B.~M., {et~al.}
  2018, \aj, 156, 123

\bibitem[{{Astropy Collaboration} {et~al.}(2013){Astropy Collaboration},
  {Robitaille}, {Tollerud}, {Greenfield}, {Droettboom}, {Bray}, {Aldcroft},
  {Davis}, {Ginsburg}, {Price-Whelan}, {Kerzendorf}, {Conley}, {Crighton},
  {Barbary}, {Muna}, {Ferguson}, {Grollier}, {Parikh}, {Nair}, {Unther},
  {Deil}, {Woillez}, {Conseil}, {Kramer}, {Turner}, {Singer}, {Fox}, {Weaver},
  {Zabalza}, {Edwards}, {Azalee Bostroem}, {Burke}, {Casey}, {Crawford},
  {Dencheva}, {Ely}, {Jenness}, {Labrie}, {Lim}, {Pierfederici}, {Pontzen},
  {Ptak}, {Refsdal}, {Servillat}, \& {Streicher}}]{astropy:2013}
{Astropy Collaboration}, {Robitaille}, T.~P., {Tollerud}, E.~J., {et~al.} 2013,
  \aap, 558, A33

\bibitem[{{Bailer-Jones} {et~al.}(2021){Bailer-Jones}, {Rybizki}, {Fouesneau},
  {Demleitner}, \& {Andrae}}]{Bailer2021AJ....161..147B}
{Bailer-Jones}, C.~A.~L., {Rybizki}, J., {Fouesneau}, M., {Demleitner}, M., \&
  {Andrae}, R. 2021, \aj, 161, 147

\bibitem[{{Banerjee} {et~al.}(2020){Banerjee}, {Belczynski}, {Fryer},
  {Berczik}, {Hurley}, {Spurzem}, \& {Wang}}]{Banerjee2020A&A...639A..41B}
{Banerjee}, S., {Belczynski}, K., {Fryer}, C.~L., {et~al.} 2020, \aap, 639, A41

\bibitem[{{Bekenstein} \& {Milgrom}(1984)}]{bekenstein1984a}
{Bekenstein}, J. \& {Milgrom}, M. 1984, \apj, 286, 7

\bibitem[{{Belczynski} {et~al.}(2016){Belczynski}, {Heger}, {Gladysz},
  {Ruiter}, {Woosley}, {Wiktorowicz}, {Chen}, {Bulik}, {O'Shaughnessy}, {Holz},
  {Fryer}, \& {Berti}}]{Belczynski2016A&A...594A..97B}
{Belczynski}, K., {Heger}, A., {Gladysz}, W., {et~al.} 2016, \aap, 594, A97

\bibitem[{{Belczynski} {et~al.}(2008){Belczynski}, {Kalogera}, {Rasio}, {Taam},
  {Zezas}, {Bulik}, {Maccarone}, \& {Ivanova}}]{Belczynski2008ApJS..174..223B}
{Belczynski}, K., {Kalogera}, V., {Rasio}, F.~A., {et~al.} 2008, \apjs, 174,
  223

\bibitem[{{Bellini} {et~al.}(2017){Bellini}, {Bianchini}, {Varri}, {Anderson},
  {Piotto}, {van der Marel}, {Vesperini}, \&
  {Watkins}}]{Bellini2017ApJ...844..167B}
{Bellini}, A., {Bianchini}, P., {Varri}, A.~L., {et~al.} 2017, \apj, 844, 167

\bibitem[{{Bertin} \& {Varri}(2008)}]{Bertin2008ApJ...689.1005B}
{Bertin}, G. \& {Varri}, A.~L. 2008, \apj, 689, 1005

\bibitem[{{Bianchini} {et~al.}(2018){Bianchini}, {van der Marel}, {del Pino},
  {Watkins}, {Bellini}, {Fardal}, {Libralato}, \&
  {Sills}}]{Bianchini2018MNRAS.481.2125B}
{Bianchini}, P., {van der Marel}, R.~P., {del Pino}, A., {et~al.} 2018, \mnras,
  481, 2125

\bibitem[{{Bovy}(2015)}]{Bovy2015ApJS..216...29B}
{Bovy}, J. 2015, \apjs, 216, 29

\bibitem[{{Braine} {et~al.}(2020){Braine}, {Hughes}, {Rosolowsky}, {Gratier},
  {Colombo}, {Meidt}, \& {Schinnerer}}]{Braine2020A&A...633A..17B}
{Braine}, J., {Hughes}, A., {Rosolowsky}, E., {et~al.} 2020, \aap, 633, A17

\bibitem[{{Chen} {et~al.}(2019){Chen}, {Pineda}, {Offner}, {Goodman},
  {Burkert}, {Friesen}, {Rosolowsky}, {Scibelli}, \&
  {Shirley}}]{Chen2019ApJ...886..119C}
{Chen}, H. H.-H., {Pineda}, J.~E., {Offner}, S. S.~R., {et~al.} 2019, \apj,
  886, 119

\bibitem[{{Choudhuri}(2010)}]{Choudhuri2010asph.book.....C}
{Choudhuri}, A.~R. 2010, {Astrophysics for Physicists} (Cambridge University
  Press)

\bibitem[{{Deng} {et~al.}(2012){Deng}, {Newberg}, {Liu}, {Carlin}, {Beers},
  {Chen}, {Chen}, {Christlieb}, {Grillmair}, {Guhathakurta}, {Han}, {Hou},
  {Lee}, {L{\'e}pine}, {Li}, {Liu}, {Pan}, {Sellwood}, {Wang}, {Wang}, {Yang},
  {Yanny}, {Zhang}, {Zhang}, {Zheng}, \& {Zhu}}]{Deng2012RAA....12..735D}
{Deng}, L.-C., {Newberg}, H.~J., {Liu}, C., {et~al.} 2012, Research in
  Astronomy and Astrophysics, 12, 735

\bibitem[{{Dinnbier} \&
  {Kroupa}(2020{\natexlab{a}})}]{Dinnbier2020A&A...640A..84D}
{Dinnbier}, F. \& {Kroupa}, P. 2020{\natexlab{a}}, \aap, 640, A84

\bibitem[{{Dinnbier} \&
  {Kroupa}(2020{\natexlab{b}})}]{Dinnbier2020A&A...640A..85D}
{Dinnbier}, F. \& {Kroupa}, P. 2020{\natexlab{b}}, \aap, 640, A85

\bibitem[{{Dinnbier} {et~al.}(2022){Dinnbier}, {Kroupa}, \&
  {Anderson}}]{Dinnbier2022A&A...660A..61D}
{Dinnbier}, F., {Kroupa}, P., \& {Anderson}, R.~I. 2022, \aap, 660, A61

\bibitem[{{Fabricius} {et~al.}(2021){Fabricius}, {Luri}, {Arenou}, {Babusiaux},
  {Helmi}, {Muraveva}, {Reyl{\'e}}, {Spoto}, {Vallenari}, {Antoja}, {Balbinot},
  {Barache}, {Bauchet}, {Bragaglia}, {Busonero}, {Cantat-Gaudin}, {Carrasco},
  {Diakit{\'e}}, {Fabrizio}, {Figueras}, {Garcia-Gutierrez}, {Garofalo},
  {Jordi}, {Kervella}, {Khanna}, {Leclerc}, {Licata}, {Lambert}, {Marrese},
  {Masip}, {Ramos}, {Robichon}, {Robin}, {Romero-G{\'o}mez}, {Rubele}, \&
  {Weiler}}]{Fabricius2021A&A...649A...5F}
{Fabricius}, C., {Luri}, X., {Arenou}, F., {et~al.} 2021, \aap, 649, A5

\bibitem[{{Fryer} {et~al.}(2012){Fryer}, {Belczynski}, {Wiktorowicz},
  {Dominik}, {Kalogera}, \& {Holz}}]{Fryer2012ApJ...749...91F}
{Fryer}, C.~L., {Belczynski}, K., {Wiktorowicz}, G., {et~al.} 2012, \apj, 749,
  91

\bibitem[{{Gaia Collaboration} {et~al.}(2018){Gaia Collaboration}, {Brown},
  {Vallenari}, {Prusti}, {de Bruijne}, {Babusiaux}, {Bailer-Jones}, {Biermann},
  {Evans}, {Eyer}, {Jansen}, {Jordi}, {Klioner}, {Lammers}, {Lindegren},
  {Luri}, {Mignard}, {Panem}, {Pourbaix}, {Randich}, {Sartoretti}, {Siddiqui},
  {Soubiran}, {van Leeuwen}, {Walton}, {Arenou}, {Bastian}, {Cropper},
  {Drimmel}, {Katz}, {Lattanzi}, {Bakker}, {Cacciari}, {Casta{\~n}eda},
  {Chaoul}, {Cheek}, {De Angeli}, {Fabricius}, {Guerra}, {Holl}, {Masana},
  {Messineo}, {Mowlavi}, {Nienartowicz}, {Panuzzo}, {Portell}, {Riello},
  {Seabroke}, {Tanga}, {Th{\'e}venin}, {Gracia-Abril}, {Comoretto},
  {Garcia-Reinaldos}, {Teyssier}, {Altmann}, {Andrae}, {Audard},
  {Bellas-Velidis}, {Benson}, {Berthier}, {Blomme}, {Burgess}, {Busso},
  {Carry}, {Cellino}, {Clementini}, {Clotet}, {Creevey}, {Davidson}, {De
  Ridder}, {Delchambre}, {Dell'Oro}, {Ducourant},
  {Fern{\'a}ndez-Hern{\'a}ndez}, {Fouesneau}, {Fr{\'e}mat}, {Galluccio},
  {Garc{\'\i}a-Torres}, {Gonz{\'a}lez-N{\'u}{\~n}ez}, {Gonz{\'a}lez-Vidal},
  {Gosset}, {Guy}, {Halbwachs}, {Hambly}, {Harrison}, {Hern{\'a}ndez},
  {Hestroffer}, {Hodgkin}, {Hutton}, {Jasniewicz}, {Jean-Antoine-Piccolo},
  {Jordan}, {Korn}, {Krone-Martins}, {Lanzafame}, {Lebzelter}, {L{\"o}ffler},
  {Manteiga}, {Marrese}, {Mart{\'\i}n-Fleitas}, {Moitinho}, {Mora}, {Muinonen},
  {Osinde}, {Pancino}, {Pauwels}, {Petit}, {Recio-Blanco}, {Richards},
  {Rimoldini}, {Robin}, {Sarro}, {Siopis}, {Smith}, {Sozzetti}, {S{\"u}veges},
  {Torra}, {van Reeven}, {Abbas}, {Abreu Aramburu}, {Accart}, {Aerts},
  {Altavilla}, {{\'A}lvarez}, {Alvarez}, {Alves}, {Anderson}, {Andrei},
  {Anglada Varela}, {Antiche}, {Antoja}, {Arcay}, {Astraatmadja}, {Bach},
  {Baker}, {Balaguer-N{\'u}{\~n}ez}, {Balm}, {Barache}, {Barata}, {Barbato},
  {Barblan}, {Barklem}, {Barrado}, {Barros}, {Barstow}, {Bartholom{\'e}
  Mu{\~n}oz}, {Bassilana}, {Becciani}, {Bellazzini}, {Berihuete}, {Bertone},
  {Bianchi}, {Bienaym{\'e}}, {Blanco-Cuaresma}, {Boch}, {Boeche}, {Bombrun},
  {Borrachero}, {Bossini}, {Bouquillon}, {Bourda}, {Bragaglia}, {Bramante},
  {Breddels}, {Bressan}, {Brouillet}, {Br{\"u}semeister}, {Brugaletta},
  {Bucciarelli}, {Burlacu}, {Busonero}, {Butkevich}, {Buzzi}, {Caffau},
  {Cancelliere}, {Cannizzaro}, {Cantat-Gaudin}, {Carballo}, {Carlucci},
  {Carrasco}, {Casamiquela}, {Castellani}, {Castro-Ginard}, {Charlot},
  {Chemin}, {Chiavassa}, {Cocozza}, {Costigan}, {Cowell}, {Crifo}, {Crosta},
  {Crowley}, {Cuypers}, {Dafonte}, {Damerdji}, {Dapergolas}, {David}, {David},
  {de Laverny}, {De Luise}, {De March}, {de Martino}, {de Souza}, {de Torres},
  {Debosscher}, {del Pozo}, {Delbo}, {Delgado}, {Delgado}, {Di Matteo},
  {Diakite}, {Diener}, {Distefano}, {Dolding}, {Drazinos}, {Dur{\'a}n},
  {Edvardsson}, {Enke}, {Eriksson}, {Esquej}, {Eynard Bontemps}, {Fabre},
  {Fabrizio}, {Faigler}, {Falc{\~a}o}, {Farr{\`a}s Casas}, {Federici},
  {Fedorets}, {Fernique}, {Figueras}, {Filippi}, {Findeisen}, {Fonti},
  {Fraile}, {Fraser}, {Fr{\'e}zouls}, {Gai}, {Galleti}, {Garabato},
  {Garc{\'\i}a-Sedano}, {Garofalo}, {Garralda}, {Gavel}, {Gavras}, {Gerssen},
  {Geyer}, {Giacobbe}, {Gilmore}, {Girona}, {Giuffrida}, {Glass}, {Gomes},
  {Granvik}, {Gueguen}, {Guerrier}, {Guiraud}, {Guti{\'e}rrez-S{\'a}nchez},
  {Haigron}, {Hatzidimitriou}, {Hauser}, {Haywood}, {Heiter}, {Helmi}, {Heu},
  {Hilger}, {Hobbs}, {Hofmann}, {Holland}, {Huckle}, {Hypki}, {Icardi},
  {Jan{\ss}en}, {Jevardat de Fombelle}, {Jonker}, {Juh{\'a}sz}, {Julbe},
  {Karampelas}, {Kewley}, {Klar}, {Kochoska}, {Kohley}, {Kolenberg},
  {Kontizas}, {Kontizas}, {Koposov}, {Kordopatis}, {Kostrzewa-Rutkowska},
  {Koubsky}, {Lambert}, {Lanza}, {Lasne}, {Lavigne}, {Le Fustec}, {Le
  Poncin-Lafitte}, {Lebreton}, {Leccia}, {Leclerc}, {Lecoeur-Taibi},
  {Lenhardt}, {Leroux}, {Liao}, {Licata}, {Lindstr{\o}m}, {Lister}, {Livanou},
  {Lobel}, {L{\'o}pez}, {Managau}, {Mann}, {Mantelet}, {Marchal}, {Marchant},
  {Marconi}, {Marinoni}, {Marschalk{\'o}}, {Marshall}, {Martino}, {Marton},
  {Mary}, {Massari}, {Matijevi{\v{c}}}, {Mazeh}, {McMillan}, {Messina},
  {Michalik}, {Millar}, {Molina}, {Molinaro}, {Moln{\'a}r}, {Montegriffo},
  {Mor}, {Morbidelli}, {Morel}, {Morris}, {Mulone}, {Muraveva}, {Musella},
  {Nelemans}, {Nicastro}, {Noval}, {O'Mullane}, {Ord{\'e}novic},
  {Ord{\'o}{\~n}ez-Blanco}, {Osborne}, {Pagani}, {Pagano}, {Pailler},
  {Palacin}, {Palaversa}, {Panahi}, {Pawlak}, {Piersimoni}, {Pineau}, {Plachy},
  {Plum}, {Poggio}, {Poujoulet}, {Pr{\v{s}}a}, {Pulone}, {Racero}, {Ragaini},
  {Rambaux}, {Ramos-Lerate}, {Regibo}, {Reyl{\'e}}, {Riclet}, {Ripepi}, {Riva},
  {Rivard}, {Rixon}, {Roegiers}, {Roelens}, {Romero-G{\'o}mez}, {Rowell},
  {Royer}, {Ruiz-Dern}, {Sadowski}, {Sagrist{\`a} Sell{\'e}s}, {Sahlmann},
  {Salgado}, {Salguero}, {Sanna}, {Santana-Ros}, {Sarasso}, {Savietto},
  {Schultheis}, {Sciacca}, {Segol}, {Segovia}, {S{\'e}gransan}, {Shih},
  {Siltala}, {Silva}, {Smart}, {Smith}, {Solano}, {Solitro}, {Sordo}, {Soria
  Nieto}, {Souchay}, {Spagna}, {Spoto}, {Stampa}, {Steele},
  {Steidelm{\"u}ller}, {Stephenson}, {Stoev}, {Suess}, {Surdej}, {Szabados},
  {Szegedi-Elek}, {Tapiador}, {Taris}, {Tauran}, {Taylor}, {Teixeira},
  {Terrett}, {Teyssandier}, {Thuillot}, {Titarenko}, {Torra Clotet}, {Turon},
  {Ulla}, {Utrilla}, {Uzzi}, {Vaillant}, {Valentini}, {Valette}, {van Elteren},
  {Van Hemelryck}, {van Leeuwen}, {Vaschetto}, {Vecchiato}, {Veljanoski},
  {Viala}, {Vicente}, {Vogt}, {von Essen}, {Voss}, {Votruba}, {Voutsinas},
  {Walmsley}, {Weiler}, {Wertz}, {Wevers}, {Wyrzykowski}, {Yoldas},
  {{\v{Z}}erjal}, {Ziaeepour}, {Zorec}, {Zschocke}, {Zucker}, {Zurbach}, \&
  {Zwitter}}]{Gaia2018A&A...616A...1G}
{Gaia Collaboration}, {Brown}, A.~G.~A., {Vallenari}, A., {et~al.} 2018, \aap,
  616, A1

\bibitem[{{Gaia Collaboration} {et~al.}(2021){Gaia Collaboration}, {Brown},
  {Vallenari}, {Prusti}, {de Bruijne}, {Babusiaux}, {Biermann}, {Creevey},
  {Evans}, {Eyer}, {Hutton}, {Jansen}, {Jordi}, {Klioner}, {Lammers},
  {Lindegren}, {Luri}, {Mignard}, {Panem}, {Pourbaix}, {Randich}, {Sartoretti},
  {Soubiran}, {Walton}, {Arenou}, {Bailer-Jones}, {Bastian}, {Cropper},
  {Drimmel}, {Katz}, {Lattanzi}, {van Leeuwen}, {Bakker}, {Cacciari},
  {Casta{\~n}eda}, {De Angeli}, {Ducourant}, {Fabricius}, {Fouesneau},
  {Fr{\'e}mat}, {Guerra}, {Guerrier}, {Guiraud}, {Jean-Antoine Piccolo},
  {Masana}, {Messineo}, {Mowlavi}, {Nicolas}, {Nienartowicz}, {Pailler},
  {Panuzzo}, {Riclet}, {Roux}, {Seabroke}, {Sordo}, {Tanga}, {Th{\'e}venin},
  {Gracia-Abril}, {Portell}, {Teyssier}, {Altmann}, {Andrae}, {Bellas-Velidis},
  {Benson}, {Berthier}, {Blomme}, {Brugaletta}, {Burgess}, {Busso}, {Carry},
  {Cellino}, {Cheek}, {Clementini}, {Damerdji}, {Davidson}, {Delchambre},
  {Dell'Oro}, {Fern{\'a}ndez-Hern{\'a}ndez}, {Galluccio}, {Garc{\'\i}a-Lario},
  {Garcia-Reinaldos}, {Gonz{\'a}lez-N{\'u}{\~n}ez}, {Gosset}, {Haigron},
  {Halbwachs}, {Hambly}, {Harrison}, {Hatzidimitriou}, {Heiter},
  {Hern{\'a}ndez}, {Hestroffer}, {Hodgkin}, {Holl}, {Jan{\ss}en}, {Jevardat de
  Fombelle}, {Jordan}, {Krone-Martins}, {Lanzafame}, {L{\"o}ffler}, {Lorca},
  {Manteiga}, {Marchal}, {Marrese}, {Moitinho}, {Mora}, {Muinonen}, {Osborne},
  {Pancino}, {Pauwels}, {Petit}, {Recio-Blanco}, {Richards}, {Riello},
  {Rimoldini}, {Robin}, {Roegiers}, {Rybizki}, {Sarro}, {Siopis}, {Smith},
  {Sozzetti}, {Ulla}, {Utrilla}, {van Leeuwen}, {van Reeven}, {Abbas}, {Abreu
  Aramburu}, {Accart}, {Aerts}, {Aguado}, {Ajaj}, {Altavilla}, {{\'A}lvarez},
  {{\'A}lvarez Cid-Fuentes}, {Alves}, {Anderson}, {Anglada Varela}, {Antoja},
  {Audard}, {Baines}, {Baker}, {Balaguer-N{\'u}{\~n}ez}, {Balbinot}, {Balog},
  {Barache}, {Barbato}, {Barros}, {Barstow}, {Bartolom{\'e}}, {Bassilana},
  {Bauchet}, {Baudesson-Stella}, {Becciani}, {Bellazzini}, {Bernet}, {Bertone},
  {Bianchi}, {Blanco-Cuaresma}, {Boch}, {Bombrun}, {Bossini}, {Bouquillon},
  {Bragaglia}, {Bramante}, {Breedt}, {Bressan}, {Brouillet}, {Bucciarelli},
  {Burlacu}, {Busonero}, {Butkevich}, {Buzzi}, {Caffau}, {Cancelliere},
  {C{\'a}novas}, {Cantat-Gaudin}, {Carballo}, {Carlucci}, {Carnerero},
  {Carrasco}, {Casamiquela}, {Castellani}, {Castro-Ginard}, {Castro Sampol},
  {Chaoul}, {Charlot}, {Chemin}, {Chiavassa}, {Cioni}, {Comoretto}, {Cooper},
  {Cornez}, {Cowell}, {Crifo}, {Crosta}, {Crowley}, {Dafonte}, {Dapergolas},
  {David}, {David}, {de Laverny}, {De Luise}, {De March}, {De Ridder}, {de
  Souza}, {de Teodoro}, {de Torres}, {del Peloso}, {del Pozo}, {Delbo},
  {Delgado}, {Delgado}, {Delisle}, {Di Matteo}, {Diakite}, {Diener},
  {Distefano}, {Dolding}, {Eappachen}, {Edvardsson}, {Enke}, {Esquej}, {Fabre},
  {Fabrizio}, {Faigler}, {Fedorets}, {Fernique}, {Fienga}, {Figueras},
  {Fouron}, {Fragkoudi}, {Fraile}, {Franke}, {Gai}, {Garabato},
  {Garcia-Gutierrez}, {Garc{\'\i}a-Torres}, {Garofalo}, {Gavras}, {Gerlach},
  {Geyer}, {Giacobbe}, {Gilmore}, {Girona}, {Giuffrida}, {Gomel}, {Gomez},
  {Gonzalez-Santamaria}, {Gonz{\'a}lez-Vidal}, {Granvik},
  {Guti{\'e}rrez-S{\'a}nchez}, {Guy}, {Hauser}, {Haywood}, {Helmi}, {Hidalgo},
  {Hilger}, {H{\l}adczuk}, {Hobbs}, {Holland}, {Huckle}, {Jasniewicz},
  {Jonker}, {Juaristi Campillo}, {Julbe}, {Karbevska}, {Kervella}, {Khanna},
  {Kochoska}, {Kontizas}, {Kordopatis}, {Korn}, {Kostrzewa-Rutkowska},
  {Kruszy{\'n}ska}, {Lambert}, {Lanza}, {Lasne}, {Le Campion}, {Le Fustec},
  {Lebreton}, {Lebzelter}, {Leccia}, {Leclerc}, {Lecoeur-Taibi}, {Liao},
  {Licata}, {Lindstr{\o}m}, {Lister}, {Livanou}, {Lobel}, {Madrero Pardo},
  {Managau}, {Mann}, {Marchant}, {Marconi}, {Marcos Santos}, {Marinoni},
  {Marocco}, {Marshall}, {Martin Polo}, {Mart{\'\i}n-Fleitas}, {Masip},
  {Massari}, {Mastrobuono-Battisti}, {Mazeh}, {McMillan}, {Messina},
  {Michalik}, {Millar}, {Mints}, {Molina}, {Molinaro}, {Moln{\'a}r},
  {Montegriffo}, {Mor}, {Morbidelli}, {Morel}, {Morris}, {Mulone}, {Munoz},
  {Muraveva}, {Murphy}, {Musella}, {Noval}, {Ord{\'e}novic}, {Orr{\`u}},
  {Osinde}, {Pagani}, {Pagano}, {Palaversa}, {Palicio}, {Panahi}, {Pawlak},
  {Pe{\~n}alosa Esteller}, {Penttil{\"a}}, {Piersimoni}, {Pineau}, {Plachy},
  {Plum}, {Poggio}, {Poretti}, {Poujoulet}, {Pr{\v{s}}a}, {Pulone}, {Racero},
  {Ragaini}, {Rainer}, {Raiteri}, {Rambaux}, {Ramos}, {Ramos-Lerate}, {Re
  Fiorentin}, {Regibo}, {Reyl{\'e}}, {Ripepi}, {Riva}, {Rixon}, {Robichon},
  {Robin}, {Roelens}, {Rohrbasser}, {Romero-G{\'o}mez}, {Rowell}, {Royer},
  {Rybicki}, {Sadowski}, {Sagrist{\`a} Sell{\'e}s}, {Sahlmann}, {Salgado},
  {Salguero}, {Samaras}, {Sanchez Gimenez}, {Sanna}, {Santove{\~n}a},
  {Sarasso}, {Schultheis}, {Sciacca}, {Segol}, {Segovia}, {S{\'e}gransan},
  {Semeux}, {Shahaf}, {Siddiqui}, {Siebert}, {Siltala}, {Slezak}, {Smart},
  {Solano}, {Solitro}, {Souami}, {Souchay}, {Spagna}, {Spoto}, {Steele},
  {Steidelm{\"u}ller}, {Stephenson}, {S{\"u}veges}, {Szabados}, {Szegedi-Elek},
  {Taris}, {Tauran}, {Taylor}, {Teixeira}, {Thuillot}, {Tonello}, {Torra},
  {Torra}, {Turon}, {Unger}, {Vaillant}, {van Dillen}, {Vanel}, {Vecchiato},
  {Viala}, {Vicente}, {Voutsinas}, {Weiler}, {Wevers}, {Wyrzykowski}, {Yoldas},
  {Yvard}, {Zhao}, {Zorec}, {Zucker}, {Zurbach}, \&
  {Zwitter}}]{Gaia2021A&A...649A...1G}
{Gaia Collaboration}, {Brown}, A.~G.~A., {Vallenari}, A., {et~al.} 2021, \aap,
  649, A1

\bibitem[{{Gaia Collaboration} {et~al.}(2016){Gaia Collaboration}, {Prusti},
  {de Bruijne}, {Brown}, {Vallenari}, {Babusiaux}, {Bailer-Jones}, {Bastian},
  {Biermann}, {Evans}, {Eyer}, {Jansen}, {Jordi}, {Klioner}, {Lammers},
  {Lindegren}, {Luri}, {Mignard}, {Milligan}, {Panem}, {Poinsignon},
  {Pourbaix}, {Randich}, {Sarri}, {Sartoretti}, {Siddiqui}, {Soubiran},
  {Valette}, {van Leeuwen}, {Walton}, {Aerts}, {Arenou}, {Cropper}, {Drimmel},
  {H{\o}g}, {Katz}, {Lattanzi}, {O'Mullane}, {Grebel}, {Holland}, {Huc},
  {Passot}, {Bramante}, {Cacciari}, {Casta{\~n}eda}, {Chaoul}, {Cheek}, {De
  Angeli}, {Fabricius}, {Guerra}, {Hern{\'a}ndez}, {Jean-Antoine-Piccolo},
  {Masana}, {Messineo}, {Mowlavi}, {Nienartowicz}, {Ord{\'o}{\~n}ez-Blanco},
  {Panuzzo}, {Portell}, {Richards}, {Riello}, {Seabroke}, {Tanga},
  {Th{\'e}venin}, {Torra}, {Els}, {Gracia-Abril}, {Comoretto},
  {Garcia-Reinaldos}, {Lock}, {Mercier}, {Altmann}, {Andrae}, {Astraatmadja},
  {Bellas-Velidis}, {Benson}, {Berthier}, {Blomme}, {Busso}, {Carry},
  {Cellino}, {Clementini}, {Cowell}, {Creevey}, {Cuypers}, {Davidson}, {De
  Ridder}, {de Torres}, {Delchambre}, {Dell'Oro}, {Ducourant}, {Fr{\'e}mat},
  {Garc{\'\i}a-Torres}, {Gosset}, {Halbwachs}, {Hambly}, {Harrison}, {Hauser},
  {Hestroffer}, {Hodgkin}, {Huckle}, {Hutton}, {Jasniewicz}, {Jordan},
  {Kontizas}, {Korn}, {Lanzafame}, {Manteiga}, {Moitinho}, {Muinonen},
  {Osinde}, {Pancino}, {Pauwels}, {Petit}, {Recio-Blanco}, {Robin}, {Sarro},
  {Siopis}, {Smith}, {Smith}, {Sozzetti}, {Thuillot}, {van Reeven}, {Viala},
  {Abbas}, {Abreu Aramburu}, {Accart}, {Aguado}, {Allan}, {Allasia},
  {Altavilla}, {{\'A}lvarez}, {Alves}, {Anderson}, {Andrei}, {Anglada Varela},
  {Antiche}, {Antoja}, {Ant{\'o}n}, {Arcay}, {Atzei}, {Ayache}, {Bach},
  {Baker}, {Balaguer-N{\'u}{\~n}ez}, {Barache}, {Barata}, {Barbier}, {Barblan},
  {Baroni}, {Barrado y Navascu{\'e}s}, {Barros}, {Barstow}, {Becciani},
  {Bellazzini}, {Bellei}, {Bello Garc{\'\i}a}, {Belokurov}, {Bendjoya},
  {Berihuete}, {Bianchi}, {Bienaym{\'e}}, {Billebaud}, {Blagorodnova},
  {Blanco-Cuaresma}, {Boch}, {Bombrun}, {Borrachero}, {Bouquillon}, {Bourda},
  {Bouy}, {Bragaglia}, {Breddels}, {Brouillet}, {Br{\"u}semeister},
  {Bucciarelli}, {Budnik}, {Burgess}, {Burgon}, {Burlacu}, {Busonero}, {Buzzi},
  {Caffau}, {Cambras}, {Campbell}, {Cancelliere}, {Cantat-Gaudin}, {Carlucci},
  {Carrasco}, {Castellani}, {Charlot}, {Charnas}, {Charvet}, {Chassat},
  {Chiavassa}, {Clotet}, {Cocozza}, {Collins}, {Collins}, {Costigan}, {Crifo},
  {Cross}, {Crosta}, {Crowley}, {Dafonte}, {Damerdji}, {Dapergolas}, {David},
  {David}, {De Cat}, {de Felice}, {de Laverny}, {De Luise}, {De March}, {de
  Martino}, {de Souza}, {Debosscher}, {del Pozo}, {Delbo}, {Delgado},
  {Delgado}, {di Marco}, {Di Matteo}, {Diakite}, {Distefano}, {Dolding}, {Dos
  Anjos}, {Drazinos}, {Dur{\'a}n}, {Dzigan}, {Ecale}, {Edvardsson}, {Enke},
  {Erdmann}, {Escolar}, {Espina}, {Evans}, {Eynard Bontemps}, {Fabre},
  {Fabrizio}, {Faigler}, {Falc{\~a}o}, {Farr{\`a}s Casas}, {Faye}, {Federici},
  {Fedorets}, {Fern{\'a}ndez-Hern{\'a}ndez}, {Fernique}, {Fienga}, {Figueras},
  {Filippi}, {Findeisen}, {Fonti}, {Fouesneau}, {Fraile}, {Fraser}, {Fuchs},
  {Furnell}, {Gai}, {Galleti}, {Galluccio}, {Garabato}, {Garc{\'\i}a-Sedano},
  {Gar{\'e}}, {Garofalo}, {Garralda}, {Gavras}, {Gerssen}, {Geyer}, {Gilmore},
  {Girona}, {Giuffrida}, {Gomes}, {Gonz{\'a}lez-Marcos},
  {Gonz{\'a}lez-N{\'u}{\~n}ez}, {Gonz{\'a}lez-Vidal}, {Granvik}, {Guerrier},
  {Guillout}, {Guiraud}, {G{\'u}rpide}, {Guti{\'e}rrez-S{\'a}nchez}, {Guy},
  {Haigron}, {Hatzidimitriou}, {Haywood}, {Heiter}, {Helmi}, {Hobbs},
  {Hofmann}, {Holl}, {Holland}, {Hunt}, {Hypki}, {Icardi}, {Irwin}, {Jevardat
  de Fombelle}, {Jofr{\'e}}, {Jonker}, {Jorissen}, {Julbe}, {Karampelas},
  {Kochoska}, {Kohley}, {Kolenberg}, {Kontizas}, {Koposov}, {Kordopatis},
  {Koubsky}, {Kowalczyk}, {Krone-Martins}, {Kudryashova}, {Kull}, {Bachchan},
  {Lacoste-Seris}, {Lanza}, {Lavigne}, {Le Poncin-Lafitte}, {Lebreton},
  {Lebzelter}, {Leccia}, {Leclerc}, {Lecoeur-Taibi}, {Lemaitre}, {Lenhardt},
  {Leroux}, {Liao}, {Licata}, {Lindstr{\o}m}, {Lister}, {Livanou}, {Lobel},
  {L{\"o}ffler}, {L{\'o}pez}, {Lopez-Lozano}, {Lorenz}, {Loureiro},
  {MacDonald}, {Magalh{\~a}es Fernandes}, {Managau}, {Mann}, {Mantelet},
  {Marchal}, {Marchant}, {Marconi}, {Marie}, {Marinoni}, {Marrese},
  {Marschalk{\'o}}, {Marshall}, {Mart{\'\i}n-Fleitas}, {Martino}, {Mary},
  {Matijevi{\v{c}}}, {Mazeh}, {McMillan}, {Messina}, {Mestre}, {Michalik},
  {Millar}, {Miranda}, {Molina}, {Molinaro}, {Molinaro}, {Moln{\'a}r},
  {Moniez}, {Montegriffo}, {Monteiro}, {Mor}, {Mora}, {Morbidelli}, {Morel},
  {Morgenthaler}, {Morley}, {Morris}, {Mulone}, {Muraveva}, {Musella},
  {Narbonne}, {Nelemans}, {Nicastro}, {Noval}, {Ord{\'e}novic},
  {Ordieres-Mer{\'e}}, {Osborne}, {Pagani}, {Pagano}, {Pailler}, {Palacin},
  {Palaversa}, {Parsons}, {Paulsen}, {Pecoraro}, {Pedrosa}, {Pentik{\"a}inen},
  {Pereira}, {Pichon}, {Piersimoni}, {Pineau}, {Plachy}, {Plum}, {Poujoulet},
  {Pr{\v{s}}a}, {Pulone}, {Ragaini}, {Rago}, {Rambaux}, {Ramos-Lerate},
  {Ranalli}, {Rauw}, {Read}, {Regibo}, {Renk}, {Reyl{\'e}}, {Ribeiro},
  {Rimoldini}, {Ripepi}, {Riva}, {Rixon}, {Roelens}, {Romero-G{\'o}mez},
  {Rowell}, {Royer}, {Rudolph}, {Ruiz-Dern}, {Sadowski}, {Sagrist{\`a}
  Sell{\'e}s}, {Sahlmann}, {Salgado}, {Salguero}, {Sarasso}, {Savietto},
  {Schnorhk}, {Schultheis}, {Sciacca}, {Segol}, {Segovia}, {Segransan},
  {Serpell}, {Shih}, {Smareglia}, {Smart}, {Smith}, {Solano}, {Solitro},
  {Sordo}, {Soria Nieto}, {Souchay}, {Spagna}, {Spoto}, {Stampa}, {Steele},
  {Steidelm{\"u}ller}, {Stephenson}, {Stoev}, {Suess}, {S{\"u}veges}, {Surdej},
  {Szabados}, {Szegedi-Elek}, {Tapiador}, {Taris}, {Tauran}, {Taylor},
  {Teixeira}, {Terrett}, {Tingley}, {Trager}, {Turon}, {Ulla}, {Utrilla},
  {Valentini}, {van Elteren}, {Van Hemelryck}, {van Leeuwen}, {Varadi},
  {Vecchiato}, {Veljanoski}, {Via}, {Vicente}, {Vogt}, {Voss}, {Votruba},
  {Voutsinas}, {Walmsley}, {Weiler}, {Weingrill}, {Werner}, {Wevers},
  {Whitehead}, {Wyrzykowski}, {Yoldas}, {{\v{Z}}erjal}, {Zucker}, {Zurbach},
  {Zwitter}, {Alecu}, {Allen}, {Allende Prieto}, {Amorim},
  {Anglada-Escud{\'e}}, {Arsenijevic}, {Azaz}, {Balm}, {Beck}, {Bernstein},
  {Bigot}, {Bijaoui}, {Blasco}, {Bonfigli}, {Bono}, {Boudreault}, {Bressan},
  {Brown}, {Brunet}, {Bunclark}, {Buonanno}, {Butkevich}, {Carret}, {Carrion},
  {Chemin}, {Ch{\'e}reau}, {Corcione}, {Darmigny}, {de Boer}, {de Teodoro}, {de
  Zeeuw}, {Delle Luche}, {Domingues}, {Dubath}, {Fodor}, {Fr{\'e}zouls},
  {Fries}, {Fustes}, {Fyfe}, {Gallardo}, {Gallegos}, {Gardiol}, {Gebran},
  {Gomboc}, {G{\'o}mez}, {Grux}, {Gueguen}, {Heyrovsky}, {Hoar}, {Iannicola},
  {Isasi Parache}, {Janotto}, {Joliet}, {Jonckheere}, {Keil}, {Kim},
  {Klagyivik}, {Klar}, {Knude}, {Kochukhov}, {Kolka}, {Kos}, {Kutka}, {Lainey},
  {LeBouquin}, {Liu}, {Loreggia}, {Makarov}, {Marseille}, {Martayan},
  {Martinez-Rubi}, {Massart}, {Meynadier}, {Mignot}, {Munari}, {Nguyen},
  {Nordlander}, {Ocvirk}, {O'Flaherty}, {Olias Sanz}, {Ortiz}, {Osorio},
  {Oszkiewicz}, {Ouzounis}, {Palmer}, {Park}, {Pasquato}, {Peltzer}, {Peralta},
  {P{\'e}turaud}, {Pieniluoma}, {Pigozzi}, {Poels}, {Prat}, {Prod'homme},
  {Raison}, {Rebordao}, {Risquez}, {Rocca-Volmerange}, {Rosen}, {Ruiz-Fuertes},
  {Russo}, {Sembay}, {Serraller Vizcaino}, {Short}, {Siebert}, {Silva},
  {Sinachopoulos}, {Slezak}, {Soffel}, {Sosnowska}, {Strai{\v{z}}ys}, {ter
  Linden}, {Terrell}, {Theil}, {Tiede}, {Troisi}, {Tsalmantza}, {Tur},
  {Vaccari}, {Vachier}, {Valles}, {Van Hamme}, {Veltz}, {Virtanen}, {Wallut},
  {Wichmann}, {Wilkinson}, {Ziaeepour}, \&
  {Zschocke}}]{Gaia2016A&A...595A...1G}
{Gaia Collaboration}, {Prusti}, T., {de Bruijne}, J.~H.~J., {et~al.} 2016,
  \aap, 595, A1

\bibitem[{{Gaia Collaboration} {et~al.}(2022){Gaia Collaboration}, {Vallenari},
  {Brown}, {Prusti}, {de Bruijne}, {Arenou}, {Babusiaux}, {Biermann},
  {Creevey}, {Ducourant}, {Evans}, {Eyer}, {Guerra}, {Hutton}, {Jordi},
  {Klioner}, {Lammers}, {Lindegren}, {Luri}, {Mignard}, {Panem}, {Pourbaix},
  {Randich}, {Sartoretti}, {Soubiran}, {Tanga}, {Walton}, {Bailer-Jones},
  {Bastian}, {Drimmel}, {Jansen}, {Katz}, {Lattanzi}, {van Leeuwen}, {Bakker},
  {Cacciari}, {Casta{\~n}eda}, {De Angeli}, {Fabricius}, {Fouesneau},
  {Fr{\'e}mat}, {Galluccio}, {Guerrier}, {Heiter}, {Masana}, {Messineo},
  {Mowlavi}, {Nicolas}, {Nienartowicz}, {Pailler}, {Panuzzo}, {Riclet}, {Roux},
  {Seabroke}, {Sordo{\o}rcit}, {Th{\'e}venin}, {Gracia-Abril}, {Portell},
  {Teyssier}, {Altmann}, {Andrae}, {Audard}, {Bellas-Velidis}, {Benson},
  {Berthier}, {Blomme}, {Burgess}, {Busonero}, {Busso}, {C{\'a}novas}, {Carry},
  {Cellino}, {Cheek}, {Clementini}, {Damerdji}, {Davidson}, {de Teodoro},
  {Nu{\~n}ez Campos}, {Delchambre}, {Dell'Oro}, {Esquej},
  {Fern{\'a}ndez-Hern{\'a}ndez}, {Fraile}, {Garabato}, {Garc{\'\i}a-Lario},
  {Gosset}, {Haigron}, {Halbwachs}, {Hambly}, {Harrison}, {Hern{\'a}ndez},
  {Hestroffer}, {Hodgkin}, {Holl}, {Jan{\ss}en}, {Jevardat de Fombelle},
  {Jordan}, {Krone-Martins}, {Lanzafame}, {L{\"o}ffler}, {Marchal}, {Marrese},
  {Moitinho}, {Muinonen}, {Osborne}, {Pancino}, {Pauwels}, {Recio-Blanco},
  {Reyl{\'e}}, {Riello}, {Rimoldini}, {Roegiers}, {Rybizki}, {Sarro}, {Siopis},
  {Smith}, {Sozzetti}, {Utrilla}, {van Leeuwen}, {Abbas}, {{\'A}brah{\'a}m},
  {Abreu Aramburu}, {Aerts}, {Aguado}, {Ajaj}, {Aldea-Montero}, {Altavilla},
  {{\'A}lvarez}, {Alves}, {Anders}, {Anderson}, {Anglada Varela}, {Antoja},
  {Baines}, {Baker}, {Balaguer-N{\'u}{\~n}ez}, {Balbinot}, {Balog}, {Barache},
  {Barbato}, {Barros}, {Barstow}, {Bartolom{\'e}}, {Bassilana}, {Bauchet},
  {Becciani}, {Bellazzini}, {Berihuete}, {Bernet}, {Bertone}, {Bianchi},
  {Binnenfeld}, {Blanco-Cuaresma}, {Blazere}, {Boch}, {Bombrun}, {Bossini},
  {Bouquillon}, {Bragaglia}, {Bramante}, {Breedt}, {Bressan}, {Brouillet},
  {Brugaletta}, {Bucciarelli}, {Burlacu}, {Butkevich}, {Buzzi}, {Caffau},
  {Cancelliere}, {Cantat-Gaudin}, {Carballo}, {Carlucci}, {Carnerero},
  {Carrasco}, {Casamiquela}, {Castellani}, {Castro-Ginard}, {Chaoul},
  {Charlot}, {Chemin}, {Chiaramida}, {Chiavassa}, {Chornay}, {Comoretto},
  {Contursi}, {Cooper}, {Cornez}, {Cowell}, {Crifo}, {Cropper}, {Crosta},
  {Crowley}, {Dafonte}, {Dapergolas}, {David}, {David}, {de Laverny}, {De
  Luise}, {De March}, {De Ridder}, {de Souza}, {de Torres}, {del Peloso}, {del
  Pozo}, {Delbo}, {Delgado}, {Delisle}, {Demouchy}, {Dharmawardena}, {Di
  Matteo}, {Diakite}, {Diener}, {Distefano}, {Dolding}, {Edvardsson}, {Enke},
  {Fabre}, {Fabrizio}, {Faigler}, {Fedorets}, {Fernique}, {Fienga}, {Figueras},
  {Fournier}, {Fouron}, {Fragkoudi}, {Gai}, {Garcia-Gutierrez},
  {Garcia-Reinaldos}, {Garc{\'\i}a-Torres}, {Garofalo}, {Gavel}, {Gavras},
  {Gerlach}, {Geyer}, {Giacobbe}, {Gilmore}, {Girona}, {Giuffrida}, {Gomel},
  {Gomez}, {Gonz{\'a}lez-N{\'u}{\~n}ez}, {Gonz{\'a}lez-Santamar{\'\i}a},
  {Gonz{\'a}lez-Vidal}, {Granvik}, {Guillout}, {Guiraud},
  {Guti{\'e}rrez-S{\'a}nchez}, {Guy}, {Hatzidimitriou}, {Hauser}, {Haywood},
  {Helmer}, {Helmi}, {Sarmiento}, {Hidalgo}, {Hilger}, {H{\l}adczuk}, {Hobbs},
  {Holland}, {Huckle}, {Jardine}, {Jasniewicz}, {Jean-Antoine Piccolo},
  {Jim{\'e}nez-Arranz}, {Jorissen}, {Juaristi Campillo}, {Julbe}, {Karbevska},
  {Kervella}, {Khanna}, {Kontizas}, {Kordopatis}, {Korn}, {K{\'o}sp{\'a}l},
  {Kostrzewa-Rutkowska}, {Kruszy{\'n}ska}, {Kun}, {Laizeau}, {Lambert},
  {Lanza}, {Lasne}, {Le Campion}, {Lebreton}, {Lebzelter}, {Leccia}, {Leclerc},
  {Lecoeur-Taibi}, {Liao}, {Licata}, {Lindstr{\o}m}, {Lister}, {Livanou},
  {Lobel}, {Lorca}, {Loup}, {Madrero Pardo}, {Magdaleno Romeo}, {Managau},
  {Mann}, {Manteiga}, {Marchant}, {Marconi}, {Marcos}, {Marcos Santos},
  {Mar{\'\i}n Pina}, {Marinoni}, {Marocco}, {Marshall}, {Polo},
  {Mart{\'\i}n-Fleitas}, {Marton}, {Mary}, {Masip}, {Massari},
  {Mastrobuono-Battisti}, {Mazeh}, {McMillan}, {Messina}, {Michalik}, {Millar},
  {Mints}, {Molina}, {Molinaro}, {Moln{\'a}r}, {Monari}, {Mongui{\'o}},
  {Montegriffo}, {Montero}, {Mor}, {Mora}, {Morbidelli}, {Morel}, {Morris},
  {Muraveva}, {Murphy}, {Musella}, {Nagy}, {Noval}, {Oca{\~n}a}, {Ogden},
  {Ordenovic}, {Osinde}, {Pagani}, {Pagano}, {Palaversa}, {Palicio},
  {Pallas-Quintela}, {Panahi}, {Payne-Wardenaar}, {Pe{\~n}alosa Esteller},
  {Penttil{\"a}}, {Pichon}, {Piersimoni}, {Pineau}, {Plachy}, {Plum}, {Poggio},
  {Pr{\v{s}}a}, {Pulone}, {Racero}, {Ragaini}, {Rainer}, {Raiteri}, {Rambaux},
  {Ramos}, {Ramos-Lerate}, {Re Fiorentin}, {Regibo}, {Richards}, {Rios Diaz},
  {Ripepi}, {Riva}, {Rix}, {Rixon}, {Robichon}, {Robin}, {Robin}, {Roelens},
  {Rogues}, {Rohrbasser}, {Romero-G{\'o}mez}, {Rowell}, {Royer}, {Ruz Mieres},
  {Rybicki}, {Sadowski}, {S{\'a}ez N{\'u}{\~n}ez}, {Sagrist{\`a} Sell{\'e}s},
  {Sahlmann}, {Salguero}, {Samaras}, {Sanchez Gimenez}, {Sanna},
  {Santove{\~n}a}, {Sarasso}, {Schultheis}, {Sciacca}, {Segol}, {Segovia},
  {S{\'e}gransan}, {Semeux}, {Shahaf}, {Siddiqui}, {Siebert}, {Siltala},
  {Silvelo}, {Slezak}, {Slezak}, {Smart}, {Snaith}, {Solano}, {Solitro},
  {Souami}, {Souchay}, {Spagna}, {Spina}, {Spoto}, {Steele},
  {Steidelm{\"u}ller}, {Stephenson}, {S{\"u}veges}, {Surdej}, {Szabados},
  {Szegedi-Elek}, {Taris}, {Taylo}, {Teixeira}, {Tolomei}, {Tonello}, {Torra},
  {Torra}, {Torralba Elipe}, {Trabucchi}, {Tsounis}, {Turon}, {Ulla}, {Unger},
  {Vaillant}, {van Dillen}, {van Reeven}, {Vanel}, {Vecchiato}, {Viala},
  {Vicente}, {Voutsinas}, {Weiler}, {Wevers}, {Wyrzykowski}, {Yoldas}, {Yvard},
  {Zhao}, {Zorec}, {Zucker}, \& {Zwitter}}]{Gaia2022arXiv220800211G}
{Gaia Collaboration}, {Vallenari}, A., {Brown}, A.~G.~A., {et~al.} 2022, arXiv
  e-prints, arXiv:2208.00211

\bibitem[{{Gilmore} {et~al.}(2012){Gilmore}, {Randich}, {Asplund}, {Binney},
  {Bonifacio}, {Drew}, {Feltzing}, {Ferguson}, {Jeffries}, {Micela},
  {Negueruela}, {Prusti}, {Rix}, {Vallenari}, {Alfaro}, {Allende-Prieto},
  {Babusiaux}, {Bensby}, {Blomme}, {Bragaglia}, {Flaccomio}, {Fran{\c{c}}ois},
  {Irwin}, {Koposov}, {Korn}, {Lanzafame}, {Pancino}, {Paunzen},
  {Recio-Blanco}, {Sacco}, {Smiljanic}, {Van Eck}, {Walton}, {Aden}, {Aerts},
  {Affer}, {Alcala}, {Altavilla}, {Alves}, {Antoja}, {Arenou}, {Argiroffi},
  {Asensio Ramos}, {Bailer-Jones}, {Balaguer-Nunez}, {Bayo}, {Barbuy},
  {Barisevicius}, {Barrado y Navascues}, {Battistini}, {Bellas Velidis},
  {Bellazzini}, {Belokurov}, {Bergemann}, {Bertelli}, {Biazzo}, {Bienayme},
  {Bland-Hawthorn}, {Boeche}, {Bonito}, {Boudreault}, {Bouvier}, {Brandao},
  {Brown}, {de Bruijne}, {Burleigh}, {Caballero}, {Caffau}, {Calura},
  {Capuzzo-Dolcetta}, {Caramazza}, {Carraro}, {Casagrande}, {Casewell},
  {Chapman}, {Chiappini}, {Chorniy}, {Christlieb}, {Cignoni}, {Cocozza},
  {Colless}, {Collet}, {Collins}, {Correnti}, {Covino}, {Crnojevic}, {Cropper},
  {Cunha}, {Damiani}, {David}, {Delgado}, {Duffau}, {Edvardsson}, {Eldridge},
  {Enke}, {Eriksson}, {Evans}, {Eyer}, {Famaey}, {Fellhauer}, {Ferreras},
  {Figueras}, {Fiorentino}, {Flynn}, {Folha}, {Franciosini}, {Frasca},
  {Freeman}, {Fremat}, {Friel}, {Gaensicke}, {Gameiro}, {Garzon}, {Geier},
  {Geisler}, {Gerhard}, {Gibson}, {Gomboc}, {Gomez}, {Gonzalez-Fernandez},
  {Gonzalez Hernandez}, {Gosset}, {Grebel}, {Greimel}, {Groenewegen},
  {Grundahl}, {Guarcello}, {Gustafsson}, {Hadrava}, {Hatzidimitriou}, {Hambly},
  {Hammersley}, {Hansen}, {Haywood}, {Heber}, {Heiter}, {Held}, {Helmi},
  {Hensler}, {Herrero}, {Hill}, {Hodgkin}, {Huelamo}, {Huxor}, {Ibata},
  {Jackson}, {de Jong}, {Jonker}, {Jordan}, {Jordi}, {Jorissen}, {Katz},
  {Kawata}, {Keller}, {Kharchenko}, {Klement}, {Klutsch}, {Knude}, {Koch},
  {Kochukhov}, {Kontizas}, {Koubsky}, {Lallement}, {de Laverny}, {van Leeuwen},
  {Lemasle}, {Lewis}, {Lind}, {Lindstrom}, {Lobel}, {Lopez Santiago}, {Lucas},
  {Ludwig}, {Lueftinger}, {Magrini}, {Maiz Apellaniz}, {Maldonado}, {Marconi},
  {Marino}, {Martayan}, {Martinez-Valpuesta}, {Matijevic}, {McMahon},
  {Messina}, {Meyer}, {Miglio}, {Mikolaitis}, {Minchev}, {Minniti}, {Moitinho},
  {Momany}, {Monaco}, {Montalto}, {Monteiro}, {Monier}, {Montes}, {Mora},
  {Moraux}, {Morel}, {Mowlavi}, {Mucciarelli}, {Munari}, {Napiwotzki},
  {Nardetto}, {Naylor}, {Naze}, {Nelemans}, {Okamoto}, {Ortolani}, {Pace},
  {Palla}, {Palous}, {Parker}, {Penarrubia}, {Pillitteri}, {Piotto}, {Posbic},
  {Prisinzano}, {Puzeras}, {Quirrenbach}, {Ragaini}, {Read}, {Read}, {Reyle},
  {De Ridder}, {Robichon}, {Robin}, {Roeser}, {Romano}, {Royer}, {Ruchti},
  {Ruzicka}, {Ryan}, {Ryde}, {Santos}, {Sanz Forcada}, {Sarro Baro},
  {Sbordone}, {Schilbach}, {Schmeja}, {Schnurr}, {Schoenrich}, {Scholz},
  {Seabroke}, {Sharma}, {De Silva}, {Smith}, {Solano}, {Sordo}, {Soubiran},
  {Sousa}, {Spagna}, {Steffen}, {Steinmetz}, {Stelzer}, {Stempels},
  {Tabernero}, {Tautvaisiene}, {Thevenin}, {Torra}, {Tosi}, {Tolstoy}, {Turon},
  {Walker}, {Wambsganss}, {Worley}, {Venn}, {Vink}, {Wyse}, {Zaggia},
  {Zeilinger}, {Zoccali}, {Zorec}, {Zucker}, {Zwitter}, \& {Gaia-ESO Survey
  Team}}]{Gilmore2012Msngr.147...25G}
{Gilmore}, G., {Randich}, S., {Asplund}, M., {et~al.} 2012, The Messenger, 147,
  25

\bibitem[{{Guilherme-Garcia} {et~al.}(2023){Guilherme-Garcia}, {Krone-Martins},
  \& {Moitinho}}]{Guilherme2023A&A...673A.128G}
{Guilherme-Garcia}, P., {Krone-Martins}, A., \& {Moitinho}, A. 2023, \aap, 673,
  A128

\bibitem[{{Hao} {et~al.}(2022){Hao}, {Xu}, {Bian}, {Hou}, {Lin}, {Li}, \&
  {Liu}}]{Hao2022ApJ...938..100H}
{Hao}, C.~J., {Xu}, Y., {Bian}, S.~B., {et~al.} 2022, \apj, 938, 100

\bibitem[{{Hao} {et~al.}(2024){Hao}, {Xu}, {Hou}, {Bian}, {Lin}, {Li}, {Dong},
  \& {Liu}}]{Hao2024arXiv240212160H}
{Hao}, C.~J., {Xu}, Y., {Hou}, L.~G., {et~al.} 2024, arXiv e-prints,
  arXiv:2402.12160

\bibitem[{{Healy} \& {McCullough}(2020)}]{Healy2020ApJ...903...99H}
{Healy}, B.~F. \& {McCullough}, P.~R. 2020, \apj, 903, 99

\bibitem[{{Healy} {et~al.}(2021){Healy}, {McCullough}, \&
  {Schlaufman}}]{Healy2021ApJ...923...23H}
{Healy}, B.~F., {McCullough}, P.~R., \& {Schlaufman}, K.~C. 2021, \apj, 923, 23

\bibitem[{{H{\'e}nault-Brunet} {et~al.}(2012){H{\'e}nault-Brunet}, {Gieles},
  {Evans}, {Sana}, {Bastian}, {Ma{\'\i}z Apell{\'a}niz}, {Taylor}, {Markova},
  {Bressert}, {de Koter}, \& {van Loon}}]{Henault2012A&A...545L...1H}
{H{\'e}nault-Brunet}, V., {Gieles}, M., {Evans}, C.~J., {et~al.} 2012, \aap,
  545, L1

\bibitem[{{Hobbs} {et~al.}(2005){Hobbs}, {Lorimer}, {Lyne}, \&
  {Kramer}}]{Hobbs2005MNRAS.360..974H}
{Hobbs}, G., {Lorimer}, D.~R., {Lyne}, A.~G., \& {Kramer}, M. 2005, \mnras,
  360, 974

\bibitem[{{Hunt} \& {Reffert}(2023)}]{Hunt2023A&A...673A.114H}
{Hunt}, E.~L. \& {Reffert}, S. 2023, \aap, 673, A114

\bibitem[{{Hurley} {et~al.}(2000){Hurley}, {Pols}, \&
  {Tout}}]{Hurley2000MNRAS.315..543H}
{Hurley}, J.~R., {Pols}, O.~R., \& {Tout}, C.~A. 2000, \mnras, 315, 543

\bibitem[{{Hurley} {et~al.}(2002){Hurley}, {Tout}, \&
  {Pols}}]{Hurley2002MNRAS.329..897H}
{Hurley}, J.~R., {Tout}, C.~A., \& {Pols}, O.~R. 2002, \mnras, 329, 897

\bibitem[{{Jadhav} \& {Subramaniam}(2021)}]{Jadhav2021MNRAS.507.1699J}
{Jadhav}, V.~V. \& {Subramaniam}, A. 2021, \mnras, 507, 1699

\bibitem[{{Jerabkova} {et~al.}(2021){Jerabkova}, {Boffin}, {Beccari}, {de
  Marchi}, {de Bruijne}, \& {Prusti}}]{Jerabkova2021A&A...647A.137J}
{Jerabkova}, T., {Boffin}, H. M.~J., {Beccari}, G., {et~al.} 2021, \aap, 647,
  A137

\bibitem[{{Kacharov} {et~al.}(2014){Kacharov}, {Bianchini}, {Koch}, {Frank},
  {Martin}, {van de Ven}, {Puzia}, {McDonald}, {Johnson}, \&
  {Zijlstra}}]{Kacharov2014A&A...567A..69K}
{Kacharov}, N., {Bianchini}, P., {Koch}, A., {et~al.} 2014, \aap, 567, A69

\bibitem[{{Kamann} {et~al.}(2019){Kamann}, {Bastian}, {Gieles}, {Balbinot}, \&
  {H{\'e}nault-Brunet}}]{Kamann2019MNRAS.483.2197K}
{Kamann}, S., {Bastian}, N.~J., {Gieles}, M., {Balbinot}, E., \&
  {H{\'e}nault-Brunet}, V. 2019, \mnras, 483, 2197

\bibitem[{{Kamann} {et~al.}(2018){Kamann}, {Husser}, {Dreizler}, {Emsellem},
  {Weilbacher}, {Martens}, {Bacon}, {den Brok}, {Giesers}, {Krajnovi{\'c}},
  {Roth}, {Wendt}, \& {Wisotzki}}]{Kamann2018MNRAS.473.5591K}
{Kamann}, S., {Husser}, T.~O., {Dreizler}, S., {et~al.} 2018, \mnras, 473, 5591

\bibitem[{{Kroupa}(1995{\natexlab{a}})}]{Kroupa1995MNRAS.277.1491K}
{Kroupa}, P. 1995{\natexlab{a}}, \mnras, 277, 1491

\bibitem[{{Kroupa}(1995{\natexlab{b}})}]{Kroupa1995MNRAS.277.1507K}
{Kroupa}, P. 1995{\natexlab{b}}, \mnras, 277, 1507

\bibitem[{{Kroupa}(2001)}]{Kroupa2001MNRAS.322..231K}
{Kroupa}, P. 2001, \mnras, 322, 231

\bibitem[{{Kroupa} {et~al.}(2001){Kroupa}, {Aarseth}, \&
  {Hurley}}]{Kroupa2001MNRAS.321..699K}
{Kroupa}, P., {Aarseth}, S., \& {Hurley}, J. 2001, \mnras, 321, 699

\bibitem[{{Kroupa} {et~al.}(2022){Kroupa}, {Jerabkova}, {Thies},
  {Pflamm-Altenburg}, {Famaey}, {Boffin}, {Dabringhausen}, {Beccari}, {Prusti},
  {Boily}, {Haghi}, {Wu}, {Haas}, {Zonoozi}, {Thomas}, {{\v{S}}ubr}, \&
  {Aarseth}}]{Kroupa2022MNRAS.517.3613K}
{Kroupa}, P., {Jerabkova}, T., {Thies}, I., {et~al.} 2022, \mnras, 517, 3613

\bibitem[{{Kuhn} {et~al.}(2019){Kuhn}, {Hillenbrand}, {Sills}, {Feigelson}, \&
  {Getman}}]{Kuhn2019ApJ...870...32K}
{Kuhn}, M.~A., {Hillenbrand}, L.~A., {Sills}, A., {Feigelson}, E.~D., \&
  {Getman}, K.~V. 2019, \apj, 870, 32

\bibitem[{{Kutner} {et~al.}(1977){Kutner}, {Tucker}, {Chin}, \&
  {Thaddeus}}]{Kutner1977ApJ...215..521K}
{Kutner}, M.~L., {Tucker}, K.~D., {Chin}, G., \& {Thaddeus}, P. 1977, \apj,
  215, 521

\bibitem[{{Lada} \& {Lada}(2003)}]{Lada2003ARA&A..41...57L}
{Lada}, C.~J. \& {Lada}, E.~A. 2003, \araa, 41, 57

\bibitem[{{Lanzoni} {et~al.}(2018){Lanzoni}, {Ferraro}, {Mucciarelli},
  {Pallanca}, {Lapenna}, {Origlia}, {Dalessandro}, {Valenti}, {Bellazzini},
  {Tiongco}, {Varri}, {Vesperini}, \& {Beccari}}]{Lanzoni2018ApJ...861...16L}
{Lanzoni}, B., {Ferraro}, F.~R., {Mucciarelli}, A., {et~al.} 2018, \apj, 861,
  16

\bibitem[{{Loktin} \& {Popov}(2020)}]{Loktin2020AN....341..638L}
{Loktin}, A.~V. \& {Popov}, A.~A. 2020, Astronomische Nachrichten, 341, 638

\bibitem[{{L{\"u}ghausen} {et~al.}(2015){L{\"u}ghausen}, {Famaey}, \&
  {Kroupa}}]{lueghausen2015a}
{L{\"u}ghausen}, F., {Famaey}, B., \& {Kroupa}, P. 2015, Canadian Journal of
  Physics, 93, 232

\bibitem[{{Lynden-Bell}(1967)}]{Lynden1967MNRAS.136..101L}
{Lynden-Bell}, D. 1967, \mnras, 136, 101

\bibitem[{{Mackey} {et~al.}(2013){Mackey}, {Da Costa}, {Ferguson}, \&
  {Yong}}]{Mackey2013ApJ...762...65M}
{Mackey}, A.~D., {Da Costa}, G.~S., {Ferguson}, A.~M.~N., \& {Yong}, D. 2013,
  \apj, 762, 65

\bibitem[{{Milgrom}(1983)}]{Milgrom1983ApJ...270..365M}
{Milgrom}, M. 1983, \apj, 270, 365

\bibitem[{{Milgrom}(2010)}]{milgrom2010a}
{Milgrom}, M. 2010, \mnras, 403, 886

\bibitem[{{Moreno} {et~al.}(2022){Moreno}, {Fern{\'a}ndez-Trincado},
  {P{\'e}rez-Villegas}, {Chaves-Velasquez}, \&
  {Schuster}}]{Moreno2022MNRAS.510.5945M}
{Moreno}, E., {Fern{\'a}ndez-Trincado}, J.~G., {P{\'e}rez-Villegas}, A.,
  {Chaves-Velasquez}, L., \& {Schuster}, W.~J. 2022, \mnras, 510, 5945

\bibitem[{{Moreno} {et~al.}(2021){Moreno}, {Fern{\'a}ndez-Trincado},
  {Schuster}, {P{\'e}rez-Villegas}, \&
  {Chaves-Velasquez}}]{Moreno2021MNRAS.506.4687M}
{Moreno}, E., {Fern{\'a}ndez-Trincado}, J.~G., {Schuster}, W.~J.,
  {P{\'e}rez-Villegas}, A., \& {Chaves-Velasquez}, L. 2021, \mnras, 506, 4687

\bibitem[{{Moreno} {et~al.}(2014){Moreno}, {Pichardo}, \&
  {Vel{\'a}zquez}}]{Moreno2014ApJ...793..110M}
{Moreno}, E., {Pichardo}, B., \& {Vel{\'a}zquez}, H. 2014, \apj, 793, 110

\bibitem[{{Pfalzner} \& {Kaczmarek}(2013)}]{Pfalzner2013A&A...559A..38P}
{Pfalzner}, S. \& {Kaczmarek}, T. 2013, \aap, 559, A38

\bibitem[{{Piatti} \& {Malhan}(2022)}]{Piatti2022MNRAS.511L...1P}
{Piatti}, A.~E. \& {Malhan}, K. 2022, \mnras, 511, L1

\bibitem[{{Read} {et~al.}(2006){Read}, {Wilkinson}, {Evans}, {Gilmore}, \&
  {Kleyna}}]{Read2006MNRAS.366..429R}
{Read}, J.~I., {Wilkinson}, M.~I., {Evans}, N.~W., {Gilmore}, G., \& {Kleyna},
  J.~T. 2006, \mnras, 366, 429

\bibitem[{{Rosolowsky} {et~al.}(2003){Rosolowsky}, {Engargiola}, {Plambeck}, \&
  {Blitz}}]{Rosolowsky2003ApJ...599..258R}
{Rosolowsky}, E., {Engargiola}, G., {Plambeck}, R., \& {Blitz}, L. 2003, \apj,
  599, 258

\bibitem[{{Rossi} \& {Hurley}(2015)}]{Rossi2015MNRAS.454.1453R}
{Rossi}, L.~J. \& {Hurley}, J.~R. 2015, \mnras, 454, 1453

\bibitem[{{Sollima} {et~al.}(2019){Sollima}, {Baumgardt}, \&
  {Hilker}}]{Sollima2019MNRAS.485.1460S}
{Sollima}, A., {Baumgardt}, H., \& {Hilker}, M. 2019, \mnras, 485, 1460

\bibitem[{{Steinmetz} {et~al.}(2020){Steinmetz}, {Matijevi{\v{c}}}, {Enke},
  {Zwitter}, {Guiglion}, {McMillan}, {Kordopatis}, {Valentini}, {Chiappini},
  {Casagrande}, {Wojno}, {Anguiano}, {Bienaym{\'e}}, {Bijaoui}, {Binney},
  {Burton}, {Cass}, {de Laverny}, {Fiegert}, {Freeman}, {Fulbright}, {Gibson},
  {Gilmore}, {Grebel}, {Helmi}, {Kunder}, {Munari}, {Navarro}, {Parker},
  {Ruchti}, {Recio-Blanco}, {Reid}, {Seabroke}, {Siviero}, {Siebert}, {Stupar},
  {Watson}, {Williams}, {Wyse}, {Anders}, {Antoja}, {Birko}, {Bland-Hawthorn},
  {Bossini}, {Garc{\'\i}a}, {Carrillo}, {Chaplin}, {Elsworth}, {Famaey},
  {Gerhard}, {Jofre}, {Just}, {Mathur}, {Miglio}, {Minchev}, {Monari},
  {Mosser}, {Ritter}, {Rodrigues}, {Scholz}, {Sharma}, {Sysoliatina}, \& {RAVE
  Collaboration}}]{Steinmetz2020AJ....160...82S}
{Steinmetz}, M., {Matijevi{\v{c}}}, G., {Enke}, H., {et~al.} 2020, \aj, 160, 82

\bibitem[{{Szigeti} {et~al.}(2021){Szigeti}, {M{\'e}sz{\'a}ros}, {Szab{\'o}},
  {Fern{\'a}ndez-Trincado}, {Lane}, \& {Cohen}}]{Szigeti2021MNRAS.504.1144S}
{Szigeti}, L., {M{\'e}sz{\'a}ros}, S., {Szab{\'o}}, G.~M., {et~al.} 2021,
  \mnras, 504, 1144

\bibitem[{{Teyssier}(2002)}]{Teyssier2002A&A...385..337T}
{Teyssier}, R. 2002, \aap, 385, 337

\bibitem[{{Toomre} \& {Toomre}(1972)}]{Toomre1972ApJ...178..623T}
{Toomre}, A. \& {Toomre}, J. 1972, \apj, 178, 623

\bibitem[{{Tsantaki} {et~al.}(2022){Tsantaki}, {Pancino}, {Marrese},
  {Marinoni}, {Rainer}, {Sanna}, {Turchi}, {Randich}, {Gallart}, {Battaglia},
  \& {Masseron}}]{Tsantaki2022A&A...659A..95T}
{Tsantaki}, M., {Pancino}, E., {Marrese}, P., {et~al.} 2022, \aap, 659, A95

\bibitem[{{Tutukov}(1978)}]{Tutukov1978A&A....70...57T}
{Tutukov}, A.~V. 1978, \aap, 70, 57

\bibitem[{{van de Ven} {et~al.}(2006){van de Ven}, {van den Bosch}, {Verolme},
  \& {de Zeeuw}}]{van2006A&A...445..513V}
{van de Ven}, G., {van den Bosch}, R.~C.~E., {Verolme}, E.~K., \& {de Zeeuw},
  P.~T. 2006, \aap, 445, 513

\bibitem[{{van Leeuwen}(2009)}]{van2009A&A...497..209V}
{van Leeuwen}, F. 2009, \aap, 497, 209

\bibitem[{{Vasiliev} \& {Baumgardt}(2021)}]{Vasiliev2021MNRAS.505.5978V}
{Vasiliev}, E. \& {Baumgardt}, H. 2021, \mnras, 505, 5978

\bibitem[{{von Steiger} \& {Zurbuchen}(2016)}]{von2016ApJ...816...13V}
{von Steiger}, R. \& {Zurbuchen}, T.~H. 2016, \apj, 816, 13

\bibitem[{{Wang} {et~al.}(2020){Wang}, {Iwasawa}, {Nitadori}, \&
  {Makino}}]{Wang2020MNRAS.497..536W}
{Wang}, L., {Iwasawa}, M., {Nitadori}, K., \& {Makino}, J. 2020, \mnras, 497,
  536

\bibitem[{{Webb}(2023)}]{Webb2023JOSS....8.4483W}
{Webb}, J. 2023, The Journal of Open Source Software, 8, 4483

\bibitem[{{Zwitter} {et~al.}(2018){Zwitter}, {Kos}, {Chiavassa}, {Buder},
  {Traven}, {{\v{C}}otar}, {Lin}, {Asplund}, {Bland-Hawthorn}, {Casey}, {De
  Silva}, {Duong}, {Freeman}, {Lind}, {Martell}, {D'Orazi}, {Schlesinger},
  {Simpson}, {Sharma}, {Zucker}, {Anguiano}, {Casagrande}, {Collet}, {Horner},
  {Ireland}, {Kafle}, {Lewis}, {Munari}, {Nataf}, {Ness}, {Nordlander},
  {Stello}, {Ting}, {Tinney}, {Watson}, {Wittenmyer}, \&
  {{\v{Z}}erjal}}]{Zwitter2018MNRAS.481..645Z}
{Zwitter}, T., {Kos}, J., {Chiavassa}, A., {et~al.} 2018, \mnras, 481, 645

\end{thebibliography}

\begin{appendix}

\section{Supplementary table and figures}

\begin{table}[!ht]
    \centering
    \begin{tabular}{ll}
    \toprule
Term	&	Definition	\\ \hline
$\alpha$, $\delta$	&	Right ascension and declination	\\
$l$, $b$	&	Galactic coordinates	\\
$\varpi$	&	Parallax	\\
$\alpha_0,\delta_0$	&	Cluster centre in equatorial coordinates	\\
$l_0,b_0$	&	Cluster centre in Galactic coordinates	\\
$\Delta\alpha$	&	Tangential projection of $\alpha$	\\
	&	$=\text{cos}(\delta) \text{sin}(\alpha - \alpha_0)$	\\
$\Delta\delta$	&	Tangential projection of $\delta$	\\
	&	$=\text{sin}(\delta) \text{cos}(\delta_0) - \text{cos}(\delta)\text{sin}(\delta_0)\text{cos}(\alpha-\alpha_0)$	\\
$\Delta l$	&	Tangential projection of $l$	\\
	&	$=\text{cos}(b) \text{sin}(l-l_0)$	\\
$\Delta b$	&	Tangential projection of $b$	\\
	&	$=\text{sin}(b) \text{cos}(b_0) - \text{cos}(b)\text{sin}(b_0)\text{cos}(l-l_0)$	\\
$R$	&	Projected angular radius from the cluster centre	\\
	&	$=\sqrt{\Delta l^2+\Delta b^2}$	\\
$\phi$	&	Azimuth in the 2D tangential plane	\\
	&	$=tan^{-1}(\Delta \delta/\Delta \alpha)$	\\
$PA$	&	Position angle of spin axis as measured plane	\\
	&	clockwise from Galactic East, in the sky	\\
$XR$	&	Projected angular distance from the spin axis	\\
	&	$=-\Delta l \text{sin}(PA) + \Delta b \text{cos}(PA)$	\\
$\Delta V$	&	The difference between the mean RVs of two	\\
	&	 cluster sections divided by the assumed spin axis	\\
$\mu_{\alpha*}$	&	PM in the right ascension direction	\\
	&	$=\mu_{\alpha}\text{cos}\delta$	\\
$\mu_{\delta}$	&	PM in the declination direction	\\
$RV$	&	Radial velocity	\\
$\mu_{\alpha*,corr}$	&	Projection corrected $\mu_{\alpha*}$ (See \S~\ref{sec:projection_correction})	\\
$\mu_{\delta,corr}$	&	Projection corrected $\mu_{\delta}$ (See \S~\ref{sec:projection_correction})	\\
$RV_{corr}$	&	Projection corrected RV (See \S~\ref{sec:projection_correction})	\\
$\mu_R$	&	Radial component of the PM	\\
	&	$=\mu_{\alpha*,corr}\text{cos}(-\phi) - \mu_{\delta,corr}\text{sin}(-\phi)$	\\
$\mu_T$	&	Tangential component of the PM	\\
	&	$=\mu_{\alpha*,corr}\text{sin}(-\phi) + \mu_{\delta,corr}\text{cos}(-\phi)$	\\ \bottomrule
    \end{tabular}
    \caption{Definitions and formulae for various observed and calculated parameters in this work.}
    \label{tab:param_definitions}
\end{table} %param_definitions

\begin{table*}
    \centering
    \begin{tabular}{lcc ccc c}
    \toprule
	Name	&	$i$	&	$\omega$	&	$A_{peak}$	&	$R_{peak}$	&	Velocity noise	&	$n\_stars$	\\
		&	[\arcdeg]	&	[cy Gyr$^{-1}$]	&	[\kms]	&	[pc]	&	[\kms]	&		\\ \midrule
\multicolumn{7}{c}{Demo clusters in Figure~\ref{fig:demo_rv}}														\\
\texttt{	demo\_i90\_sb	}&	90	&	40	&	-	&	-	&	0	&	500	\\
\texttt{	demo\_i90\_dr\_unvir	}&	90	&	-	&	2	&	2	&	0	&	500	\\
\texttt{	demo\_i90\_dr\_vir	}&	90	&	-	&	2	&	10	&	0	&	500	\\
\texttt{	demo\_i90\_dr\_unvir\_n	}&	90	&	-	&	2	&	2	&	$\sqrt{0.5^2+1.3^2}$	&	500	\\
\texttt{	demo\_i90\_dr\_vir\_n	}&	90	&	-	&	2	&	10	&	$\sqrt{0.5^2+1.3^2}$	&	500	\\ \hline
\multicolumn{7}{c}{Demo clusters in Figure~\ref{fig:demo_pm}}														\\
\texttt{	demo\_i0\_sb	}&	0	&	40	&	-	&	-	&	0	&	100	\\
\texttt{	demo\_i0\_dr\_unvir	}&	0	&	-	&	2	&	2	&	0	&	100	\\
\texttt{	demo\_i0\_dr\_vir	}&	0	&	-	&	2	&	10	&	0	&	100	\\
\texttt{	demo\_i0\_dr\_unvir\_n	}&	0	&	-	&	2	&	2	&	$\sqrt{0.5^2+0.02^2}$	&	100	\\
\texttt{	demo\_i0\_dr\_vir\_n	}&	0	&	-	&	2	&	10	&	$\sqrt{0.5^2+0.02^2}$	&	100	\\ \bottomrule

    \end{tabular}
    \caption{Cluster parameters of the synthetic demo clusters. The cluster position is fixed at (-8300,0,0) pc with zero velocity. 
    The inclination ($i$) is measured from the line of sight. All rotation axes are in the Galactocentric XY plane.
    The velocity noise is composed of typical internal velocity dispersion in OCs (0.5 \kms; \citealt{Choudhuri2010asph.book.....C}) and typical errors in the \textit{Gaia} DR3 observations for a cluster at 300 pc (\citealt{Gaia2021A&A...649A...1G}: RV\_error $\approx 1.3$ \kms; \citealt{Fabricius2021A&A...649A...5F}: PM\_error $\approx 0.02$ \kms).
    The meanings of the suffixes is as follows: \texttt{sb}: solid body, \texttt{dr}: differential rotation, \texttt{vir}: virialised, \texttt{unvir}: non-virialised, \texttt{n}: noise included.}
    \label{tab:demo_clusters}
\end{table*}  %demo_clusters

\begin{figure}
    \centering
    \includegraphics[width=0.49\textwidth]{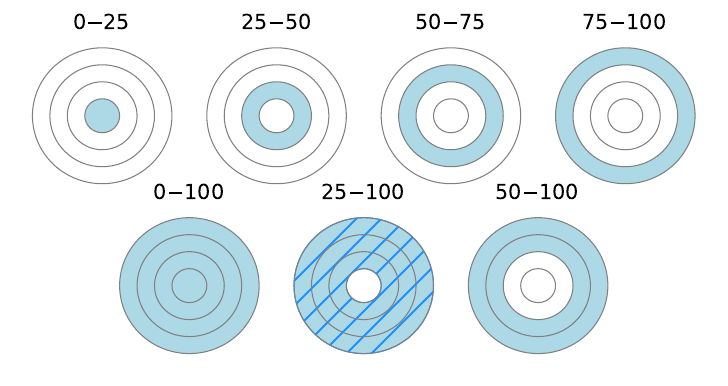}
    \caption{Demonstration of various radial slices used while calculating the $PA$. The grey circles denote the radii of the circles within which 25th, 50th, 75th and 100th percentile of the members reside. The 25th--100th percentile region is used for the RV based analysis as it avoids the noisy velocities in the cluster centre.}
    \label{fig:roi}
\end{figure}

\begin{figure*}
    \centering
    \includegraphics[width=0.98\textwidth]{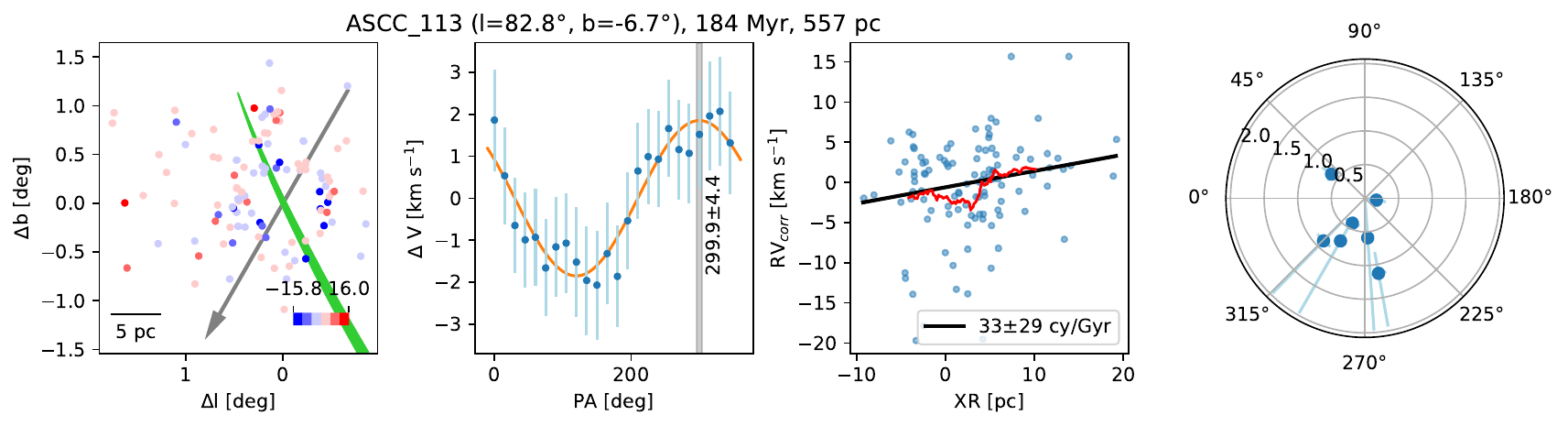}
    \includegraphics[width=0.98\textwidth]{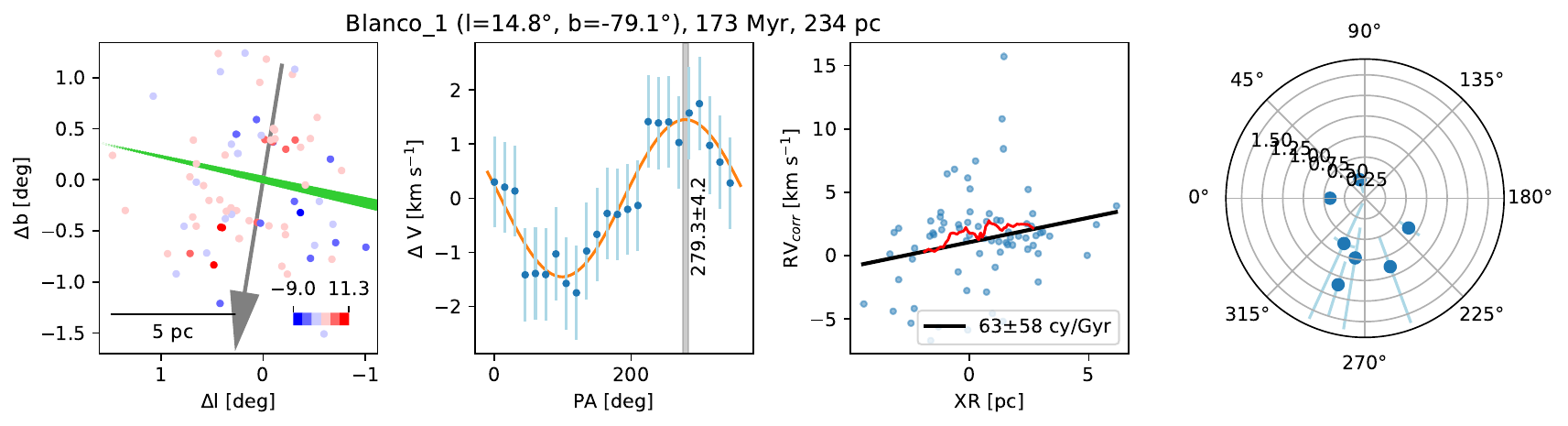}
    \includegraphics[width=0.98\textwidth]{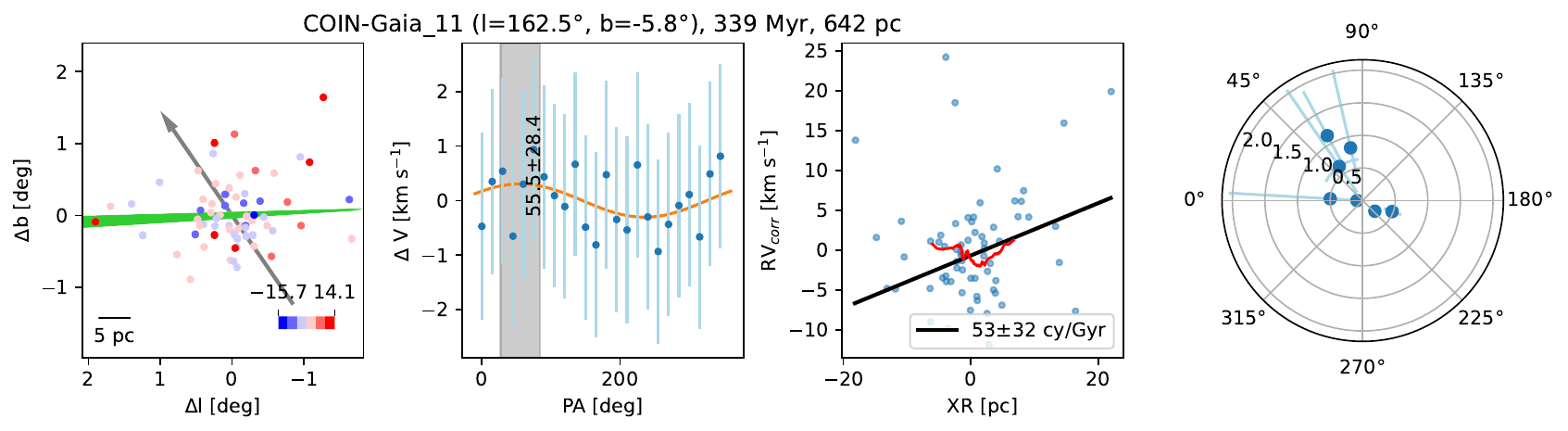}
    \includegraphics[width=0.98\textwidth]{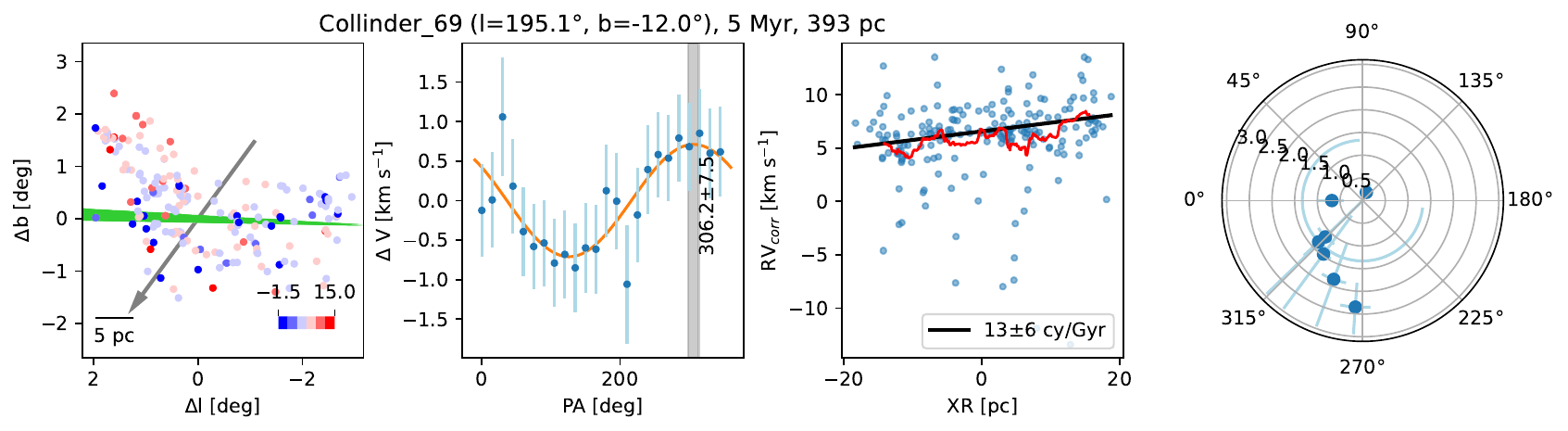}
    \caption{Diagnostic plots for RV based spinning clusters. All the subplots are similar to Figure~\ref{fig:comb_2099}.}
    \label{fig:comb_rv}
\end{figure*}

\renewcommand{\thefigure}{A.\arabic{figure} (Continued...)}
\addtocounter{figure}{-1}

\begin{figure*}
    \centering
    \includegraphics[width=0.98\textwidth]{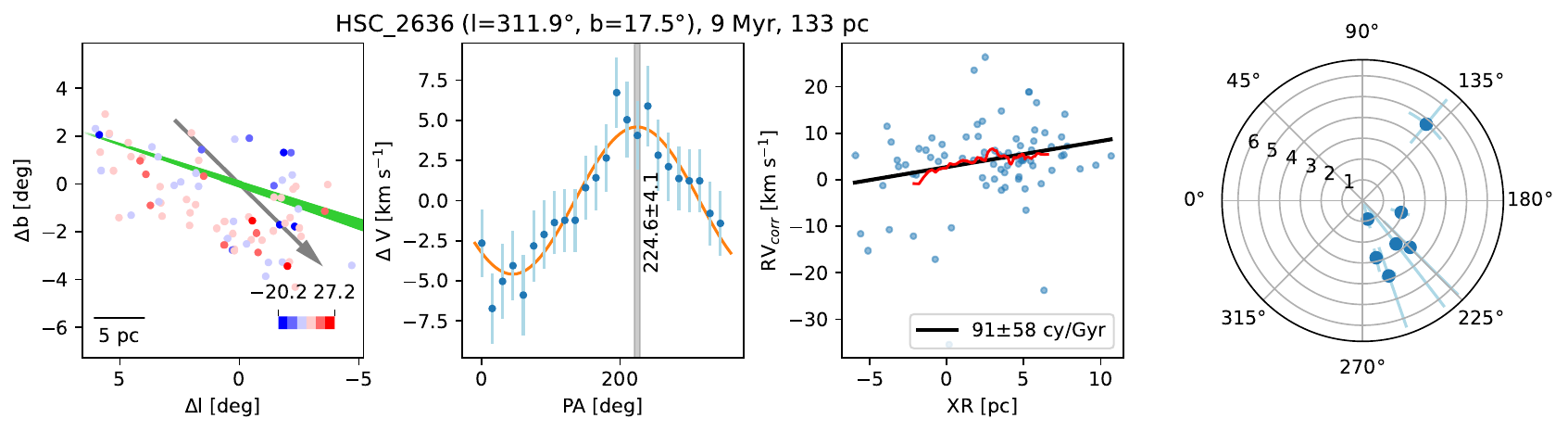}
    \includegraphics[width=0.98\textwidth]{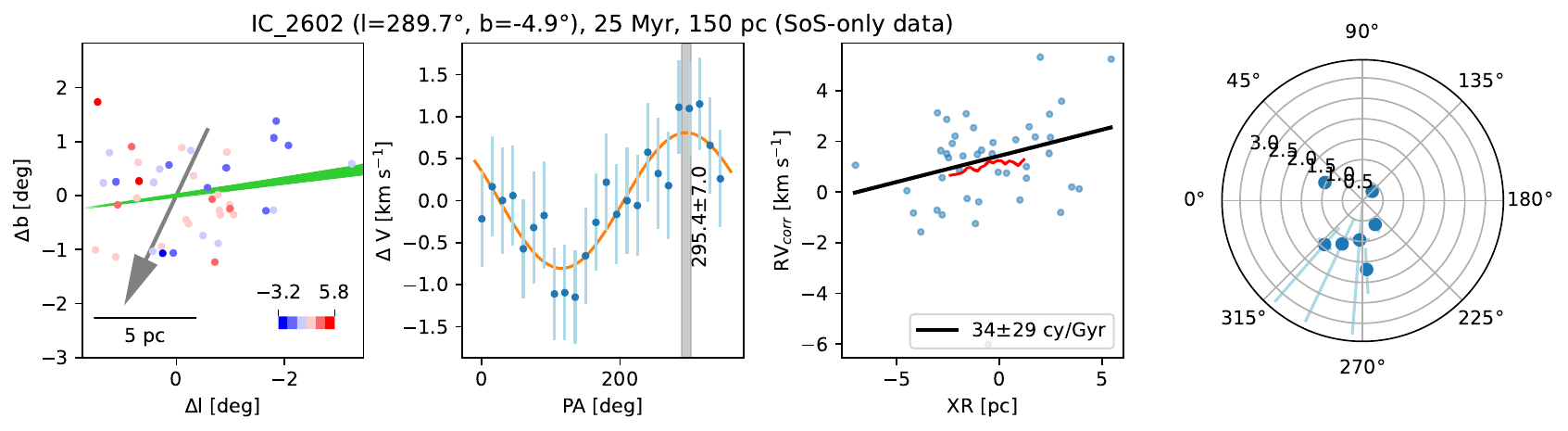}
    \includegraphics[width=0.98\textwidth]{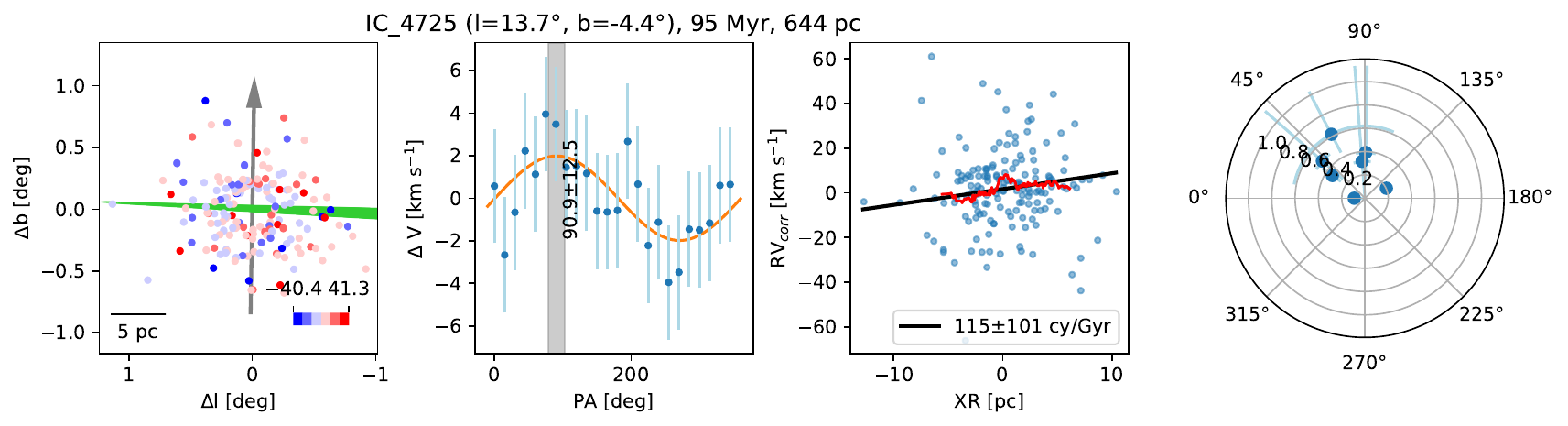}
    \includegraphics[width=0.98\textwidth]{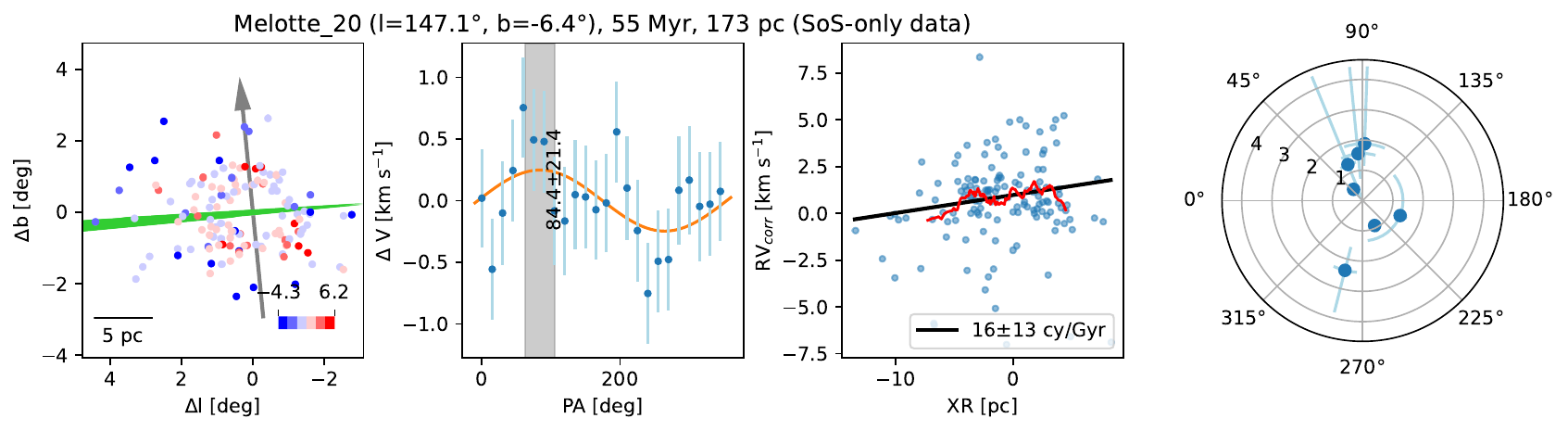}
    \caption{Diagnostic plots for RV based spinning clusters. All the subplots are similar to Figure~\ref{fig:comb_2099}.}
\end{figure*}

\addtocounter{figure}{-1}

\begin{figure*}
    \centering
    \includegraphics[width=0.98\textwidth]{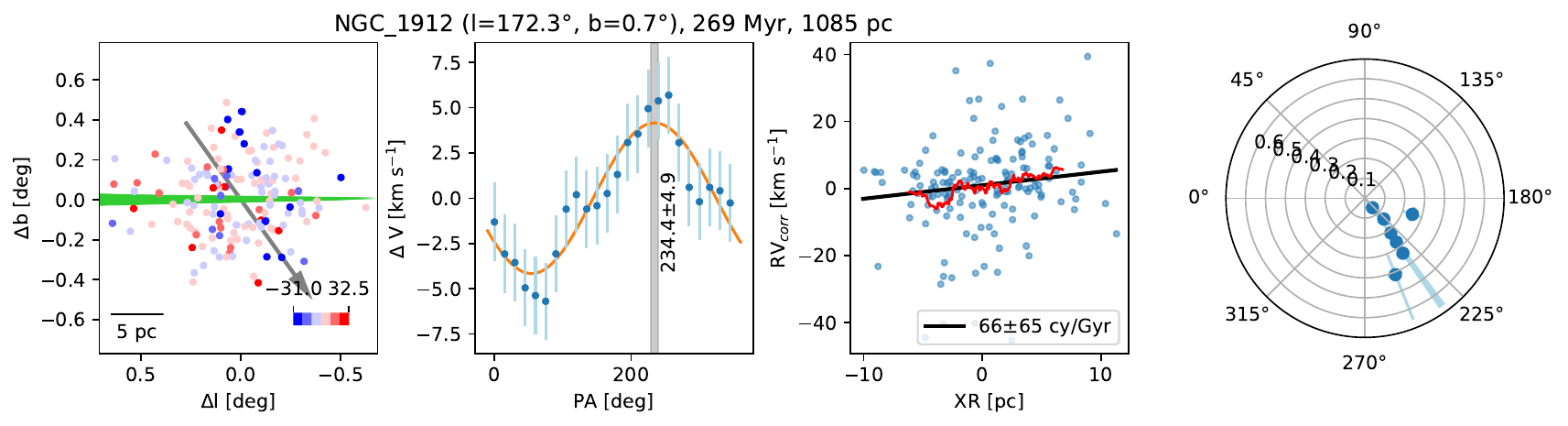}
    \includegraphics[width=0.98\textwidth]{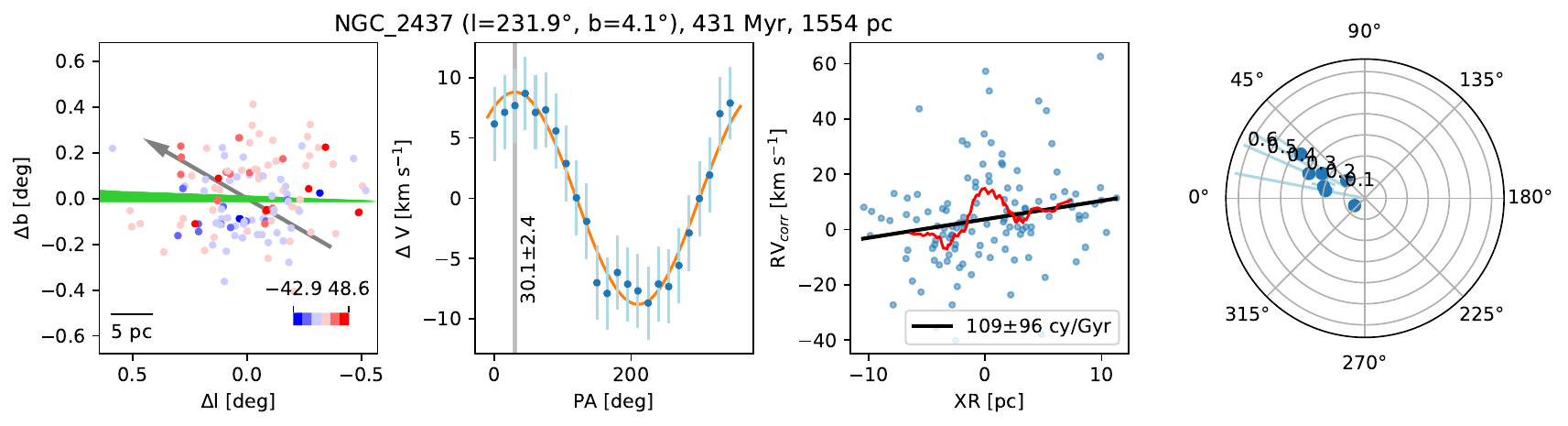}
    \includegraphics[width=0.98\textwidth]{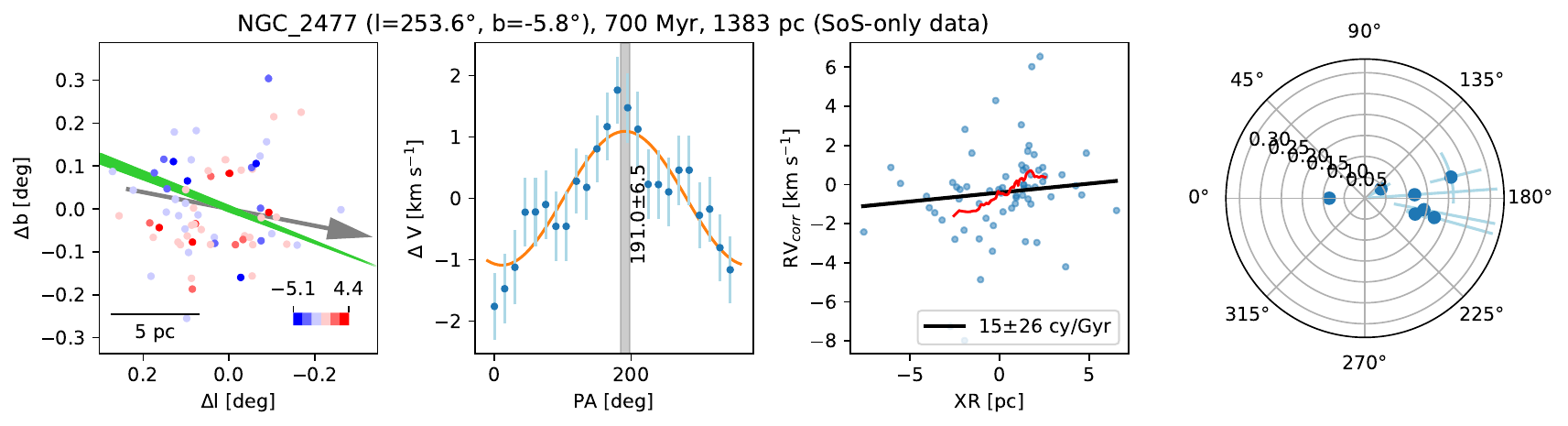}
    \includegraphics[width=0.98\textwidth]{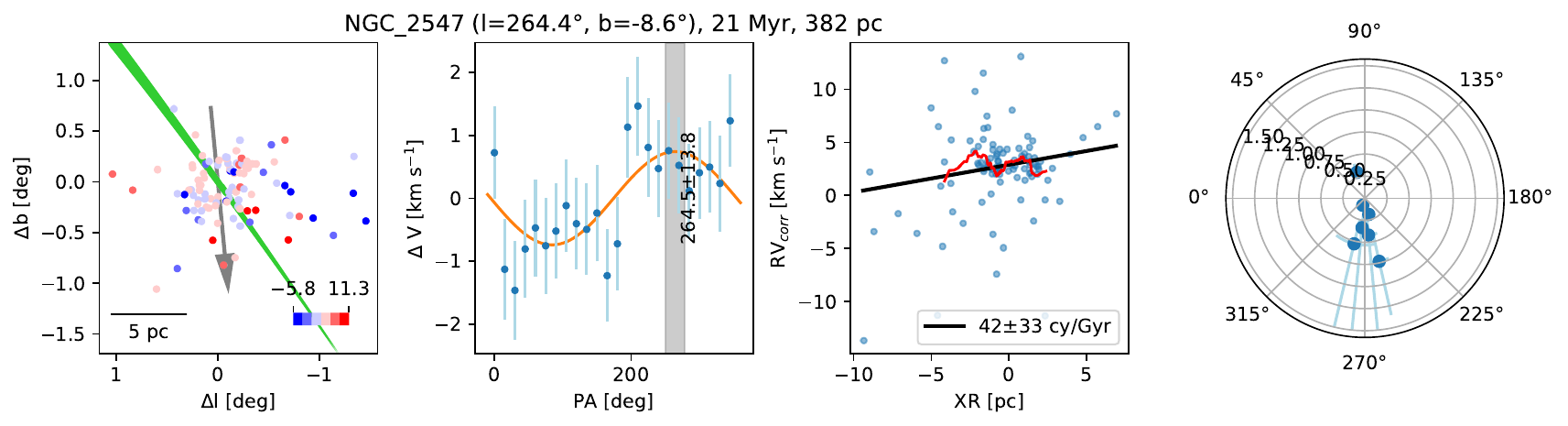}
    \caption{Diagnostic plots for RV based spinning clusters. All the subplots are similar to Figure~\ref{fig:comb_2099}.}
\end{figure*}

\addtocounter{figure}{-1}

\begin{figure*}
    \centering
    \includegraphics[width=0.98\textwidth]{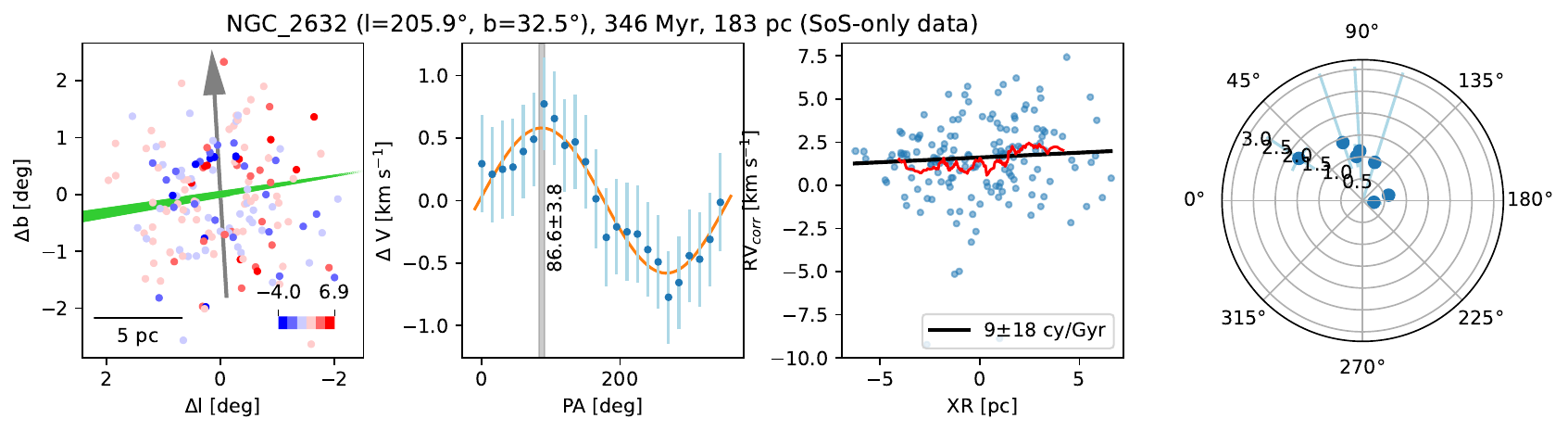}
    \includegraphics[width=0.98\textwidth]{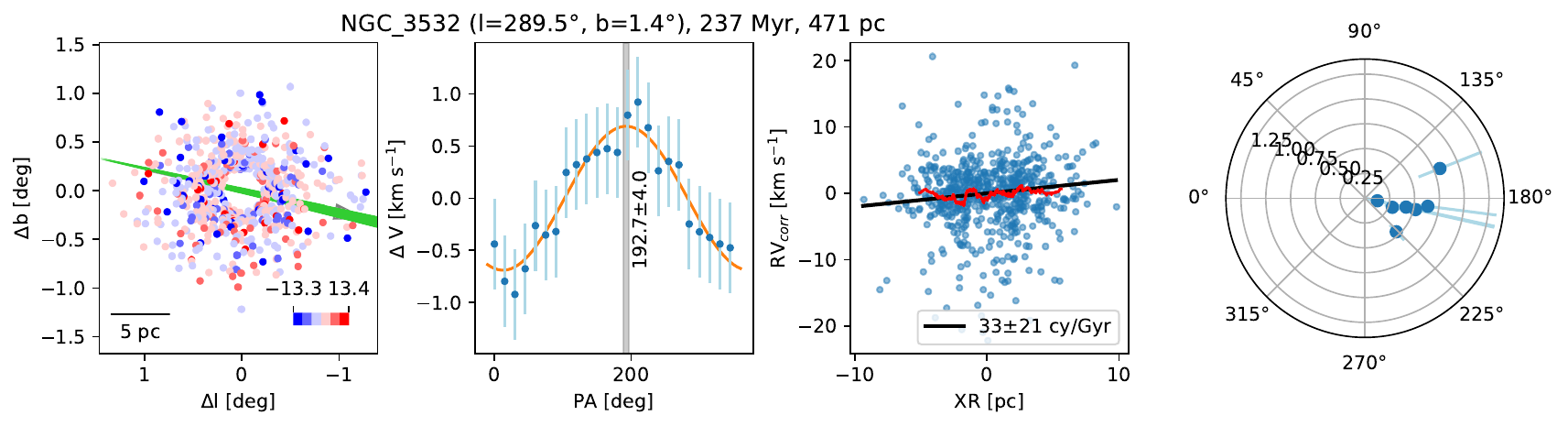}
    \includegraphics[width=0.98\textwidth]{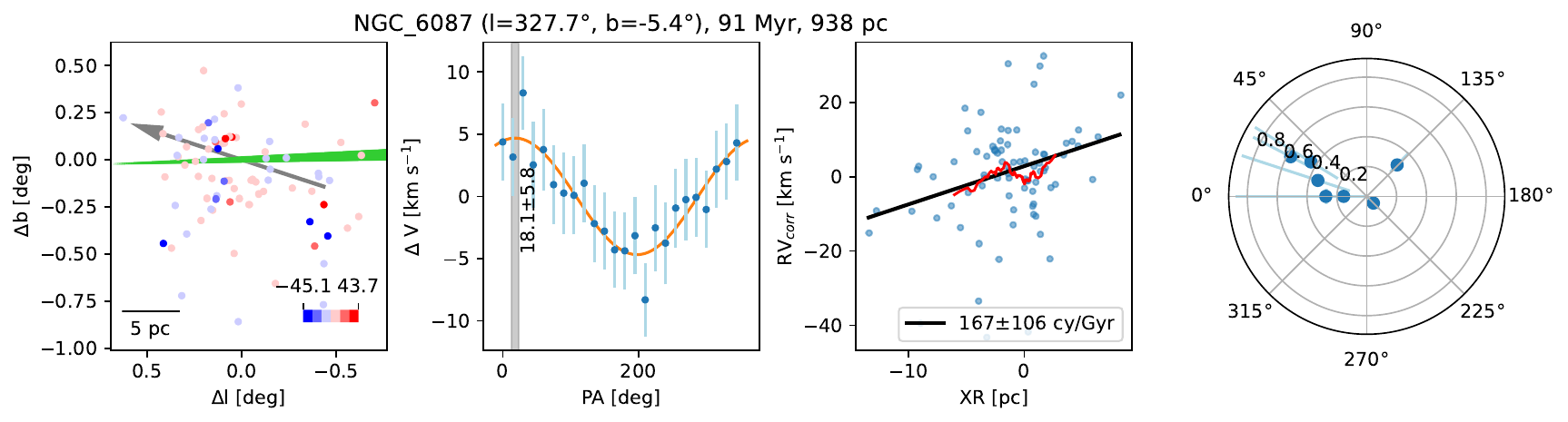}
    \includegraphics[width=0.98\textwidth]{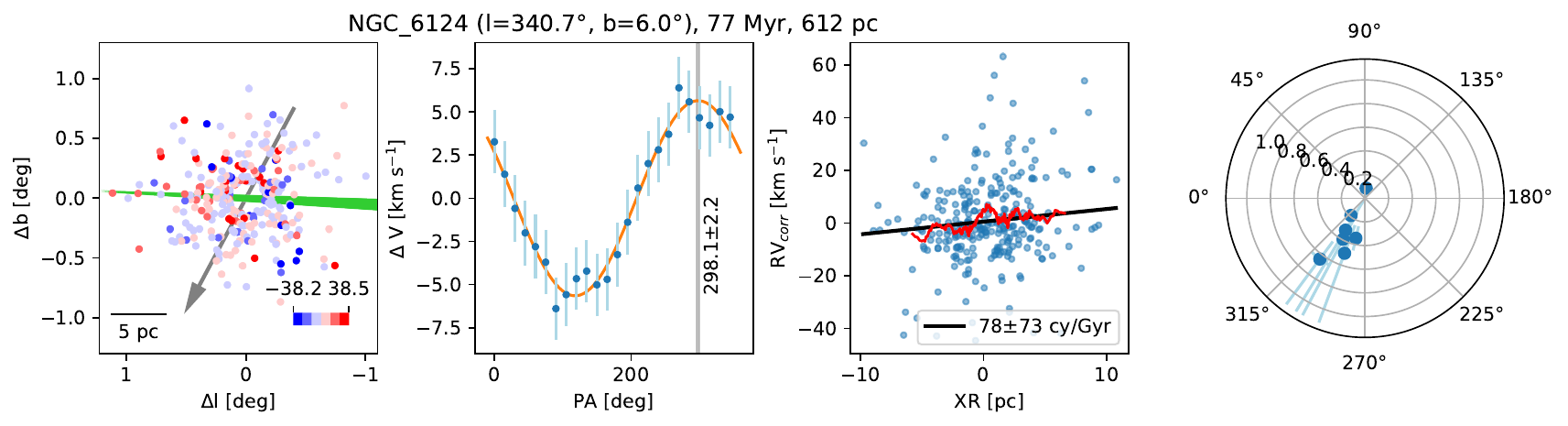}
    \caption{Diagnostic plots for RV based spinning clusters. All the subplots are similar to Figure~\ref{fig:comb_2099}.}
\end{figure*}

\addtocounter{figure}{-1}

\begin{figure*}
    \centering
    \includegraphics[width=0.98\textwidth]{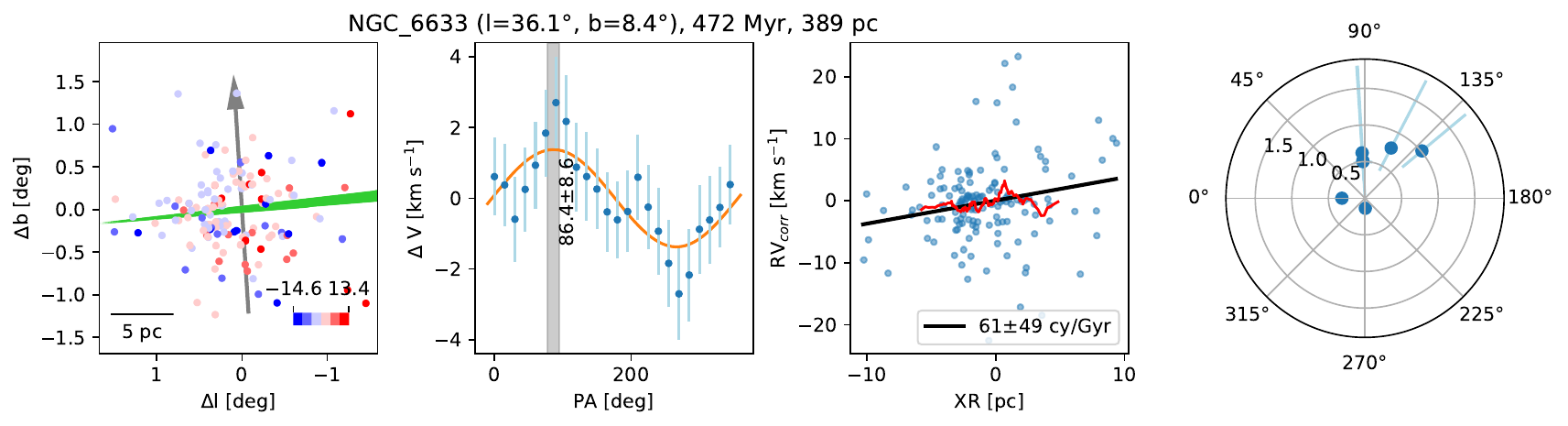}
    \includegraphics[width=0.98\textwidth]{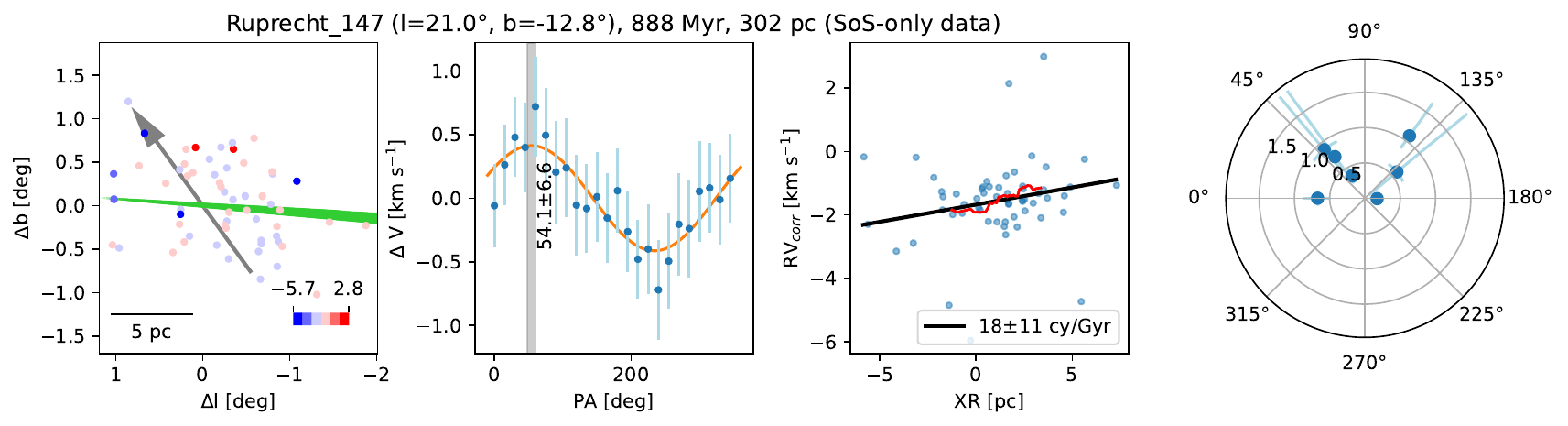}
    \includegraphics[width=0.98\textwidth]{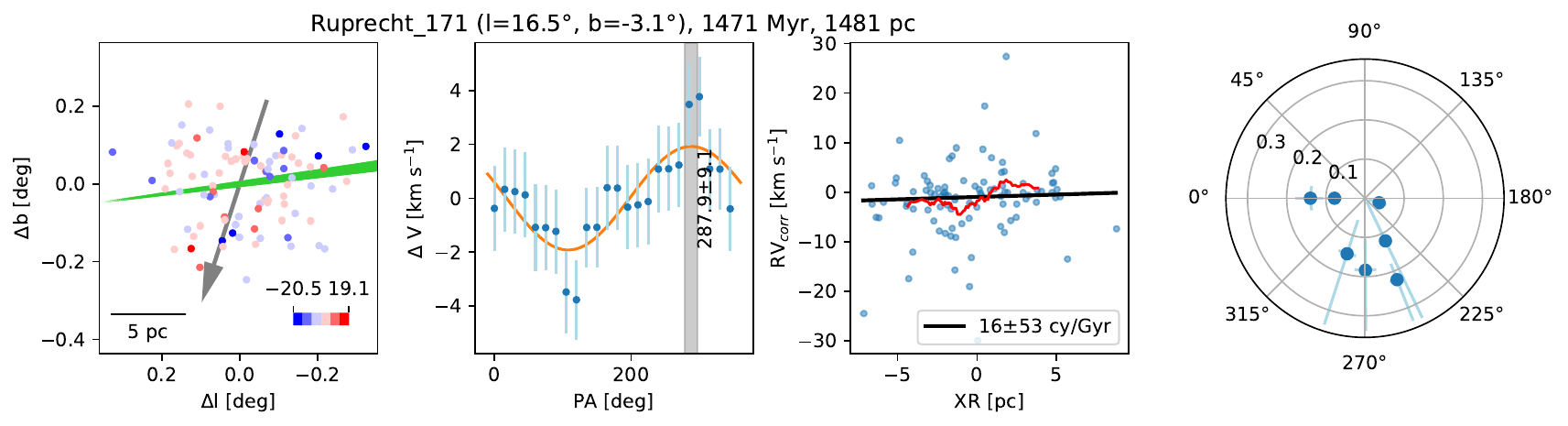}
    \includegraphics[width=0.98\textwidth]{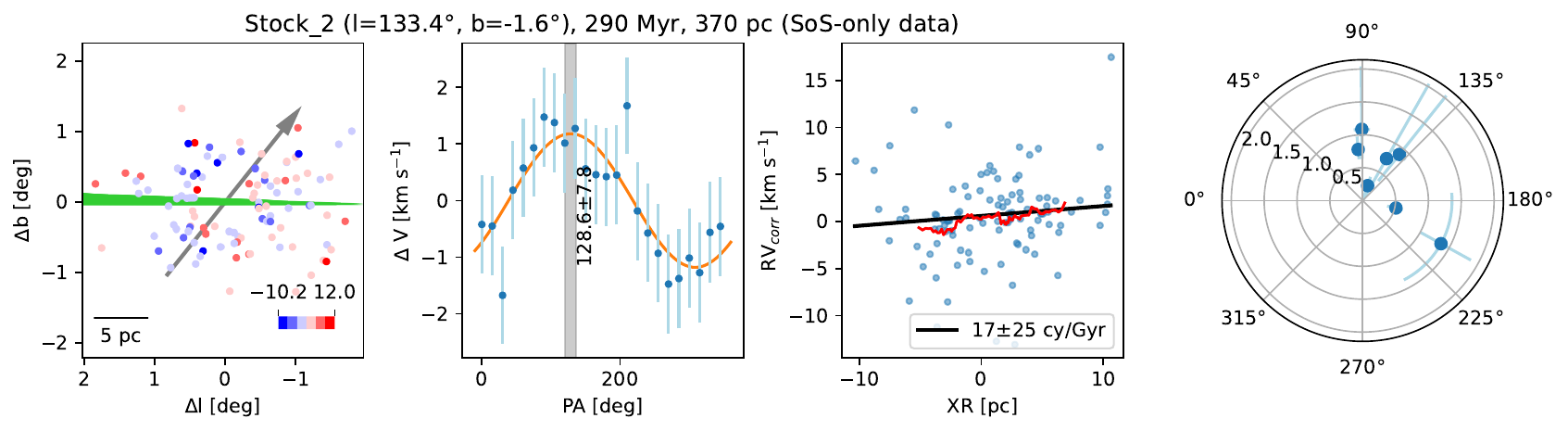}
    \caption{Diagnostic plots for RV based spinning clusters. All the subplots are similar to Figure~\ref{fig:comb_2099}.}
\end{figure*}

\addtocounter{figure}{-1}

\begin{figure*}
    \centering
    \includegraphics[width=0.98\textwidth]{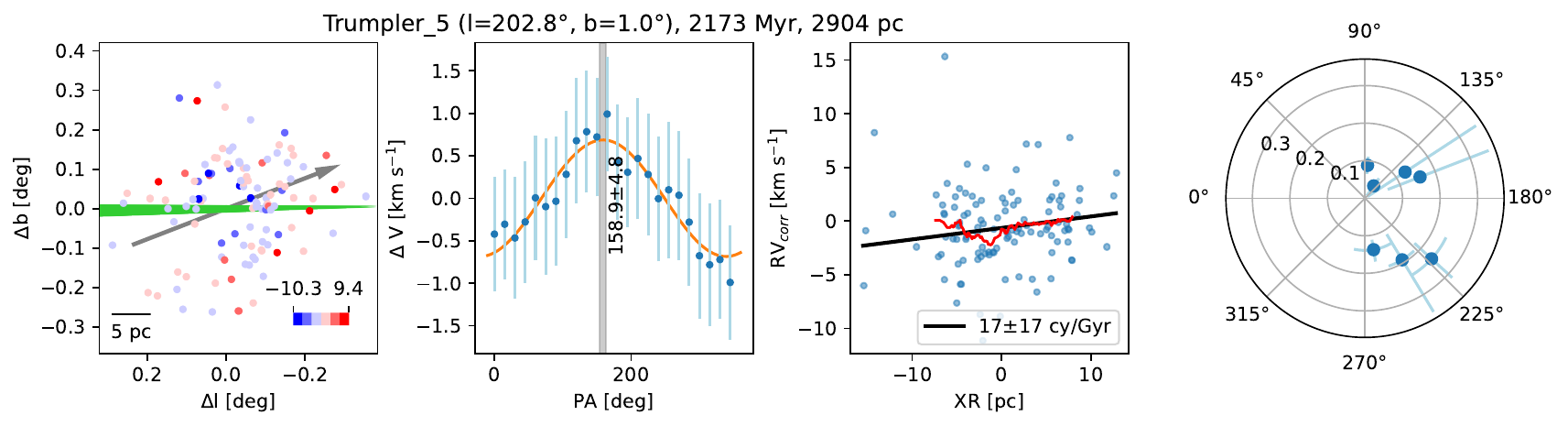}
    \caption{Diagnostic plots for RV based spinning clusters. All the subplots are similar to Figure~\ref{fig:comb_2099}.}
\end{figure*}

\renewcommand{\thefigure}{A.\arabic{figure}}

\begin{figure*}
    \centering
    \includegraphics[width=0.49\textwidth]{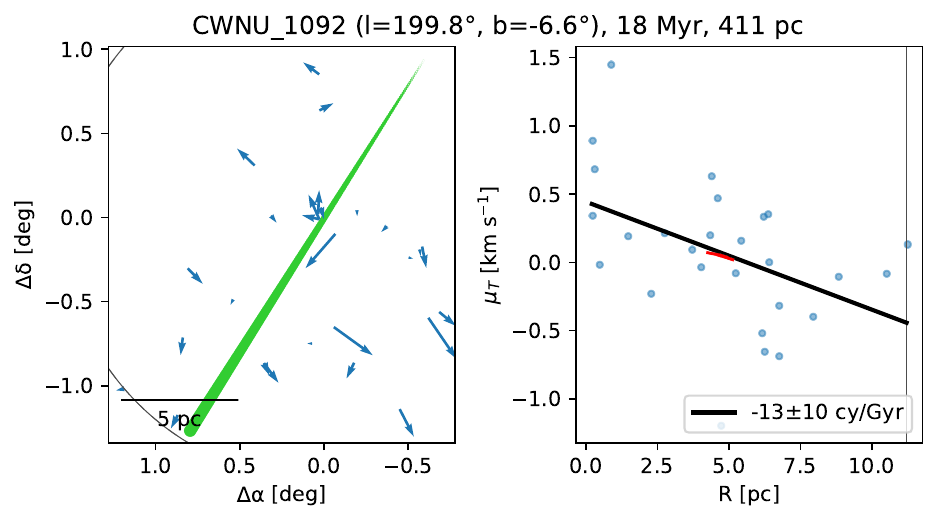}
    \includegraphics[width=0.49\textwidth]{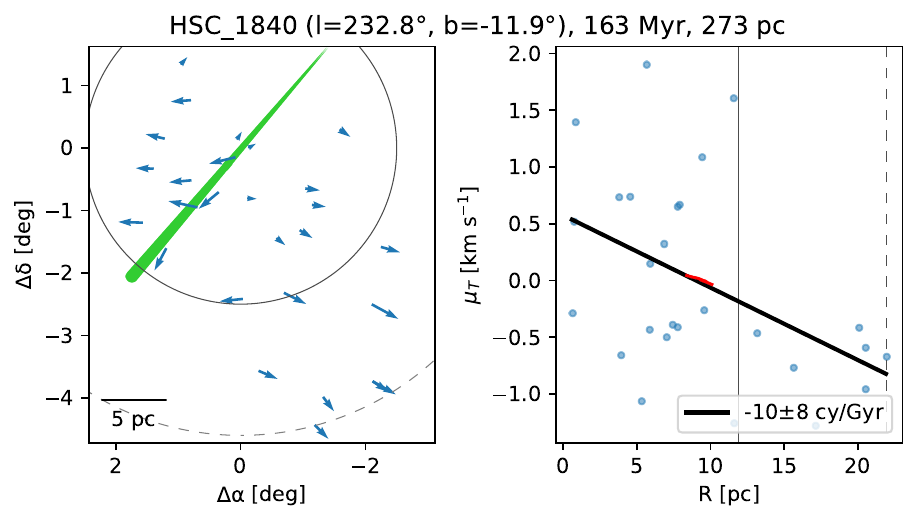}
    \includegraphics[width=0.49\textwidth]{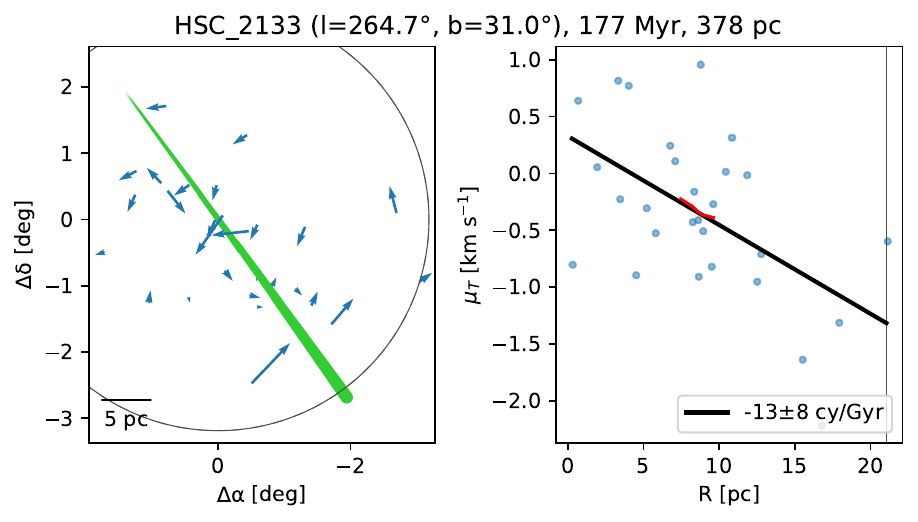}
    \includegraphics[width=0.49\textwidth]{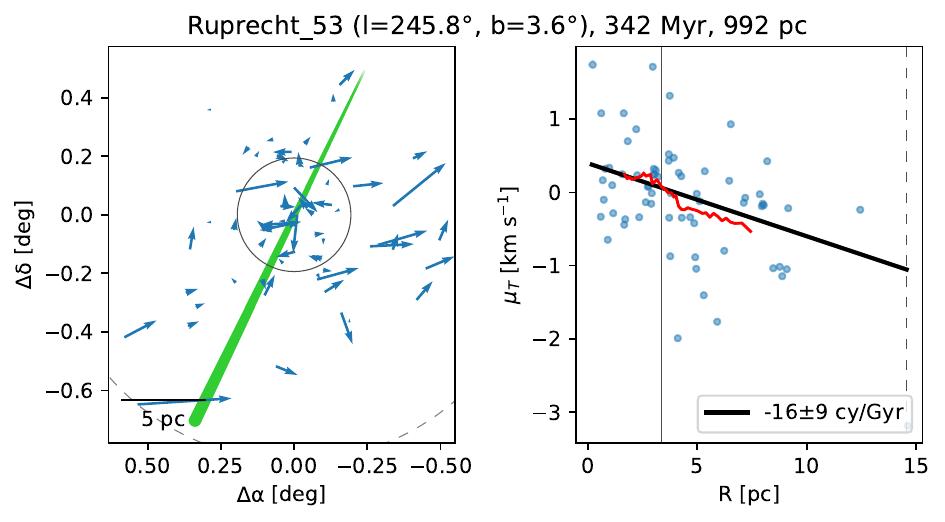}
    \caption{Diagnostic plots for PM based spinning clusters. All the subplots are similar to Figure~\ref{fig:comb_2099}.}
    \label{fig:comb_pm}
\end{figure*}

\begin{figure*}
    \centering
    \includegraphics[width=0.49\textwidth]{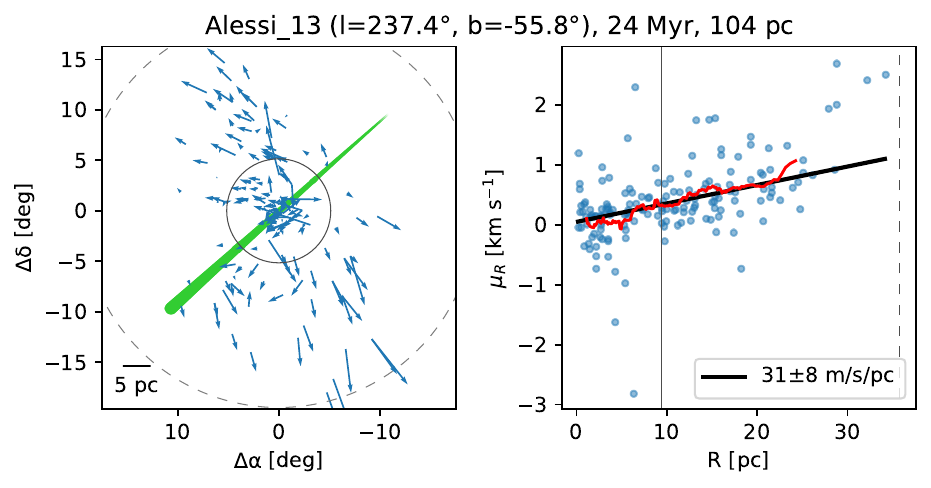}
    \includegraphics[width=0.49\textwidth]{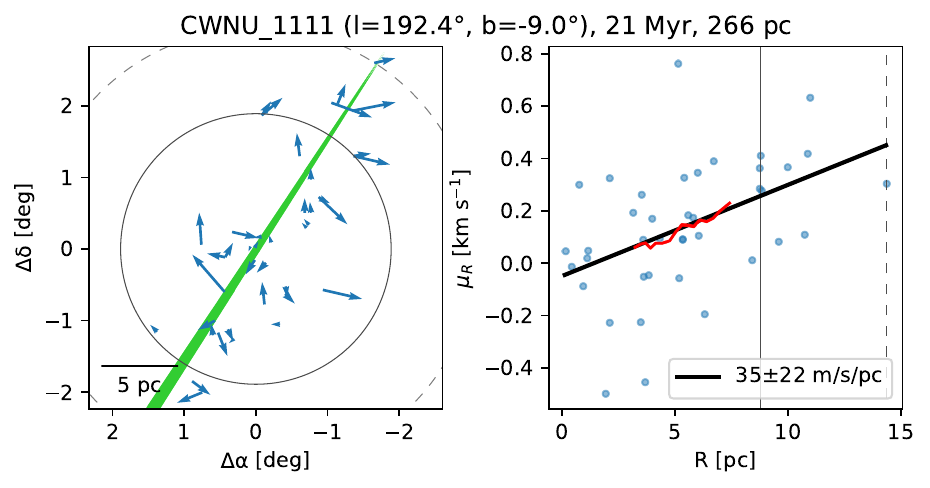}
    \includegraphics[width=0.49\textwidth]{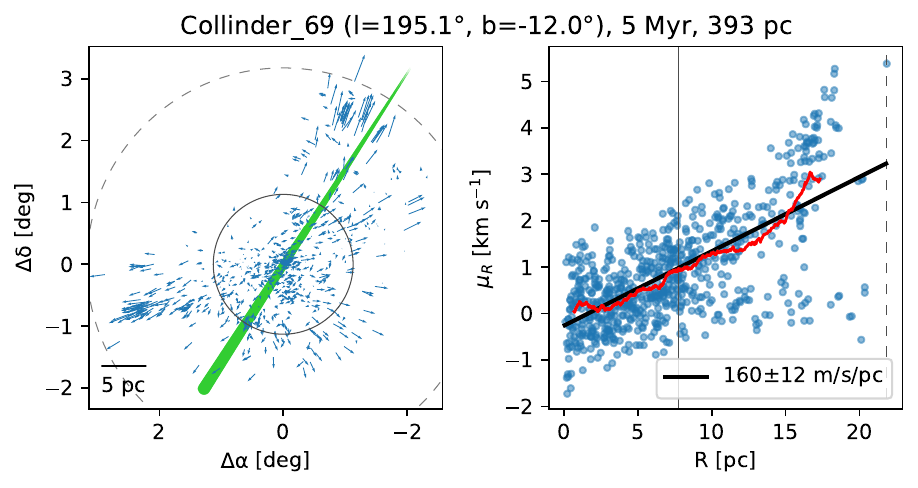}
    \includegraphics[width=0.49\textwidth]{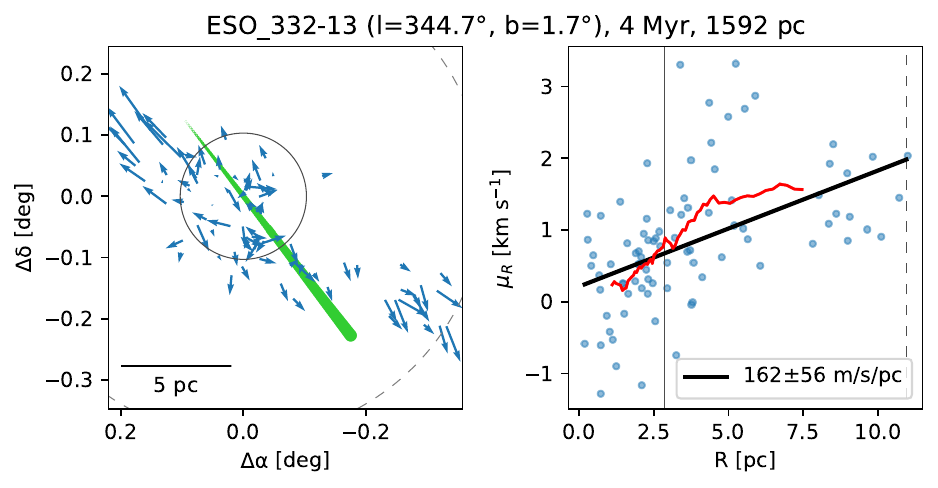}
    \includegraphics[width=0.49\textwidth]{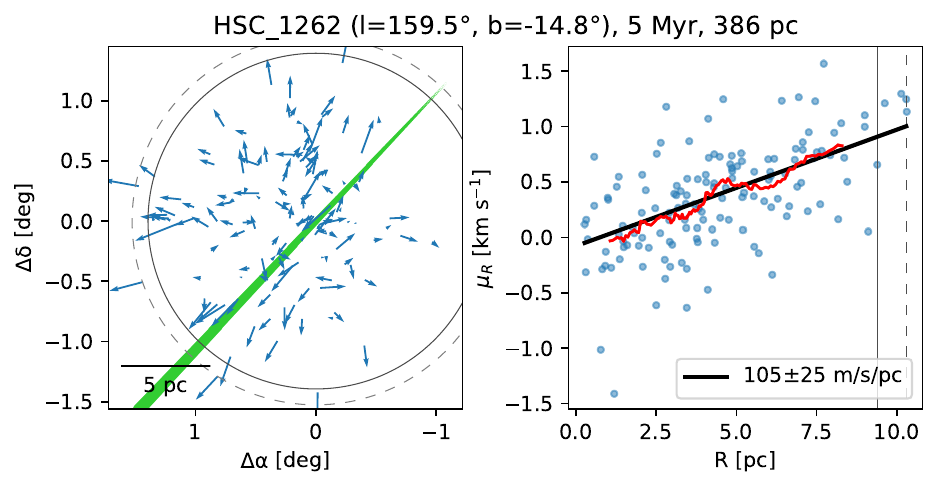}
    \includegraphics[width=0.49\textwidth]{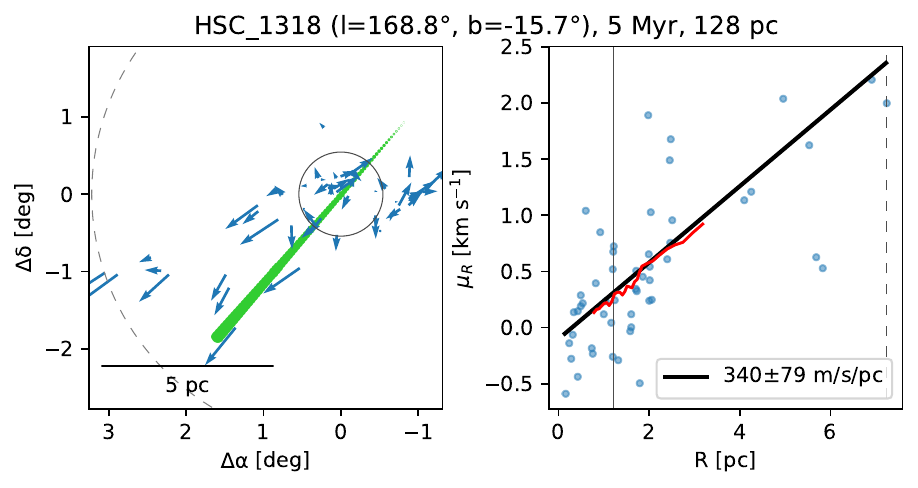}
    \includegraphics[width=0.49\textwidth]{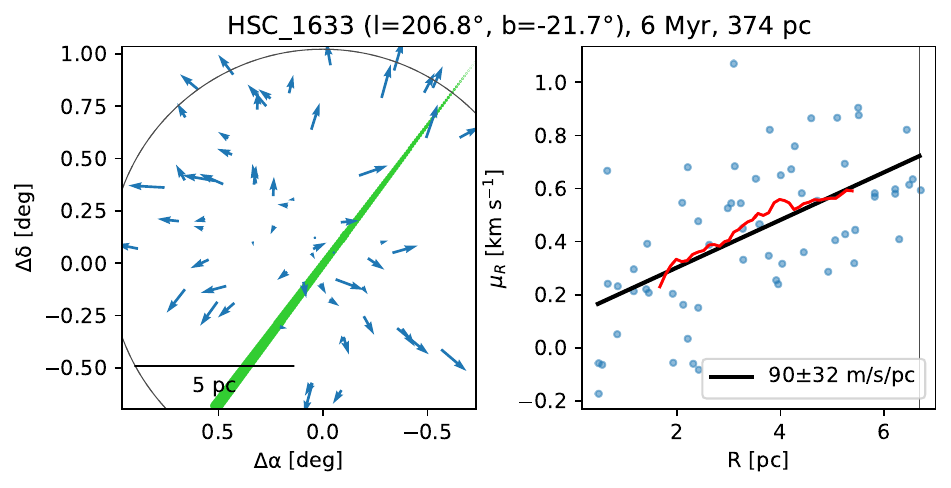}
    \includegraphics[width=0.49\textwidth]{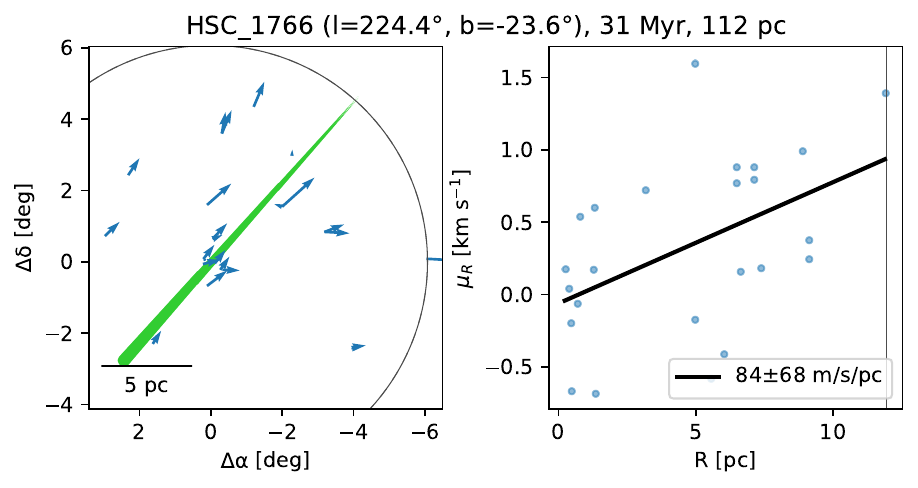}
    \includegraphics[width=0.49\textwidth]{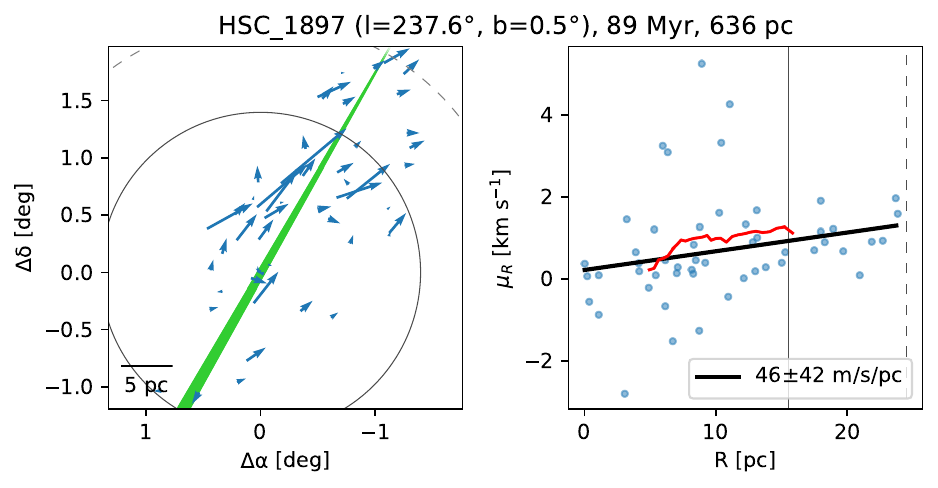}
    \includegraphics[width=0.49\textwidth]{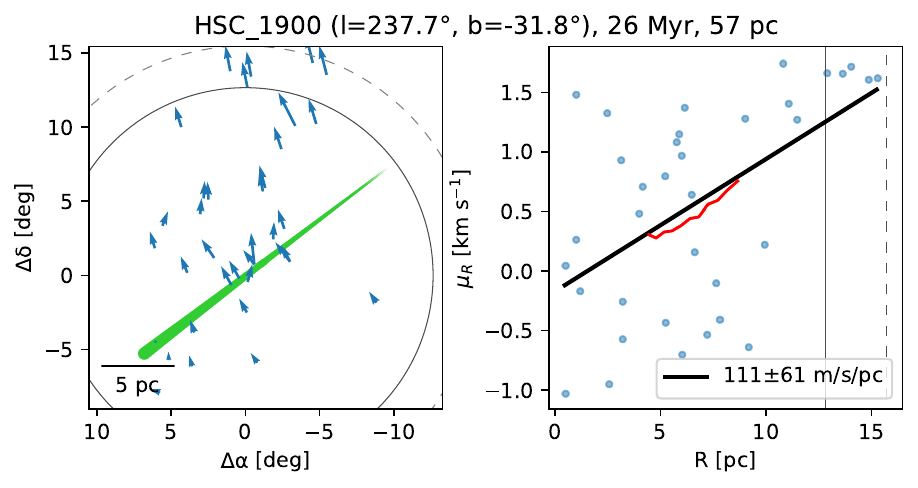}
    \caption{Diagnostic plots for expanding clusters. All the subplots are similar to Figure~\ref{fig:comb_2099}.}
    \label{fig:comb_expanding}
\end{figure*}

\renewcommand{\thefigure}{A.\arabic{figure} (Continued...)}
\addtocounter{figure}{-1}

\begin{figure*}
    \centering
    \includegraphics[width=0.49\textwidth]{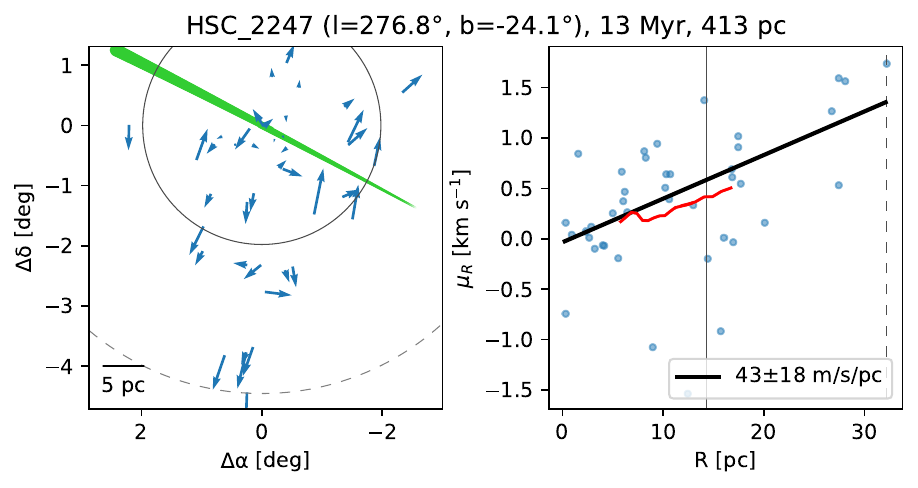}
    \includegraphics[width=0.49\textwidth]{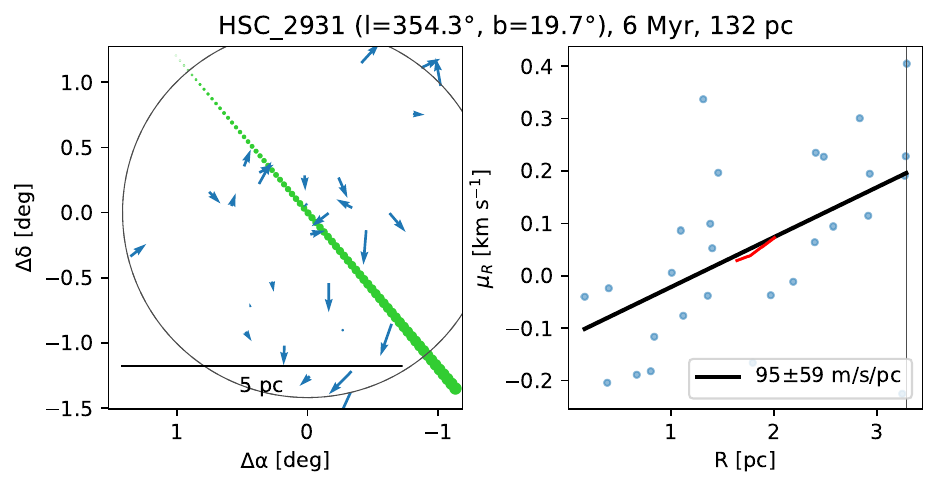}
    \includegraphics[width=0.49\textwidth]{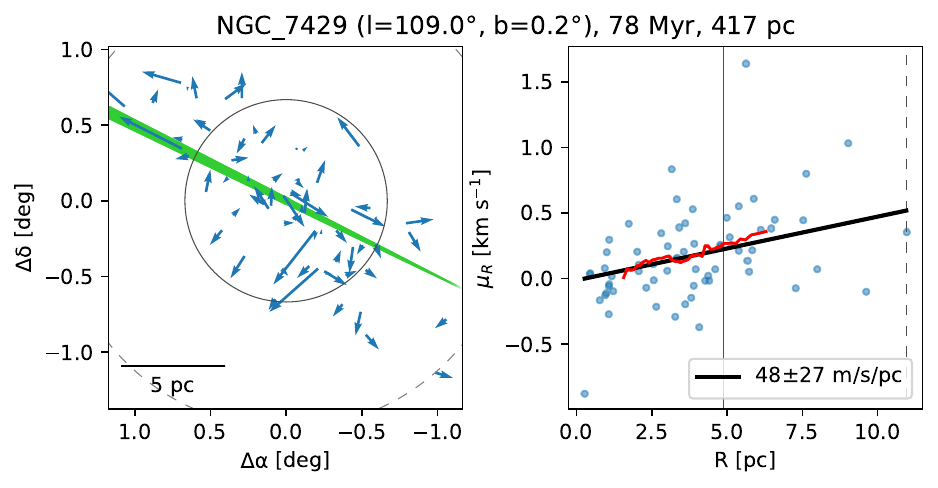}
    \includegraphics[width=0.49\textwidth]{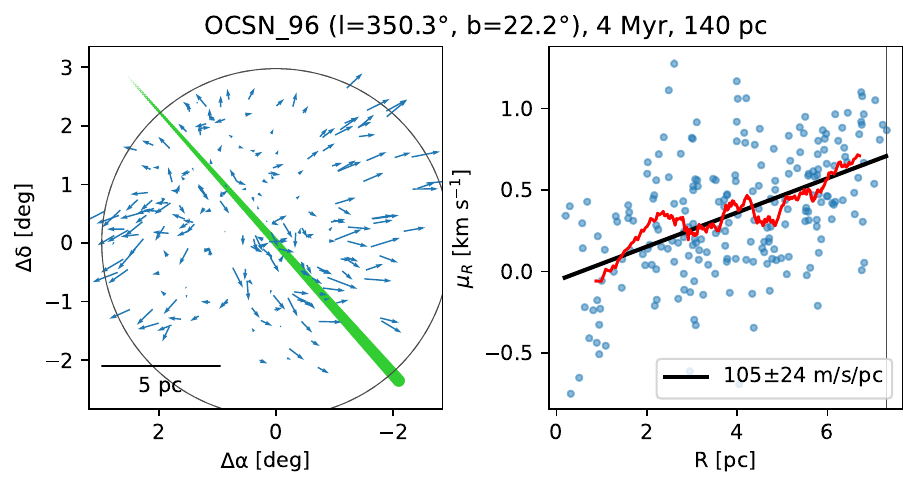}
    \includegraphics[width=0.49\textwidth]{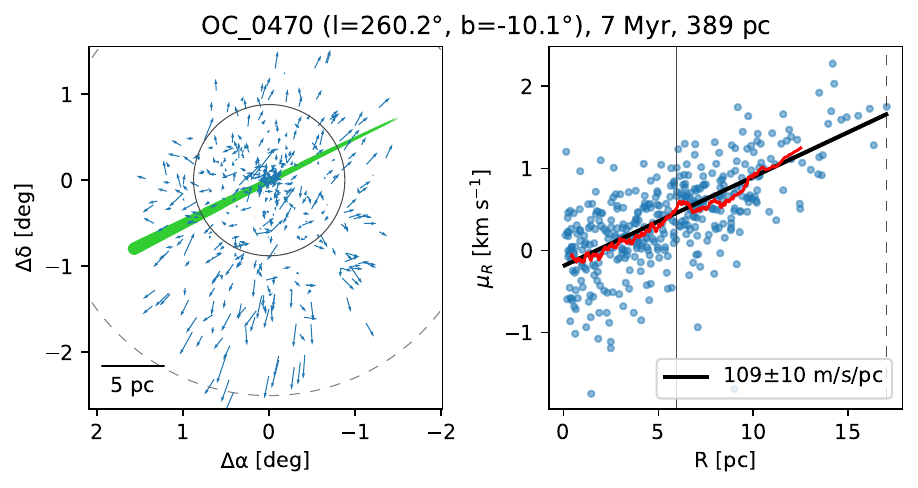}
    \includegraphics[width=0.49\textwidth]{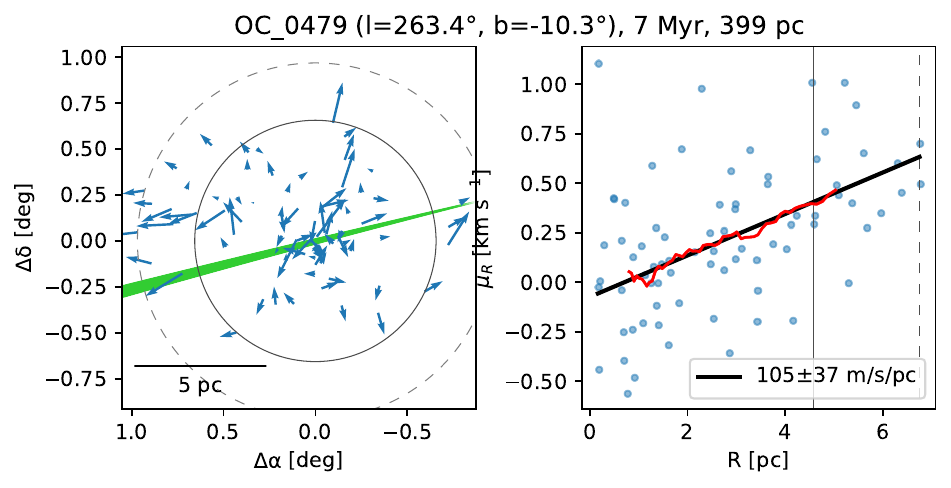}
    \includegraphics[width=0.49\textwidth]{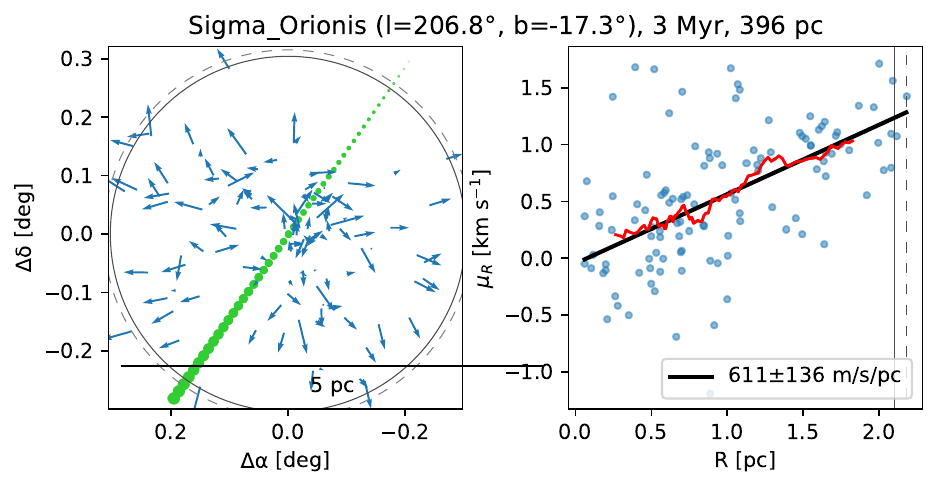}
    \includegraphics[width=0.49\textwidth]{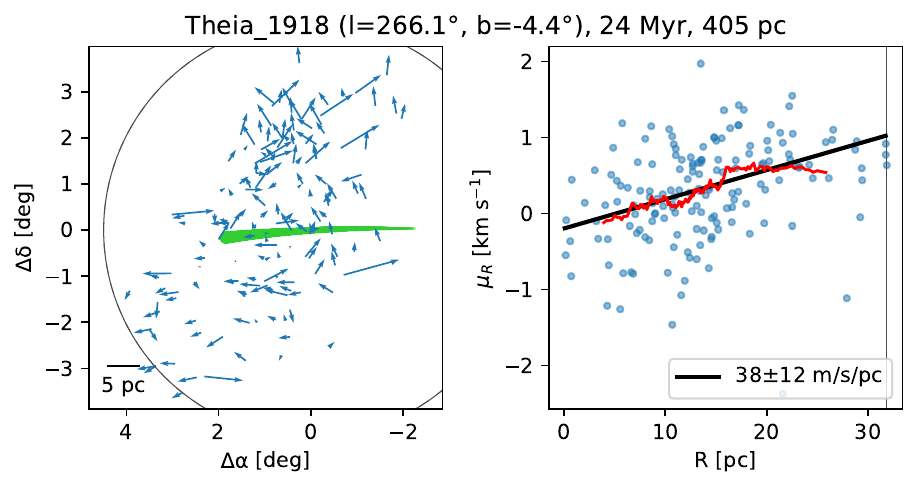}
    \caption{Diagnostic plots for expanding clusters. All the subplots are similar to Figure~\ref{fig:comb_2099}.}
\end{figure*}

\renewcommand{\thefigure}{A.\arabic{figure}}

\begin{figure*}
    \centering
    \includegraphics[width=0.49\textwidth]{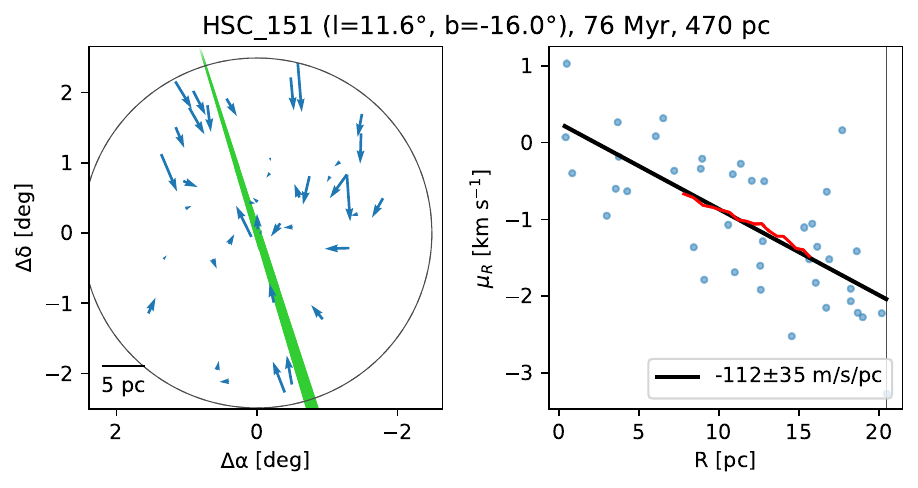}
    \includegraphics[width=0.49\textwidth]{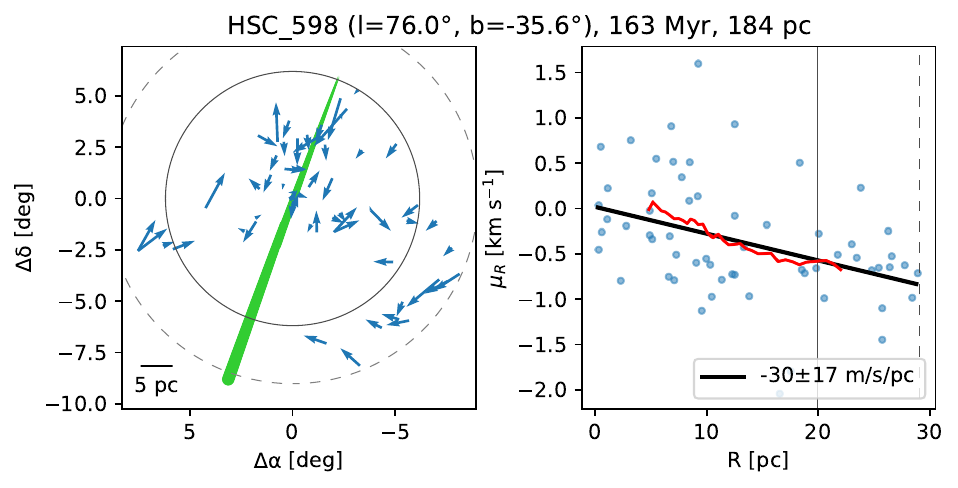}
    \includegraphics[width=0.49\textwidth]{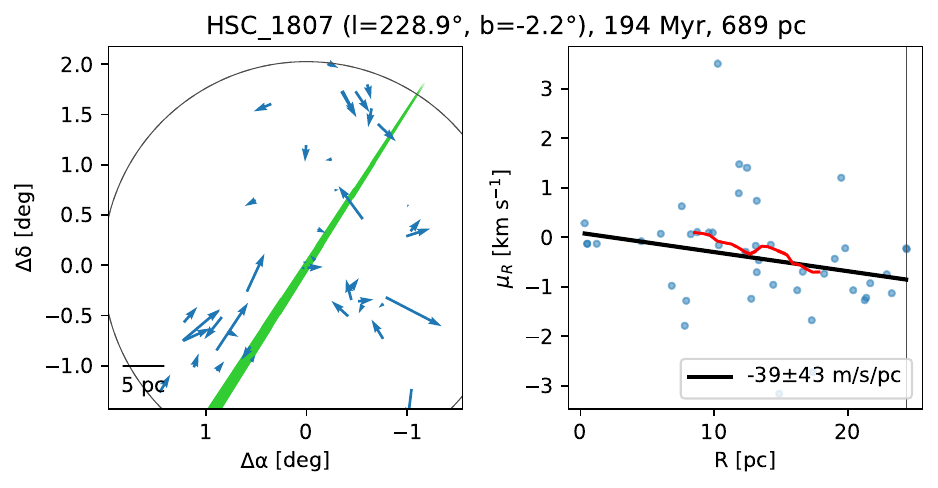}
    \caption{Diagnostic plots for contracting clusters. All the subplots are similar to Figure~\ref{fig:comb_2099}.}
    \label{fig:comb_contracting}
\end{figure*}

\end{appendix}

\end{document}